\documentclass[twocolumn,trackchanges]{aastex631}
\usepackage[utf8]{inputenc}
\usepackage{amsmath}
\usepackage{graphicx}
\usepackage{tabularx}
\usepackage{multirow}
\usepackage{etoolbox}
\makeatletter
\patchcmd{\ltx@foottext}{%
  .5\textwidth\advance\hsize-18pt}{%
  \linewidth\advance\hsize-1.8em%
}{}{}
\makeatother

\defcitealias{NG15}{NG15}
\defcitealias{NG15_gwb}{NG15\_GWB}
\defcitealias{NG15detchar}{NG15\_Noise}
\defcitealias{NG15_astro}{NG15\_SMBHBs}
\defcitealias{EPTA_data}{EPTA DR2}
\defcitealias{EPTA_noise}{EPTA\_Noise}
\defcitealias{3P+paper}{IPTA\_comp}

\newcommand\Comment{}

\shorttitle{Chromatic GP models for 6 NANOGrav pulsars}
\shortauthors{Larsen, Mingarelli, Hazboun et al.}

\begin{document}

%\linenumbers

\title{The NANOGrav 15 yr Data Set: Chromatic Gaussian Process Noise Models for Six Pulsars}
\author[0000-0001-6436-8216]{Bjorn Larsen}
\affiliation{Department of Physics, Yale University, New Haven, CT 06520, USA}
\author[0000-0002-4307-1322]{Chiara M. F. Mingarelli}
\affiliation{Department of Physics, Yale University, New Haven, CT 06520, USA}
\author[0000-0003-2742-3321]{Jeffrey S. Hazboun}
\affiliation{Department of Physics, Oregon State University, Corvallis, OR 97331, USA}
\author[0000-0003-2111-1001]{Aur\'elien Chalumeau}
\affiliation{Dipartimento di Fisica ``G. Occhialini", Universit{\'a} degli Studi di Milano-Bicocca, Piazza della Scienza 3, I-20126 Milano, Italy}
\affiliation{INFN, Sezione di Milano-Bicocca, Piazza della Scienza 3, I-20126 Milano, Italy}
\author[0000-0003-1884-348X]{Deborah C. Good}
\affiliation{Department of Physics and Astronomy, University of Montana, 32 Campus Drive, Missoula, MT 59812}
\author[0000-0003-1407-6607]{Joseph Simon}
\altaffiliation{NSF Astronomy and Astrophysics Postdoctoral Fellow}
\affiliation{Department of Astrophysical and Planetary Sciences, University of Colorado, Boulder, CO 80309, USA}
\author[0000-0001-5134-3925]{Gabriella Agazie}
\affiliation{Center for Gravitation, Cosmology and Astrophysics, Department of Physics, University of Wisconsin-Milwaukee,\\ P.O. Box 413, Milwaukee, WI 53201, USA}
\author[0000-0002-8935-9882]{Akash Anumarlapudi}
\affiliation{Center for Gravitation, Cosmology and Astrophysics, Department of Physics, University of Wisconsin-Milwaukee,\\ P.O. Box 413, Milwaukee, WI 53201, USA}
\author[0000-0003-0638-3340]{Anne M. Archibald}
\affiliation{Newcastle University, NE1 7RU, UK}
\author{Zaven Arzoumanian}
\affiliation{X-Ray Astrophysics Laboratory, NASA Goddard Space Flight Center, Code 662, Greenbelt, MD 20771, USA}
\author[0000-0003-2745-753X]{Paul T. Baker}
\affiliation{Department of Physics and Astronomy, Widener University, One University Place, Chester, PA 19013, USA}
\author[0000-0003-3053-6538]{Paul R. Brook}
\affiliation{Institute for Gravitational Wave Astronomy and School of Physics and Astronomy, University of Birmingham, Edgbaston, Birmingham B15 2TT, UK}
\author[0000-0002-6039-692X]{H. Thankful Cromartie}
\affiliation{National Research Council Postdoctoral Associate, National Academy of Sciences, Washington, DC 20001, USA resident at Naval Research Laboratory, Washington, DC 20375, USA}
\author[0000-0002-1529-5169]{Kathryn Crowter}
\affiliation{Department of Physics and Astronomy, University of British Columbia, 6224 Agricultural Road, Vancouver, BC V6T 1Z1, Canada}
\author[0000-0002-2185-1790]{Megan E. DeCesar}
\affiliation{George Mason University, resident at the Naval Research Laboratory, Washington, DC 20375, USA}
\author[0000-0002-6664-965X]{Paul B. Demorest}
\affiliation{National Radio Astronomy Observatory, 1003 Lopezville Rd., Socorro, NM 87801, USA}
\author[0000-0001-8885-6388]{Timothy Dolch}
\affiliation{Department of Physics, Hillsdale College, 33 E. College Street, Hillsdale, MI 49242, USA}
\affiliation{Eureka Scientific, 2452 Delmer Street, Suite 100, Oakland, CA 94602-3017, USA}
\author[0000-0001-7828-7708]{Elizabeth C. Ferrara}
\affiliation{Department of Astronomy, University of Maryland, College Park, MD 20742, USA}
\affiliation{Center for Research and Exploration in Space Science and Technology, NASA/GSFC, Greenbelt, MD 20771}
\affiliation{NASA Goddard Space Flight Center, Greenbelt, MD 20771, USA}
\author[0000-0001-5645-5336]{William Fiore}
\affiliation{Department of Physics and Astronomy, West Virginia University, P.O. Box 6315, Morgantown, WV 26506, USA}
\affiliation{Center for Gravitational Waves and Cosmology, West Virginia University, Chestnut Ridge Research Building, Morgantown, WV 26505, USA}
\author[0000-0001-8384-5049]{Emmanuel Fonseca}
\affiliation{Department of Physics and Astronomy, West Virginia University, P.O. Box 6315, Morgantown, WV 26506, USA}
\affiliation{Center for Gravitational Waves and Cosmology, West Virginia University, Chestnut Ridge Research Building, Morgantown, WV 26505, USA}
\author[0000-0001-7624-4616]{Gabriel E. Freedman}
\affiliation{Center for Gravitation, Cosmology and Astrophysics, Department of Physics, University of Wisconsin-Milwaukee,\\ P.O. Box 413, Milwaukee, WI 53201, USA}
\author[0000-0001-6166-9646]{Nate Garver-Daniels}
\affiliation{Department of Physics and Astronomy, West Virginia University, P.O. Box 6315, Morgantown, WV 26506, USA}
\affiliation{Center for Gravitational Waves and Cosmology, West Virginia University, Chestnut Ridge Research Building, Morgantown, WV 26505, USA}
\author[0000-0001-8158-683X]{Peter A. Gentile}
\affiliation{Department of Physics and Astronomy, West Virginia University, P.O. Box 6315, Morgantown, WV 26506, USA}
\affiliation{Center for Gravitational Waves and Cosmology, West Virginia University, Chestnut Ridge Research Building, Morgantown, WV 26505, USA}
\author[0000-0003-4090-9780]{Joseph Glaser}
\affiliation{Department of Physics and Astronomy, West Virginia University, P.O. Box 6315, Morgantown, WV 26506, USA}
\affiliation{Center for Gravitational Waves and Cosmology, West Virginia University, Chestnut Ridge Research Building, Morgantown, WV 26505, USA}
\author[0000-0003-1082-2342]{Ross J. Jennings}
\altaffiliation{NANOGrav Physics Frontiers Center Postdoctoral Fellow}
\affiliation{Department of Physics and Astronomy, West Virginia University, P.O. Box 6315, Morgantown, WV 26506, USA}
\affiliation{Center for Gravitational Waves and Cosmology, West Virginia University, Chestnut Ridge Research Building, Morgantown, WV 26505, USA}
\author[0000-0001-6607-3710]{Megan L. Jones}
\affiliation{Center for Gravitation, Cosmology and Astrophysics, Department of Physics, University of Wisconsin-Milwaukee,\\ P.O. Box 413, Milwaukee, WI 53201, USA}
\author[0000-0001-6295-2881]{David L. Kaplan}
\affiliation{Center for Gravitation, Cosmology and Astrophysics, Department of Physics, University of Wisconsin-Milwaukee,\\ P.O. Box 413, Milwaukee, WI 53201, USA}
\author[0000-0002-0893-4073]{Matthew Kerr}
\affiliation{Space Science Division, Naval Research Laboratory, Washington, DC 20375-5352, USA}
\author[0000-0003-0721-651X]{Michael T. Lam}
\affiliation{SETI Institute, 339 N Bernardo Ave Suite 200, Mountain View, CA 94043, USA}
\affiliation{School of Physics and Astronomy, Rochester Institute of Technology, Rochester, NY 14623, USA}
\affiliation{Laboratory for Multiwavelength Astrophysics, Rochester Institute of Technology, Rochester, NY 14623, USA}
\author[0000-0003-1301-966X]{Duncan R. Lorimer}
\affiliation{Department of Physics and Astronomy, West Virginia University, P.O. Box 6315, Morgantown, WV 26506, USA}
\affiliation{Center for Gravitational Waves and Cosmology, West Virginia University, Chestnut Ridge Research Building, Morgantown, WV 26505, USA}
\author[0000-0001-5373-5914]{Jing Luo}
\altaffiliation{Deceased}
\affiliation{Department of Astronomy \& Astrophysics, University of Toronto, 50 Saint George Street, Toronto, ON M5S 3H4, Canada}
\author[0000-0001-5229-7430]{Ryan S. Lynch}
\affiliation{Green Bank Observatory, P.O. Box 2, Green Bank, WV 24944, USA}
\author[0000-0001-5481-7559]{Alexander McEwen}
\affiliation{Center for Gravitation, Cosmology and Astrophysics, Department of Physics, University of Wisconsin-Milwaukee,\\ P.O. Box 413, Milwaukee, WI 53201, USA}
\author[0000-0001-7697-7422]{Maura A. McLaughlin}
\affiliation{Department of Physics and Astronomy, West Virginia University, P.O. Box 6315, Morgantown, WV 26506, USA}
\affiliation{Center for Gravitational Waves and Cosmology, West Virginia University, Chestnut Ridge Research Building, Morgantown, WV 26505, USA}
\author[0000-0002-4642-1260]{Natasha McMann}
\affiliation{Department of Physics and Astronomy, Vanderbilt University, 2301 Vanderbilt Place, Nashville, TN 37235, USA}
\author[0000-0001-8845-1225]{Bradley W. Meyers}
\affiliation{Department of Physics and Astronomy, University of British Columbia, 6224 Agricultural Road, Vancouver, BC V6T 1Z1, Canada}
\affiliation{International Centre for Radio Astronomy Research, Curtin University, Bentley, WA 6102, Australia}
\author[0000-0002-3616-5160]{Cherry Ng}
\affiliation{Dunlap Institute for Astronomy and Astrophysics, University of Toronto, 50 St. George St., Toronto, ON M5S 3H4, Canada}
\author[0000-0002-6709-2566]{David J. Nice}
\affiliation{Department of Physics, Lafayette College, Easton, PA 18042, USA}
\author[0000-0001-5465-2889]{Timothy T. Pennucci}
\affiliation{Institute of Physics and Astronomy, E\"{o}tv\"{o}s Lor\'{a}nd University, P\'{a}zm\'{a}ny P. s. 1/A, 1117 Budapest, Hungary}
\author[0000-0002-8509-5947]{Benetge B. P. Perera}
\affiliation{Arecibo Observatory, HC3 Box 53995, Arecibo, PR 00612, USA}
\author[0000-0002-8826-1285]{Nihan S. Pol}
\affiliation{Department of Physics and Astronomy, Vanderbilt University, 2301 Vanderbilt Place, Nashville, TN 37235, USA}
\author[0000-0002-2074-4360]{Henri A. Radovan}
\affiliation{Department of Physics, University of Puerto Rico, Mayag\"{u}ez, PR 00681, USA}
\author[0000-0001-5799-9714]{Scott M. Ransom}
\affiliation{National Radio Astronomy Observatory, 520 Edgemont Road, Charlottesville, VA 22903, USA}
\author[0000-0002-5297-5278]{Paul S. Ray}
\affiliation{Space Science Division, Naval Research Laboratory, Washington, DC 20375-5352, USA}
\author[0000-0003-4391-936X]{Ann Schmiedekamp}
\affiliation{Department of Physics, Penn State Abington, Abington, PA 19001, USA}
\author[0000-0002-1283-2184]{Carl Schmiedekamp}
\affiliation{Department of Physics, Penn State Abington, Abington, PA 19001, USA}
\author[0000-0002-7283-1124]{Brent J. Shapiro-Albert}
\affiliation{Department of Physics and Astronomy, West Virginia University, P.O. Box 6315, Morgantown, WV 26506, USA}
\affiliation{Center for Gravitational Waves and Cosmology, West Virginia University, Chestnut Ridge Research Building, Morgantown, WV 26505, USA}
\affiliation{Giant Army, 915A 17th Ave, Seattle WA 98122}
\author[0000-0001-9784-8670]{Ingrid H. Stairs}
\affiliation{Department of Physics and Astronomy, University of British Columbia, 6224 Agricultural Road, Vancouver, BC V6T 1Z1, Canada}
\author[0000-0002-7261-594X]{Kevin Stovall}
\affiliation{National Radio Astronomy Observatory, 1003 Lopezville Rd., Socorro, NM 87801, USA}
\author[0000-0002-2820-0931]{Abhimanyu Susobhanan}
\affiliation{Center for Gravitation, Cosmology and Astrophysics, Department of Physics, University of Wisconsin-Milwaukee,\\ P.O. Box 413, Milwaukee, WI 53201, USA}
\author[0000-0002-1075-3837]{Joseph K. Swiggum}
\altaffiliation{NANOGrav Physics Frontiers Center Postdoctoral Fellow}
\affiliation{Department of Physics, Lafayette College, Easton, PA 18042, USA}
\author[0000-0001-9678-0299]{Haley M. Wahl}
\affiliation{Department of Physics and Astronomy, West Virginia University, P.O. Box 6315, Morgantown, WV 26506, USA}
\affiliation{Center for Gravitational Waves and Cosmology, West Virginia University, Chestnut Ridge Research Building, Morgantown, WV 26505, USA}
\author[0000-0003-1361-7723]{David J. Champion}
\affiliation{Max-Planck-Institut f{\"u}r Radioastronomie, Auf dem H\"ugel 69, 53121, Bonn, Germany}
\author[0000-0002-1775-9692]{Isma\"el Cognard}
\affiliation{Laboratoire de Physique et Chimie de l'Environnement et de l'Espace, Universit\'e d’Orl\'eans/CNRS, 45071 Orl\'eans Cedex 02, France}
\affiliation{Observatoire Radioastronomique de Nan\c{c}ay, Observatoire de Paris, Universit\'e PSL, Universit\'e d’Orl\'{e}ans, CNRS, 18330 Nan\c{c}ay, France}
\author[0000-0002-9049-8716]{Lucas Guillemot}
\affiliation{Laboratoire de Physique et Chimie de l'Environnement et de l'Espace, Universit\'e d’Orl\'eans/CNRS, 45071 Orl\'eans Cedex 02, France}
\affiliation{Observatoire Radioastronomique de Nan\c{c}ay, Observatoire de Paris, Universit\'e PSL, Universit\'e d’Orl\'{e}ans, CNRS, 18330 Nan\c{c}ay, France}
\author[0000-0002-3407-8071]{Huanchen Hu}
\affiliation{Max-Planck-Institut f{\"u}r Radioastronomie, Auf dem H\"ugel 69, 53121, Bonn, Germany}
\author[0000-0001-5567-5492]{Michael J. Keith}
\affiliation{Jodrell Bank Centre for Astrophysics, Department of Physics and Astronomy, University of Manchester, Manchester M13 9PL, UK}
\author[0000-0002-2953-7376]{Kuo Liu}
\affiliation{Shanghai Astronomical Observatory, Chinese Academy of Sciences, 80 Nandan Road, Shanghai 200030, China}
\affiliation{Max-Planck-Institut f{\"u}r Radioastronomie, Auf dem H\"ugel 69, 53121, Bonn, Germany}
\author[0000-0002-2885-8485]{James W. McKee}
\affiliation{E.A. Milne Centre for Astrophysics, University of Hull, Cottingham Road, Kingston-upon-Hull, HU6 7RX, UK}
\affiliation{Centre of Excellence for Data Science, Artificial Intelligence and Modelling (DAIM), University of Hull, Cottingham Road, Kingston-upon-Hull, HU6 7RX, UK}
\author[0000-0002-4140-5616]{Aditya Parthasarathy}
\affiliation{ASTRON, Netherlands Institute for Radio Astronomy, Oude Hoogeveensedijk 4, 7991 PD Dwingeloo, The Netherlands}
\affiliation{Anton Pannekoek Institute for Astronomy, University of Amsterdam, Science Park 904, 1098 XH Amsterdam, The Netherlands}
\affiliation{Max-Planck-Institut f{\"u}r Radioastronomie, Auf dem H\"ugel 69, 53121, Bonn, Germany}
\author[0000-0002-1806-2483]{Delphine Perrodin}
\affiliation{INAF-Osservatorio Astronomico di Cagliari, via della Scienza 5, 09047 Selargius, Italy}
\author[0000-0001-5902-3731]{Andrea Possenti}
\affiliation{INAF-Osservatorio Astronomico di Cagliari, via della Scienza 5, 09047 Selargius, Italy}
\author[0000-0002-8452-4834]{Golam M. Shaifullah}
\affiliation{Dipartimento di Fisica ``G. Occhialini", Universit{\'a} degli Studi di Milano-Bicocca, Piazza della Scienza 3, I-20126 Milano, Italy}
\affiliation{INFN, Sezione di Milano-Bicocca, Piazza della Scienza 3, I-20126 Milano, Italy}
\affiliation{INAF-Osservatorio Astronomico di Cagliari, via della Scienza 5, 09047 Selargius, Italy}
\author[0000-0002-3649-276X]{Gilles Theureau}
\affiliation{Laboratoire de Physique et Chimie de l'Environnement et de l'Espace, Universit\'e d’Orl\'eans/CNRS, 45071 Orl\'eans Cedex 02, France}
\affiliation{Observatoire Radioastronomique de Nan\c{c}ay, Observatoire de Paris, Universit\'e PSL, Universit\'e d’Orl\'{e}ans, CNRS, 18330 Nan\c{c}ay, France}
\affiliation{Laboratoire Univers et Th\'eories LUTh, Observatoire de Paris, Universit\'e PSL, CNRS, Universit\'e de Paris, 92190 Meudon, France}
\correspondingauthor{Bjorn Larsen}
\email{bjorn.larsen@yale.edu}

\begin{abstract}
Pulsar timing arrays (PTAs) are designed to detect low-frequency gravitational waves (GWs).
GWs induce achromatic signals in PTA data, meaning that the timing delays do not depend on radio-frequency.
However, pulse arrival times are also affected by radio-frequency dependent ``chromatic’’ noise from sources such as dispersion measure (DM) and scattering delay variations.
Furthermore, the characterization of GW signals may be influenced by the choice of chromatic noise model for each pulsar.
To better understand this effect, we assess if and how different chromatic noise models affect achromatic noise properties in each pulsar.
The models we compare include existing DM models used by NANOGrav and noise models used for the European PTA Data Release 2 (EPTA DR2).
We perform this comparison using a subsample of six pulsars from the NANOGrav 15 yr data set, selecting the same six pulsars as from the EPTA DR2 six-pulsar dataset.
We find that the choice of chromatic noise model noticeably affects the achromatic noise properties of several pulsars.
This is most dramatic for PSR J1713+0747, where the amplitude of its achromatic red noise lowers from $\log_{10}A_{\text{RN}} = -14.1^{+0.1}_{-0.1}$ to $-14.7^{+0.3}_{-0.5}$, \Comment{and the spectral index broadens from $\gamma_{\text{RN}} = 2.6^{+0.5}_{-0.4}$ to $\gamma_{\text{RN}} = 3.5^{+1.2}_{-0.9}$}.
We also compare each pulsar’s noise properties with those inferred from the EPTA DR2, using the same models.
From the discrepancies, we identify potential areas where the noise models could be improved.
These results highlight the potential for custom chromatic noise models to improve PTA sensitivity to GWs.
\end{abstract}


\section{Introduction}
\label{sec:intro}
Pulsar timing arrays (PTAs) are designed to detect low-frequency gravitational waves (GWs).
GWs induce small shifts in pulse times-of-arrival (TOAs), which can be measured using a network of millisecond pulsars \citep{Sahzin1978, Detweiler1979, HD1983}.
Several collaborations around the globe carry out searches for GW signals using PTAs, including the North American Nanohertz Observatory for Gravitational waves (NANOGrav, \citealt{Ransom2019}), the European PTA (EPTA, \citealt{Desvignes2016}), the Parkes PTA (PPTA, \citealt{Manchester2013}), the Indian PTA (InPTA, \citealt{Joshi2018}), the Chinese PTA (CPTA, \citealt{Lee2016}), and the MeerKAT PTA (MPTA, \citealt{Miles2023}). Together, NANOGrav, EPTA, PPTA, and InPTA form the International PTA (IPTA, e.g., \citealt{Verbiest2016, IPTADR2}).

Evidence for a stochastic GW background (GWB) at nanohertz frequencies has recently been presented by NANOGrav (\citealt{NG15_gwb}, henceforth \citetalias{NG15_gwb}), EPTA + InPTA \citep{EPTA_gwb}, PPTA \citep{PPTA_gwb}, and CPTA \citep{CPTA2023}, with varying levels of significance but broadly consistent properties across data sets (\citealt{3P+paper}, henceforth \citetalias{3P+paper}).
This evidence is based on the presence of a time-correlated, low-frequency (``red'') noise process common to, and spatially correlated between, all pulsars across the sky.
These spatial correlations follow the Hellings and Downs (HD) curve, which is the definitive signature of an isotropic GWB \citep{HD1983}.

Among the next major milestones in PTA science is to identify and characterize the source of the GWB, which may be astrophysical, cosmological, or a combination of both \citep{Burke-Spolaor2019, MingarelliCasey-Clyde2022}.
A likely source of the GWB is the incoherent superposition of GWs produced by hundreds of thousands of slowly inspiralling supermassive black hole binaries (SMBHBs, e.g., \citealt{RajagopalRomani1995, Kelley2017, NG15_astro}, henceforth \citetalias{NG15_astro}).
More exotic sources of the background have also been proposed, such as early-universe phase transitions, cosmic strings, and relic GWs from inflation (e.g., \citealt{CapriniFigueroa2018, NG15_new_physics, Vagnozzi2023}).

Each potential source may be distinguished using the shape of the GWB spectrum inferred from PTA timing residuals \citep{Lasky+2016, Kaiser2022}.
For instance, the simplest analytic model of the GWB from SMBHB populations predicts a power-law timing residual spectrum (equation~\ref{eq:rn_pl}) with a spectral index $\gamma_{\text{GWB}} = 13/3$ \citep{Phinney2001}.
However, deviations from this simplified spectrum may result e.g., from more complicated models of SMBHB evolution (\citealt{Sesana2013, KocsisSesana2011}; \citetalias{NG15_astro}), discreteness of the SMBHB population (\citealt{Sesana2008}; \citetalias{NG15_astro}), or resolvable single sources within the PTA data set \citep{Becsy+2023}.

In order to measure the GWB spectrum as accurately as possible, it is important to account for different sources of noise affecting individual pulsars.
Specifically, either overfitting \citep{HazbounSimon2020} or underfitting \citep{Hazboun2020, Zic2022} for pulsar noise processes may bias inferences of a common \emph{uncorrelated} red noise (CURN) process, which encodes the spectrum of the GWB without including interpulsar HD correlations \citep{NG12p5_gwb, Romano+2021, Taylor2022}.
A promising approach to accurately model pulsar noise, first employed by \citet{Lentati2016}, is the creation of custom noise models for each pulsar using Bayesian model selection methods.
Recently, \citet{PPTA_noise} found use of custom pulsar noise models to significantly influence the recovered spectral characteristics of the CURN in PPTA DR3 \citep{PPTA_data}.
Conversely, \citet{Chalumeau+2022} found that custom pulsar noise models have a minimal effect on both the spectral characterization of the CURN and the detection statistics for HD correlations using the EPTA DR2 six pulsar data set \citep{Chen+2021}.
These differences suggest the importance of custom pulsar noise models for GWB analyses may vary depending on the properties of the data set.

We turn our attention now to the NANOGrav 15 yr data set (\citealt{NG15}, henceforth \citetalias{NG15}).
\citetalias{NG15_gwb, NG15_astro} compared the inferred GWB spectral parameters when changing the dispersion measure (DM) model applied to all pulsars in \citetalias{NG15}, where DM variations introduce \emph{chromatic} (radio frequency dependent) timing noise.
The choice of DM model in \citetalias{NG15} was found to affect spectral parameter inferences of the GWB, with a slightly higher $A_{\text{GWB}}$ and lower $\gamma_{\text{GWB}}$ predicted using the standard DM model, ``DMX,'' than the alternative model, ``DMGP''.
While the effect is minor (the 68\% credible regions of the 2D GWB posteriors overlap using both models, \citetalias{NG15_gwb}), the choice of model could still be consequential for astrophysical inferences.
It is therefore important to investigate the effect of these different models on a deeper level.

\begin{table*}[ht!]
    %\vspace{-10.cm}
    \centering
    %\vspace{-0.5\baselineskip}
    \begin{tabularx}{\linewidth}{c | c | l}
        \hline\hline
        \textbf{Category} & \textbf{Term} & \textbf{Definition} \\ \hline
        %\textbf{Term} & \textbf{Definition} \\
        \multirow{3}{2cm}[1.8mm]{\begin{tabular}{c} Acronyms \\ (General) \end{tabular}} & PTA, TOA, GWB & Pulsar timing array, Time of arrival, Gravitational wave background \\
        & GP, MCMC & Gaussian process, Markov chain Monte Carlo \\
        & PSD, \citetalias{NG15} & Power Spectral Density, NANOGrav 15 yr dataset \citep{NG15} \\ \hline
        & DM, $\Delta$DM & Dispersion measure, Deviation from fiducial DM value \\
        & DMX/DMGP & Piecewise-constant model/Fourier-basis GP model for DM estimation \\
        Acronyms & ISM, SW & Interstellar medium, Solar wind \\
        (Noise processes) & ARN, RN, WN & Achromatic red noise, red noise, white noise \\
        & \texttt{DMX}/\texttt{DMGP}/ & Labels full pulsar noise model using DMX/DMGP/ \\
        & \texttt{CustomGP} & \;\;\;\; DMGP plus additional chromatic terms \\ \hline
        & $N_{\text{ToA}}$, $T_{\text{psr}}$ & Number of TOAs, Total observation timespan of pulsar \\
        & $\Delta t$, $\delta t$ & Time delay (generic), Estimated time delay (Appendix~\ref{appendix:chromatic_modeling} only) \\
        Symbols & $\delta\vec{t}$, $\rho$ & Timing residual vector (Section~\ref{sec:GP models} only), Timing residual power \\
        (General) & $\vec{b}$/$\vec{\eta}$, $N_{\text{freqs}}$ & GP coefficient/hyperparameter vector, Number of frequencies in GP Fourier basis \\
        & $\mathcal{B}$, $\mathcal{N}$/$\mathcal{U}$ & Bayes factor, Normal/Uniform distribution \\
        & $E_{1,2}$ & 1st and 2nd exponential timing events in PSR J1713+0747 \\ \hline
        & $A$/$\gamma$/$f$ & Spectral amplitude/index/frequency \\
        Symbols & $\nu$, $\chi$ & Radio frequency, Chromatic radio-frequency scaling index \\
        (Model parameters) & $n_{\text{Earth}}$ & Estimated local electron number density \\
        & $A_{\text{E}}$/$\tau_{\text{E}}$/$t_{\text{E}}$ & Amplitude/timescale/initial time of decaying exponential in PSR J1713+0747 \\
        \hline\hline
    \end{tabularx}
    \caption{\Comment{Definitions of terms and parameters commonly used throughout this work.}}
    \label{tab:glossary}
    \vspace{-0.5\baselineskip}
\end{table*}

Here we investigate how the choice of chromatic noise model affects the achromatic red noise (ARN) in six pulsars from \citetalias{NG15}, noting that a GWB signal contributes to a component of the ARN in each pulsar.
We specifically compare the following three models: 1) the standard noise model used by NANOGrav (\texttt{DMX}), 2) a noise model using Gaussian Processes (GPs) for DM variations (\texttt{DMGP}), and 3) a new noise model using GPs to account for additional chromatic noise alongside DM (\texttt{CustomGP}), similar to the models used in EPTA DR2 (\citealt{EPTA_noise}, henceforth \citetalias{EPTA_noise}).
We select PSRs J0613$-$0200, J1012+5307, J1600$-$3053, J1713+0737, J1744$-$1134, and J1909$-$3744 as our focus for this study since they are the same pulsars from the EPTA DR2 six-pulsar data set \citep{Chalumeau+2022}.
As a consistency check, we compare each pulsar's noise properties as inferred under the \texttt{CustomGP} model using both \citetalias{NG15} and EPTA DR2.
Since the astrophysical noise in each pulsar ought to be consistent regardless of the data set \citepalias{3P+paper}, we use the inconsistencies to identify potential improvements to these noise models.

The paper is laid out as follows.
In section~\ref{sec:data} we describe the data used here.
In section~\ref{sec:noise_budget} we present details on relevant noise processes affecting single pulsars.
In section~\ref{sec:GP models} we describe the GP models we use.
In section~\ref{sec:results} we present our results, starting with an overview of how (and if) each pulsar's ARN changes as a function of the noise model, followed by a pulsar-by-pulsar noise breakdown including our comparisons with EPTA DR2.
Finally, in section~\ref{sec:discussion} we discuss our results and provide recommendations for future analyses.
Table~\ref{tab:glossary} shows the acronyms and symbols used in this paper.

\section{Data}
\label{sec:data}

\subsection{The NANOGrav 15 yr data set}

The NANOGrav 15 yr data set \citepalias{NG15} contains observations of 68 millisecond pulsars with timespans ranging from 3 to 15 years.
\citetalias{NG15} is comprised of observations from three radio observatories: the Green Bank Telescope (GBT), the Arecibo Observatory (AO), and the Very Large Array (VLA).
All six pulsars studied here are observed by the GBT.
PSR J1713+0747 includes additional observations from the AO, while PSRs J1600$-$3053, J1713+0747, and J1909$-$3744 also include observations from the VLA.
The observations were collected, reduced, and analyzed to produce a best fit timing model, a set of narrowband and wideband TOAs, and a configuration file for each pulsar \citepalias{NG15}.
Here we use the narrowband TOAs, which are derived from many subbands of the radio observing bands.
\citetalias{NG15} uses the JPL DE440 solar system ephemeris \citep{Park2021} and the TT(BIPM2019) timescale in order to correct observatory and terrestrial clocks to an inertial reference frame at the solar system barycenter.

\subsection{EPTA DR2}

\begin{figure}[h!]
    %\vspace{-0.5\baselineskip}
    \centering
    \includegraphics[width=\linewidth]{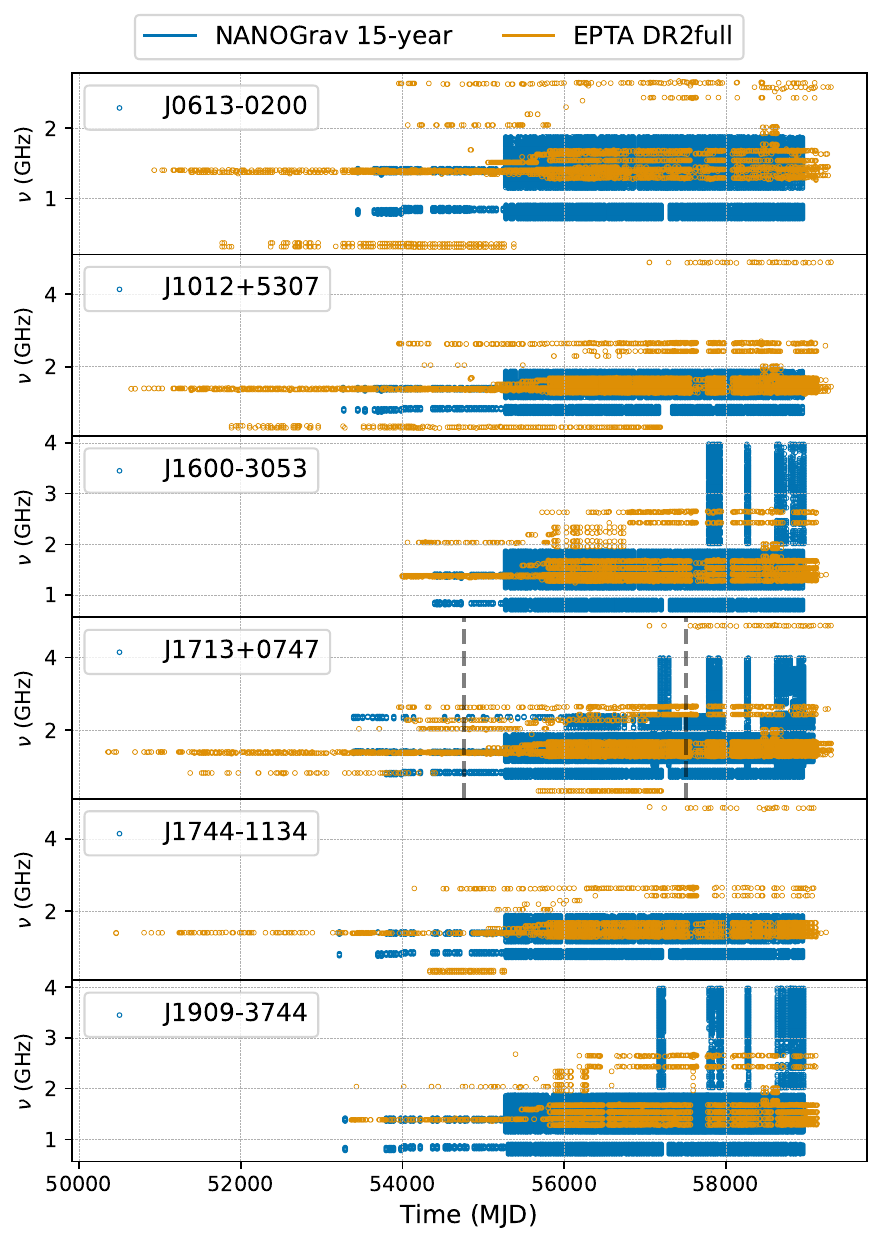}
    %\vspace{-0.5\baselineskip}
    \caption{TOAs for six pulsars from the NANOGrav 15 yr data set \citep{NG15} and EPTA DR2 \citep{EPTA_data}, visualized as a function of time and radio frequency.
    The radio frequencies used to collect TOAs are important for determining how well each pulsar's chromatic noise can be constrained.
    Dashed lines are used to mark the times of two known chromatic timing events in PSR J1713+0747 \citep{Lam2018}.}
    \label{fig:datasets}
    %\vspace{-0.5\baselineskip}
\end{figure}

\citetalias{EPTA_data} \citep{EPTA_data} was first made up of 6 millisecond pulsars \citep{Chen+2021} and later upgraded with more pulsars for a total of 25 millisecond pulsars and a maximum timespan of 24 years.
\citetalias{EPTA_data} is comprised of observations from six radio telescopes: the Effelsberg Radio Telescope (EFF), Lovell Telescope (LT), Mark II Telescope (MK2), Nançay Radio Telescope (NRT), Westerbork Synthesis Radio Telescope (WSRT), and the Sardinia Radio Telescope (SRT).
All telescopes are also used in tied-array mode to make observations as the Large European Array for Pulsars (LEAP, \citealt{Bassa2016}).
PSR J1909$-$3744 is only observed by the NRT and SRT due to its lower declination, but the remaining five pulsars are observed by all six telescopes. 
The \citetalias{EPTA_data} observations undergo data reduction, combination, outlier, and timing analyses to produce a final timing model and set of narrowband TOAs for each pulsar. 
\citetalias{EPTA_data} uses the JPL DE440 solar system ephemeris (same as \citetalias{NG15}) and the TT(BIPM2021) timescale.

\citet{EPTA_data} also presents multiple dataset versions, one of which excluded TOAs which had not undergone coherent dedispersion and another which included TOAs from InPTA DR1 \citep{Tarafdar+2022}.
While there are differences between the noise properties inferred using each dataset (see \citetalias{EPTA_noise}), we found the differences were not substantial enough to change the conclusions of our comparison with \citetalias{NG15}.
As such, we compare our results to \citetalias{EPTA_data}.

\subsection{Differences Between Data Sets}

Figure~\ref{fig:datasets} shows each pulsar's TOAs from \citetalias{NG15} and \citetalias{EPTA_data}, visualized as a function of time and frequency.
\citetalias{NG15} has fairly consistent multi-frequency coverage over time and a low frequency floor of 724 MHz for all six pulsars.
Since \citetalias{EPTA_data} has a longer timespan and is comprised of more telescopes, its level of radio frequency coverage is more varied. 
In particular, \citetalias{EPTA_data} includes TOAs down to 323 MHz from WSRT for PSRs J0613$-$0200, J1012+5307, J1713+0747, and J1744$-$1134, although not across the entire timespan.
These differences in radio-frequency coverage will become relevant when accounting for discrepancies in chromatic noise characterization between the two data sets, since chromatic processes induced from the interstellar medium introduce larger delays at low radio frequencies.

\section{Single Pulsar Noise Budget}
\label{sec:noise_budget}
Here we discuss various sources of noise relevant to millisecond pulsars, which form the basis of \citetalias{NG15}.
Most pulsars experience rotational irregularities which manifest as spin noise, an ARN process \citep{Verbiest2009, CordesShannon2010, ShannonCordes2010}.
Spin noise is found to be small in millisecond pulsars due to their very small spin frequency derivatives, with notable exceptions such as PSR B1937+21 \citep{ShannonCordes2010, NG15detchar}. 
Nevertheless, accounting for spin noise is very important since the GWB also manifests as an ARN process in single pulsars.
\citet{NG15detchar} (henceforth \citetalias{NG15detchar}) finds that 12 pulsars in \citetalias{NG15} still have significant detections of ARN in the presence of a GWB signal.
One of these 12, PSR J1012+5307, is among the six we include here.
The remaining five pulsars display significant ARN which does \emph{not} persist in the presence of a GWB signal using \citetalias{NG15}, i.e., these pulsars currently show little evidence for both intrinsic spin noise and a GWB signal \citepalias{NG15detchar}.
\Comment{As such, changes to these five pulsars' ARN properties could impact GWB inferences.}

To maximize PTA sensitivity to GWs, it is also important to account for \emph{chromatic} noise.
A major source of chromatic noise comes from dispersion measure (DM) variations \citep{RankinRoberts1971, You2007, Jones2017}. 
DM is defined as the integrated electron column density between the Earth and the pulsar,
\begin{eqnarray}
    \text{DM} = \int_0^dn_e(l)dl,
\end{eqnarray}
where $n_e$ is the free electron density, $l$ defines the Earth-pulsar line of sight, and $d$ is the distance to the pulsar \citep{Lorimerkramer2012}.
DM may undergo linear, annual, and/or stochastic variations due to the changing line of sight through the ionized interstellar medium (ISM, \citealt{Lam2016, Jones2017}) and the solar wind \citep{Madison2019, Tiburzi+2021}. 
DM introduces the following timing delay,
\begin{eqnarray}
    \label{eq:DM_delay}
    \Delta t_{\text{DM}} &=& \frac{e^2}{2\pi m_ec}\frac{\text{DM}}{\nu^2} \\
    &=& (4.15\text{ $\mu$s})\left(\frac{\text{DM}}{10^{-3}\text{ pc cm$^{-3}$}}\right)\left(\frac{1\text{ GHz}}{\nu}\right)^2,
\end{eqnarray}
where $\nu$ is the radio frequency of the pulse.
Millisecond pulsars from \citetalias{NG15} have \Comment{peak-to-peak DM variations ranging from} $\Delta$DM $ \sim 10^{-4} - 10^{-2}$ pc cm$^{-3}$ over the full data timespan \citep{NG15}.
As such, the time delays introduced by DM variations often dominate over ARN processes in millisecond pulsars. 

The $\nu^{-2}$ dependence in equation~\ref{eq:DM_delay} allows DM variations to be decoupled from other processes affecting pulsar timing.
However, a number of factors can systematically bias DM measurements, including asynchronous measurements across radio frequency bands \citep{Lam2015}, finite observing bandwidths \citep{Sosa2024}, or the presence of additional chromatic effects \citepalias{NG15detchar}.
Inaccurate DM values may result in ARN with a low spectral index $\gamma_{\text{RN}}$ (\citealt{CordesShannon2010, Lam2015}; \citetalias{NG15detchar}), which may reduce the pulsar's sensitivity to GW signals.

A secondary source of chromatic noise is interstellar scattering, which results from frequency-dependent refraction of radio pulses through an inhomogeneous ISM \citep{CordesRickett1998, Hemberger2008, Lorimerkramer2012}.
\Comment{The first order effect of scattering is to delay the TOA} by $\Delta t \propto \nu^{-4}$ for a Gaussian inhomogeneity \citep{Lang1971} or $\Delta t \propto \nu^{-4.4}$ for a Kolmogorov-turbulent medium \citep{Romani1986}.
However, this frequency scaling may vary more substantially depending on the geometry of the scattering medium, \Comment{with observed scalings ranging from $\nu^{-0.7}$ to $\nu^{-5.6}$ for different pulsar lines of sight \citep{Lewandowski2015, Turner2021}, and scalings predicted as high as $\nu^{-6.4}$ from simulations \citep{ShannonCordes2017}}. 
While noise from time-variable scattering is not expected to affect pulsar timing as strongly as DM variations, many high-DM pulsars have been observed by NANOGrav with large scattering tails (\citealt{Alam2021_wb}; \citetalias{NG15detchar}).
Unmitigated scattering variations \Comment{may be absorbed as excess WN, DM noise, ARN, or some combination thereof} (\citealt{Lentati2016, Shapiro-Albert2021}; \citetalias{NG15detchar}).

\Comment{Changes to the pulse profile itself also present a potential source of noise in millisecond pulsars.
These may result e.g., from polarization calibration errors, scatter-broadening of the pulse profile, or intrinsic changes in the pulsar magnetosphere \citepalias{NG15detchar}.
Frequency-dependence of pulse profiles is already accounted for in timing analyses using log-polynomial functions in frequency, parameterized by FD (``frequency-dependent'') parameters \citepalias{NG15}.
However, these do not have built in time-dependence.
\citet{Brook2018} observed long-term pulse profile variability in several pulsars from the NANOGrav 11 yr dataset \citep{NG11}.
Of the pulsars we study here, PSR J1713+0747 was identified to have high profile variability on short timescales.
Pulsars may also experience transient changes to their pulse profiles, with associated delays to their timing residuals \citep{Shannon2016, Goncharov2021}.
A dramatic pulse profile change took place for PSR J1713+0747 in early 2021 \citep{Singha2021}, which introduced chromatic timing delays scaling non-monotonically with radio frequency \citep{Jennings2022}. 
While this event is not in \citetalias{NG15}, PSR J1713+0747 features two weaker chromatic timing events at earlier times \citep{Lam2018}, with evidence of associated changes to the pulse profile found in one or both events \citep{Goncharov2021, Lin2021}.}

Chromatic noise is not typically mitigated prior to TOA generation in PTA pipelines, but it can be included in the noise model as a GP.
\Comment{Existing alternatives include wideband \citep{Pennucci2014, Liu2014} and profile domain timing \citep{Lentati2017}, where chromatic noise mitigation is applied at a different stage}.
NANOGrav's standard noise model mitigates DM variations using DMX timing model parameters, which fit for the DM value at each observation epoch comprised of multi-band observations \citep{Demorest2013, NG9_dataset, NG15}. 
An alternative to DMX is to treat DM variations as a red noise process, using the formalism of Gaussian Processes (``DMGP'', e.g., \citealt{Lentati2014, vanHaasterenVallisneri2014}; \citetalias{3P+paper}), alongside a solar wind model (e.g., \citealt{Hazboun2022}).
Often a power-law prior is imposed on the DM power spectral density (PSD), constraining how the DM variations may vary over time.
This choice is physically motivated by Kolmogorov turbulence in the ISM, which predicts a power-law PSD for DM variations with $\gamma_{\text{DM}} = 8/3$ \citep{Keith2013}.

Meanwhile, time-variable scattering is not always modeled explicitly.
A power-law GP model with a $\nu^{-4}$ frequency scaling was first introduced by \citet{Lam2018} to mitigate possible scattering delays in PSR J1713+0747.
Since then, this GP model has been commonly used as a first-order correction for scattering delays (e.g., \citealt{Goncharov2021, Chalumeau+2022, Srivastava2023}).
\Comment{Hereon, we refer to the $\nu^{-4}$ GP as a ``scattering-like'' chromatic noise process, since other unmodeled chromatic processes could hypothetically result in preference for this model during the Bayesian analysis.
Alternatively, one could attempt to mitigate scattering without assuming a particular frequency scaling, or search for a generic chromatic noise process, by fitting red noise processes isolated to single radio frequency observing bands (i.e., ``band'' noise; \citealt{Lentati2016, Goncharov2021, Chalumeau+2022}.)}

Additional chromatic noise processes not considered here include decorrelation of pulse jitter over radio frequency \citep{Lam2019}, frequency-dependent DM due to multipath propagation effects \citep{Cordes2016}, and low-level radio frequency interference.
While these processes may introduce additional timing errors, they are also difficult to measure and expected to primarily affect the white noise level in most pulsars \citepalias{NG15detchar}.
The ionosphere may also become a significant source of noise at very low radio frequencies ($\nu < 100$ MHz, \citealt{deGasperin2018}).

\section{Gaussian Process Models}
\label{sec:GP models}
We contextualize our models throughout this work in the framework of GPs.
GPs are flexible mathematical models which represent a series of values $\vec{y}$ (e.g., pulsar timing residuals) as samples from a multivariate Gaussian distribution,
\begin{equation}
    p(\vec{y}) = \mathcal{N}(\vec{m}, \mathbf{C}),
\end{equation}
where $\vec{m}$ is the mean vector and $\mathbf{C}$ is the covariance matrix \citep{RasmussenWilliams2006}.
GPs are particularly useful for modeling stochastic signals in astronomical time series, as otherwise unknown information about the functional form of the stochastic signal is represented by the off-diagonal elements of $\mathbf{C}$ \citep{AigrainForeman-Mackey2022}.

We summarize the implementation of GPs used in PTA analyses (e.g., \citealt{vanHaasterenVallisneri2014, Taylor2021}; \citetalias{NG15detchar}).
Our set of timing residuals $\delta\vec{t}$ are represented as a combination of deterministic terms (comprising the mean vector $\vec{m}$), white noise terms, and rank-reduced terms (which are themselves GPs).
The white noise and rank-reduced GPs are used to build the covariance matrix as
\begin{equation}
    \mathbf{C} = \mathbf{N} + \mathbf{T}\mathbf{B}\mathbf{T}^T.
\end{equation}
Here $\mathbf{N}$ is a block-diagonal white noise matrix.
$\mathbf{T}$ is a ($N_{\text{TOA}} \times N_{\text{b}}$) design matrix containing a series of $N_{\text{b}}$ basis functions.
$\mathbf{B} = \langle\vec{b}\vec{b}^T\rangle$ is a diagonal matrix encoding the variance of GP coefficients $\vec{b}$, which are given Gaussian (hyper)priors $p(\vec{b}|\vec{\eta}) = \mathcal{N}(0,\mathbf{B}(\vec{\eta}))$ with hyperparameters $\vec{\eta}$.
We first obtain hyperparameter posteriors $p(\vec{\eta}|\delta\vec{t})$ using MCMC sampling, while marginalizing over $p(\vec{b}|\vec{\eta})$.
We subsequently draw samples of our GP coefficients from the conditional probability distribution $p(\vec{b}|\vec{\eta},\delta\vec{t})$ (see e.g., \citealt{Laal2023, Meyers2023}).

Next we define the components of our noise model which we apply to all six pulsars.
Additional deterministic signals applied only to PSR J1713+0747 are presented in section~\ref{subsec:J1713_model}.

\subsection{Red noise}

Chromatic and achromatic RN processes are defined as rank-reduced GPs using a sine-cosine Fourier basis \citep{Lentati2013}.
The PSD of the Fourier coefficients $\vec{b}$ are parameterized by a power law prior with hyperparameters $\vec{\eta} = A,\gamma$ \citep{vanHaasterenLevin2013},
\begin{eqnarray}
    \label{eq:rn_pl}
    S_{\text{RN}}(f_i) &=& \frac{A_{\text{RN}}^2}{12\pi^2}\left(\frac{f_i}{\text{yr}^{-1}}\right)^{-\gamma_{\text{RN}}}\text{ yr}^{3}.
\end{eqnarray}
Here we use a log-uniform prior $\log_{10}\mathcal{U}(10^{-18},10^{-10})$ to sample $A$, and a uniform prior $\mathcal{U}(0,7)$ to sample $\gamma$.
We set the frequencies of the Fourier basis as integer multiples of the reciprocal of the pulsar's observation time $f_i = i/T_{\text{psr}}$, where $i = 1,2,3,...,N_{\text{freqs}}$, where $2N_{\text{freqs}}$ is the size of the Fourier basis, and $T_{\text{psr}}$ is the pulsar's observation timespan.

We also model ARN using a ``free-spectral'' PSD, where the power at each frequency $f_i$ is a separate parameter $\rho_i$ \citep{Lentati2013}.
To match \citetalias{EPTA_noise}, we use a log-uniform prior $\log_{10}\mathcal{U}(10^{-10},10^{-4})$ to sample each $\rho_i$ (in units of s).
This method is useful to gauge the presence of noise across the spectrum, without making any assumptions about the form of the PSD. 

To specify chromatic noise, we scale the Fourier basis by a frequency-dependent factor,
\begin{eqnarray}
    \mathbf{T}_{\text{RN}} \to \mathbf{T}_{\text{RN}}\left(\frac{\nu}{1400\text{ MHz}}\right)^{-\chi},
\end{eqnarray}
where $\nu$ is the radio frequency and $\chi$ is the chromatic index \citep{Goncharov2021}.
We select larger values of $N_\text{freqs}$ for chromatic processes, which generally have shallow spectra and are more easily decoupled from white noise at higher frequencies than ARN.
We use the following values of these parameters by default: $\chi=0$, $N_{\text{freqs}}=30$ for ARN; $\chi=2$, $N_{\text{freqs}}=100$ for DM noise; and $\chi=4$, $N_{\text{freqs}}=150$ for \Comment{scattering-like chromatic} noise. 
\Comment{The base values for $N_{\text{freqs}}$ are chosen for consistency with other PTA literature, where $N_{\text{freqs}}=30 (100)$ for ARN (DM noise) is a common, albeit arbitrary, choice, and $N_{\text{freqs}}=150$ was the only value favored for scattering-like noise in \citet{Chalumeau+2022}.}

We also calculate Bayes factors \citep{KassRaftery1995} comparing a model with red noise versus a model without red noise.
We use these red noise Bayes factors, $\mathcal{B}^{\text{RN}}$, to quantify the statistical evidence for each red noise process under the given modeling assumptions.
We calculate each $\mathcal{B}^{\text{RN}}$ using the Savage-Dickey density ratio \citep{Dickey1971}, approximated as the prior-to-posterior ratio at the lower bound of the prior distribution ($\log_{10}A_{\text{RN}} = -18$).
In many cases, $\mathcal{B}^{\text{RN}}$ cannot be calculated using the Savage-Dickey approximation due to lack of MCMC samples consistent with $\log_{10}A_{\text{RN}} = -18$.
In these cases we place a lower limit of $\log_{10}\mathcal{B}^{\text{RN}} > 3$, as they correspond to a statistically significant detection of red noise.

\subsection{Solar wind}

The solar wind (SW) may contribute substantially to DM variations as the Earth-Pulsar line of sight cuts through different regions of the heliosphere over the course of each year \citep{Lommen2006}. 
For a time-independent $1/r^2$ SW density profile, the SW's contribution to DM variations goes as
\begin{equation}
    \label{eq:SW}
    \text{DM}_{\text{SW}} = n_{\text{Earth}}(\text{1 AU})\frac{\pi - \theta_i}{\sin\theta_i},
\end{equation}
where $n_{\text{Earth}}$ is the SW free electron density at 1 AU, and $\theta_i$ is the angle between the Earth-Sun line of sight and the Earth-Pulsar line of sight \citep{Splaver2005}. 
For pulsars close to the ecliptic, $\sin\theta_i$ may become very small at the nearest conjunction of the Sun and the pulsar, leading to larger and more peaked annual spikes in the DM time series \citep{Madison2019, Hazboun2022}. 

We include this SW model as a deterministic signal in our Bayesian analysis, with $n_{\text{Earth}}$ fit independently for each pulsar using a uniform prior $\mathcal{U}(0,30)$ in units of cm$^{-3}$. 
Modifications to this model can be made to account for time-dependent or non-spherical SW density profiles \citep{You2007, Hazboun2022}.
As such, we emphasize this model acts only as a first-order correction for SW effects.
Similarly as for red noise, we use the Savage-Dickey density ratio to calculate Bayes factors, gauging how sensitive each pulsar is to detecting the SW (see Appendix~\ref{appendix:SW}).

\subsection{Timing model perturbations}

To account for covariances between noise model parameters and timing model parameters, we vary a linear approximation of the timing model \citep{vanHaasterenLevin2013}. 
The linearized timing response from perturbations to each of the best fit timing model parameters make up the timing model design matrix \citep{vanHaasteren2013, Taylor2021}.
The coefficients $\vec{b}$ corresponding to the amplitude of these perturbations are given Gaussian priors of effectively infinite variance to mimic improper uniform priors.

When using DMX timing model parameters, these priors ensure the DM estimated at each epoch is constrained only by the fit to the data.
When instead modeling DM variations as a red noise process, we remove the DMX parameters and replace them with the DM1 and DM2 timing model parameters, which parameterize a linear and quadratic trend in DM over time \citep{Lentati2014}. 
These are needed to account for long timescale DM variations below the fundamental frequency $f_1 = 1/T_{\text{psr}}$ of the DM red noise process.

\subsection{White noise}

TOA uncertainties are initially estimated based on radiometer noise, which affects the pulse profile signal-to-noise ratio \citep{Lorimerkramer2012}.
We model white noise by modifying these estimated uncertainties using three parameters: EFAC ($\mathcal{F}$), EQUAD ($\mathcal{Q}$), and ECORR ($\mathcal{J}$).
These parameters are designed to model errors in estimates of template-matching uncertainties, independent measurement noise, and pulse jitter \citep{CordesDowns1985, CordesShannon2010} respectively.
An independent set of these parameters are fit for each unique receiver/backend pair \citepalias{NG15detchar}.
Mathematically, these effects are represented in the following elements of $\mathbf{N}$,
\begin{eqnarray}
    \langle n_{i,\mu}n_{j,\nu}\rangle &=& \mathcal{F}_\mu^2(\sigma_i^2\delta_{i,j}\delta_{\mu,\nu} + \mathcal{Q}^2_\mu\delta_{i,j}\delta_{\mu,\nu}) \notag \\
    &\;&\;\;\; + \mathcal{J}_\mu^2\delta_{e(i),e(j)}\delta_{\mu,\nu}\ ,
\end{eqnarray}
where $i,j$ label each TOA, $\mu,\nu$ label each receiver/backend pair, $e(i),e(j)$ label all TOAs within the same observation epoch, and $\sigma_i$ are the original TOA errors.
The $\mathcal{J}_\mu$ terms make $\mathbf{N}$ a block-diagonal matrix, whose inverse we calculate using the Sherman-Morrison formula.
To match \citetalias{EPTA_noise}, we use a uniform prior $\mathcal{U}(0.1,5)$ for all $\mathcal{F}_\mu$ and a log-uniform prior $\log_{10}\mathcal{U}(10^{-9},10^{-5})$ in units of s for all $\mathcal{Q}_\mu$ and $\mathcal{J}_\mu$.

\subsection{PSR J1713+0747 chromatic events}
\label{subsec:J1713_model}

PSR J1713+0747 has exhibited unusual timing events near MJDs 54750 and 57510 \citep{Keith2013, Demorest2013, Lam2018}.
These events each manifest as sudden ``dip'' in \Comment{the apparent DM} value which gradually returns back to a previous level. 
GP analyses of PSR J1713+0747 (e.g., \citealt{Lam2018, Hazboun2020, Goncharov2021}) have modeled these noise transients using decaying exponential functions,
\begin{equation}
    \Delta t = -A_{\text{E}}\Theta(t_E)\exp\left(-\frac{t}{\tau_{\text{E}}}\right)\left(\frac{\nu}{\text{1400 MHz}}\right)^{-\chi_{\text{E}}},
\end{equation}
where $A_{\text{E}}$ is the amplitude, $\Theta(t_E)$ is a Heaviside step function centered at the initial time of the event, $\tau_{\text{E}}$ is the decay timescale, $\chi_{\text{E}}$ is the chromatic scaling index, and $\nu$ is the radio frequency.
We further refer to each dip as $E_1$ and $E_2$ respectively.
We use log-uniform priors $\log_{10}\mathcal{U}(10^{-10},10^{-2})$ for $A_{E_{1,2}}$ and \Comment{$\log_{10}\mathcal{U}(10^{0},10^{3.5})$} for $\tau_{E_{1,2}}$.
We use uniform priors $\mathcal{U}(54650,54850)$ for $t_{E_1}$ and $\mathcal{U}(57490,57530)$ for $t_{E_2}$.
We treat the chromatic indices $\chi_{E_{1,2}}$ in two ways: 1) we hold them at fixed $\chi_{E_{1,2}} = 2$ to model $E_{1,2}$ as DM events \citep{Lam2018}, and 2) we sample $\chi_{E_{1,2}}$ with a uniform prior $\mathcal{U}(0,7)$; see e.g., \citet{Goncharov2021, Chalumeau+2022}.

\subsection{Composite Models}

\begin{table}[ht!]
    \centering
    \vspace{-1\baselineskip}
    \begin{tabularx}{0.9\columnwidth}{c|c}
        \hline
        %\hline \texttt{DMX} & TM + WN + RN + DMX \\
        \hline \texttt{DMX} & RN (achromatic); TM (\textbf{DMX}); WN \\
        \hline & RN (achromatic, \textbf{DM}); \\
        \texttt{DMGP} & \textbf{SW}; TM (\textbf{DM1, DM2}); WN; \\
        & \textbf{DM exp. dips} (J1713+0747 only) \\ 
        \hline & RN (achromatic, DM, \textbf{scattering-like}); \\
        \texttt{CustomGP} & SW; TM (DM1, DM2); WN; \\
        & \textbf{custom exp. dips} (J1713+0747 only) \\
        \hline
    \end{tabularx}
    \caption{The three noise models we use for each pulsar from \citetalias{NG15}.
    Bolded terms indicate key changes from one noise model to the next.
    Model components are detailed throughout section~\ref{sec:GP models}.}
    \vspace{-1\baselineskip}
    \label{tab:models}
\end{table}

Table~\ref{tab:models} summarizes the set of three composite noise models we apply to these six pulsars in \citetalias{NG15}, labeled \texttt{DMX}, \texttt{DMGP}, \texttt{CustomGP}.
\texttt{DMX} labels the standard NANOGrav noise model which includes white noise, ARN, and DMX parameters.
\texttt{DMGP} labels a model in which DMX parameters are removed and replaced with the following components: a DM red noise GP, the DM1 \& DM2 parameters, and the deterministic solar wind model, as well as deterministic exponential dips scaling as $\Delta t \sim \nu^{-2}$ for PSR J1713+0747.
We use the comparison of \texttt{DMX} and \texttt{DMGP} to assess if DMX parameters produce similar results as time-correlated DM models for these pulsars.

%\vspace{-1\baselineskip}
\begin{table*}[ht!]
    \vspace{-0.5\baselineskip}
    \centering
    \begin{tabular}{ c | c | c | c c c c c c @{}}
        \hline\hline \multicolumn{3}{c|}{} & \multicolumn{6}{c}{Pulsar} \\
        \hline \multicolumn{2}{c|}{Signal/Parameter} & Model & J0613$-$0200 & J1012+5307 & J1600$-$3053 & J1713+0747 & J1744$-$1134 & J1909$-$3744 \\
        \hline &  & $\texttt{DMX}$ & $-13.8_{-0.4}^{+0.3}$ & $-12.64_{-0.06}^{+0.06}$ & $-13.5_{-0.6}^{+0.2}$ & $-14.1_{-0.1}^{+0.1}$ & $-14.1_{-0.6}^{+0.4}$ & $-14.5_{-0.4}^{+0.3}$ \\
         & $\log_{10}A$ & $\texttt{DMGP}$ & $-13.8_{-0.3}^{+0.2}$ & $\mathbf{-12.81_{-0.07}^{+0.07}}$ & $-14.5_{-0.9}^{+0.8}$ & $-14.1_{-0.2}^{+0.1}$ & $-14.3_{-0.6}^{+0.4}$ & $-14.6_{-0.4}^{+0.3}$ \\
         &  & $\texttt{CustomGP}$ & $-14.2_{-0.6}^{+0.4}$ & $-12.85_{-0.05}^{+0.05}$ & $-14.2_{-0.7}^{+0.5}$ & $\mathbf{-14.7_{-0.5}^{+0.3}}$ & $-15.2_{-1.8}^{+1.0}$ & $-14.7_{-0.4}^{+0.3}$ \\\cline{2-9} %\multirow{3}{*}{\minitab[c]{Achromatic \\ Red Noise}} &  & $\texttt{DMX}$ & $3.1_{-0.7}^{+0.9}$ & $0.8_{-0.3}^{+0.3}$ & $1.7_{-0.8}^{+1.6}$ & $2.6_{-0.4}^{+0.5}$ & $3.6_{-1.2}^{+1.4}$ & $4.1_{-0.9}^{+1.0}$ \\
        \multirow{3}{2cm}[1.8mm]{\begin{tabular}{c}Achromatic \\ Red Noise \end{tabular}} &  & $\texttt{DMX}$ & $3.1_{-0.7}^{+0.9}$ & $0.8_{-0.3}^{+0.3}$ & $1.7_{-0.8}^{+1.6}$ & $2.6_{-0.4}^{+0.5}$ & $3.6_{-1.2}^{+1.4}$ & $4.1_{-0.9}^{+1.0}$ \\
         & $\gamma$ & $\texttt{DMGP}$ & $3.1_{-0.6}^{+0.8}$ & $1.1_{-0.3}^{+0.3}$ & $3.8_{-1.9}^{+2.0}$ & $2.6_{-0.4}^{+0.5}$ & $3.8_{-1.2}^{+1.4}$ & $4.1_{-0.8}^{+1.0}$ \\
         &  & $\texttt{CustomGP}$ & $4.0_{-1.0}^{+1.3}$ & $1.2_{-0.2}^{+0.2}$ & $3.9_{-1.4}^{+1.7}$ & $3.5_{-0.9}^{+1.2}$ & $3.5_{-2.1}^{+1.9}$ & $4.4_{-0.8}^{+1.0}$ \\\cline{2-9}
         &  & $\texttt{DMX}$ & $>3$ & $>3$ & $2.4$ & $>3$ & $2.1$ & $>3$ \\
         & $\log_{10}\mathcal{B}$ & $\texttt{DMGP}$ & $>3$ & $>3$ & $>3$ & $>3$ & $1.6$ & $>3$ \\
         &  & $\texttt{CustomGP}$ & $>3$ & $>3$ & $\mathbf{1.7}$ & $>3$ & $\mathbf{-0.1}$ & $>3$ \\
        \hline & \multirow{2}{*}{$\log_{10}A$} & $\texttt{DMGP}$ & $-13.38_{-0.04}^{+0.04}$ & $-13.17_{-0.04}^{+0.05}$ & $-13.12_{-0.05}^{+0.05}$ & $-13.82_{-0.05}^{+0.05}$ & $-13.46_{-0.04}^{+0.04}$ & $-13.63_{-0.04}^{+0.04}$ \\
         &  & $\texttt{CustomGP}$ & $\mathbf{-13.7_{-0.3}^{+0.1}}$ & $\mathbf{-14.1_{-2.3}^{+0.6}}$ & $-13.20_{-0.11}^{+0.08}$ & $-13.80_{-0.05}^{+0.05}$ & $-13.55_{-0.11}^{+0.07}$ & $-13.67_{-0.04}^{+0.04}$ \\\cline{2-9}
        DM & \multirow{2}{*}{$\gamma$} & $\texttt{DMGP}$ & $1.9_{-0.2}^{+0.2}$ & $1.4_{-0.2}^{+0.2}$ & $2.2_{-0.2}^{+0.2}$ & $1.8_{-0.2}^{+0.2}$ & $1.5_{-0.2}^{+0.2}$ & $1.5_{-0.1}^{+0.1}$ \\
        Noise &  & $\texttt{CustomGP}$ & $\mathbf{2.8_{-0.4}^{+0.7}}$ & $\mathbf{2.6_{-0.9}^{+1.9}}$ & $2.6_{-0.2}^{+0.3}$ & $1.9_{-0.2}^{+0.2}$ & $1.7_{-0.2}^{+0.4}$ & $1.6_{-0.1}^{+0.2}$ \\\cline{2-9}
         & \multirow{2}{*}{$\log_{10}\mathcal{B}$} & $\texttt{DMGP}$ & $>3$ & $>3$ & $>3$ & $>3$ & $>3$ & $>3$ \\
         &  & $\texttt{CustomGP}$ & $>3$ & $\mathbf{0.1}$ & $>3$ & $>3$ & $>3$ & $>3$ \\
        \hline Scattering-like & $\log_{10}A$ & $\texttt{CustomGP}$ & $-14.01_{-0.06}^{+0.05}$ & $-13.75_{-0.06}^{+0.05}$ & $-13.59_{-0.05}^{+0.05}$ & $-14.22_{-0.05}^{+0.06}$ & $-14.23_{-0.13}^{+0.09}$ & $-14.67_{-0.09}^{+0.09}$ \\
        Chromatic & $\gamma$ & $\texttt{CustomGP}$ & $1.5_{-0.2}^{+0.2}$ & $1.5_{-0.2}^{+0.2}$ & $1.6_{-0.2}^{+0.2}$ & $1.3_{-0.2}^{+0.2}$ & $1.3_{-0.3}^{+0.4}$ & $0.5_{-0.3}^{+0.3}$ \\
        Noise & $\log_{10}\mathcal{B}$ & $\texttt{CustomGP}$ & $>3$ & $>3$ & $>3$ & $>3$ & $3.0$ & $2.6$ \\
        \hline \hline
    \end{tabular}
    \caption{\textbf{Estimated noise parameters and Bayes factors for six pulsars in \citetalias{NG15} under all three modeling assumptions.}
    Noise parameters are presented using the median and 68.3\% Bayesian credible intervals (referenced here as 1-$\sigma$ regions), and Bayes factors indicating statistical detection significance of the given signal are calculated from our posterior distributions using the Savage-Dickey approximation.
    If a parameter is bolded, that means the parameter’s 1-$\sigma$ region estimated under the current model is inconsistent with the 1-$\sigma$ region estimated under the previous model (from one row above).
    If a Bayes factor is bolded, that means the Bayes factor estimated under the current model is at least an order of magnitude different from the Bayes factor estimated under the previous model.}
    \vspace{-0.5\baselineskip}
    \label{tab:noise_params}
\end{table*}

\texttt{CustomGP} \Comment{extends the \texttt{DMGP} model} by including additional \Comment{non-dispersive} chromatic noise processes used by \citetalias{EPTA_noise} for pulsars from \citetalias{EPTA_data}. 
Namely, \texttt{CustomGP} includes the addition of a $\chi = 4$ \Comment{scattering-like chromatic} red noise process for all pulsars.
It also uses $N_{\text{freqs}}=150$ for ARN in PSR J1012+5307 and for DM noise in PSR J1909$-$3744 as these processes favored a large number of Fourier modes in \citetalias{EPTA_noise}.
Furthermore, the chromatic indices $\chi_{E_{1,2}}$ of the deterministic dips in PSR J1713+0747 are allowed to vary as free parameters instead of being fixed to $\chi_{E_{1,2}} = 2$. 

\begin{figure*}[ht!]
    \centering
    \vspace{-0.5\baselineskip}
    \includegraphics[width=\textwidth]{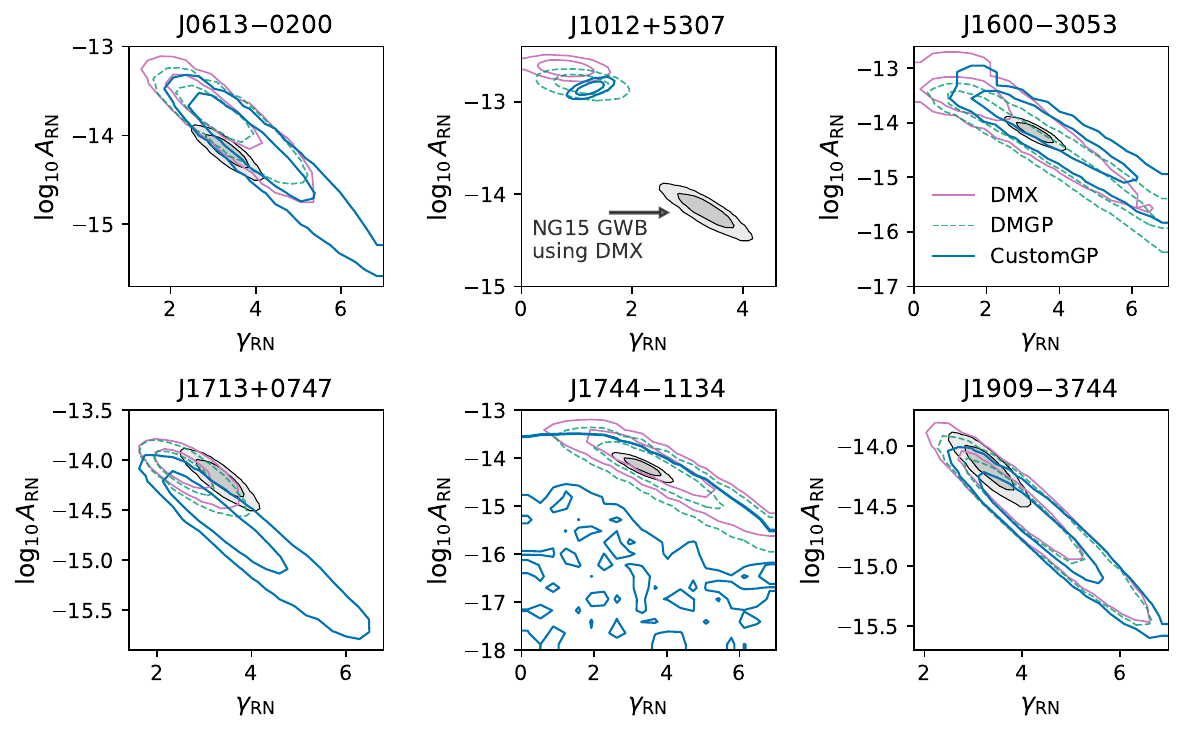}
    %\vspace{-1\baselineskip}
    \caption{\textbf{Accounting for non-dispersive chromatic noise using the \texttt{CustomGP} model noticeably affects achromatic red noise posteriors for multiple pulsars within \citetalias{NG15}.} 
    Posterior distributions for the six pulsars' ARN parameters using the \texttt{DMX} (solid purple), \texttt{DMGP} (dashed green), and \texttt{CustomGP} (solid blue) models.
    \Comment{Since the GWB makes up a portion of all pulsars ARN budget, the GWB parameters inferred from \citetalias{NG15_gwb} using the \texttt{DMX} model for all pulsars is included in gray.}
    The inferred ARN posteriors change the most substantially for PSRs J1713+0747 and J1012+5307 using \texttt{CustomGP}. 
    PSRs J0613$-$0200, J1600$-$3053, and J1744$-$1134 also feature noticeable changes to their ARN parameters at a less significant level.
    PSR J1909$-$3744's ARN parameters are the \Comment{least affected} the choice of noise model.
    \Comment{The apparent shift of PSR J1713+0747's ARN parameters away from the inferred GWB parameters indicates its choice of noise model is likely to affect GWB spectral characterization.}}
    \label{fig:red_noise}
\end{figure*}

Accounting for all of these effects, our \texttt{CustomGP} model tests for the same extent of noise processes as included in the \citetalias{EPTA_noise}. 
However, \texttt{CustomGP} is not quite equivalent to the models from \citetalias{EPTA_noise}.
This is partly due to intrinsic differences between \citetalias{NG15} and \citetalias{EPTA_data}. 
For instance, we do not use the same values of $N_{\text{freqs}}$ for red noise processes as were selected in \citetalias{EPTA_noise}.
Each data set features a different observation timespan and cadence for each pulsar, therefore the values $N_{\text{freqs}}$ favored for pulsars from \citetalias{EPTA_data} are unlikely to be optimal for the same pulsars from \citetalias{NG15}.
Instead, we ensure the ARN and chromatic noise spectra extend to at least the same high-frequency cutoff $f_{\text{max}} = N_{\text{freqs}}/T_{\text{psr}}$ as the favored models from \citetalias{EPTA_noise}.
\Comment{One additional difference is, where \citetalias{EPTA_noise} fixes $n_{\text{Earth}} = 7.9$ cm$^{-3}$ \citep{Madison2019}, we allow it to vary as free parameter for each pulsar separately.
Furthermore, \citetalias{EPTA_noise} only includes \Comment{scattering-like chromatic} noise for PSR J1600$-$3053, and does not include ARN for PSR J1600$-$3053, while we have both ARN and \Comment{scattering-like chromatic} noise processes in the six \citetalias{NG15} pulsars using \texttt{CustomGP}.}
To account for these differences, we modified the favored models from \citetalias{EPTA_noise} to include ARN, include \Comment{scattering-like chromatic} noise, and fit $n_{\text{Earth}}$ as a free parameter for all six pulsars using \citetalias{EPTA_data}.
These modifications were made to ensure fair comparison with \citetalias{NG15} using \texttt{CustomGP}, but this does not noticeably alter the ARN parameter estimation results from \citetalias{EPTA_noise}.

\section{Results}
\label{sec:results} 
Table~\ref{tab:noise_params} presents the medians and 68.3\% Bayesian credible intervals (1-$\sigma$ regions) of inferred noise parameters and Bayes Factors for each noise process.
These are tabulated for our six pulsars in \citetalias{NG15} under the three different noise models: \texttt{DMX}, \texttt{DMGP}, and \texttt{CustomGP}.
Bolded parameter values indicate cases where going from one model to the next results in a discrepancy (significant at a $>$1-$\sigma$ level) between noise parameters.
Bolded Bayes factors indicate cases where the Bayes factor changed by over an order of magnitude, i.e., if the detection significance of a noise process has substantially dropped.
The inferred solar wind electron density from each pulsar is reported in Appendix~\ref{appendix:SW}.
Interestingly, using \citetalias{NG15} we find the presence of $\chi = 4$ \Comment{scattering-like chromatic} noise is supported by a Bayes factor $\log_{10}\mathcal{B}^{\text{chrom}} \ge 2.6$ for all six pulsars using model \texttt{CustomGP}, despite a significant \Comment{scattering-like chromatic} noise detection in only one pulsar from \citetalias{EPTA_data} (PSR J1600$-$3053, \citealt{Chalumeau+2022}; \citetalias{EPTA_noise}).
\Comment{For several pulsars, these chromatic noise amplitudes are substantially higher using \citetalias{NG15} than the upper limits set using \citetalias{EPTA_data}.}
Including \Comment{scattering-like chromatic} noise in \texttt{CustomGP} changes the estimated DM noise parameters by $>$1-$\sigma$ for PSRs J0613$-$0200 and J1012+5307.
\Comment{A deeper investigation comparing with measurements of pulse broadening or scintillation would be needed to confirm the origin of these variations.}

To complement Table~\ref{tab:noise_params}, Figure~\ref{fig:red_noise} shows the 2D posterior distributions for $\log_{10}A_{\text{RN}}$ and $\gamma_{\text{RN}}$ for our six pulsar sample of \citetalias{NG15}, under all three modeling assumptions.
All contours enclose 68.3\% (1-$\sigma$) and 95.4\% (2-$\sigma$) 2D credible intervals.
\citetalias{NG15detchar} showed that out of these six pulsars, only PSR J1012+5307 shows evidence for additional ARN on top of a GWB signal.
To highlight this, the GWB parameters, \Comment{inferred from \citetalias{NG15_gwb} using the \texttt{DMX} model}, are also shown in Figure~\ref{fig:red_noise}.
\Comment{With the exception of PSR J1012+5307, the ARN and GWB parameters are similar for every pulsar, indicating the GWB makes up a substantial portion of these pulsars' ARN budget.}

\begin{figure*}[ht!]
    \centering
    \vspace{-0.5\baselineskip}
    \begin{tabularx}{\textwidth}{lll}
        \includegraphics[width=0.24\linewidth]{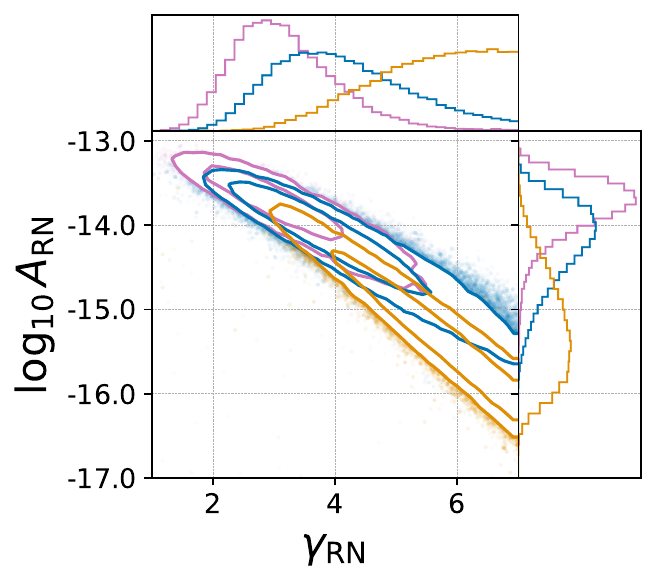} &
        \multirow{3}{*}[35mm]{\includegraphics[width=0.285\linewidth]{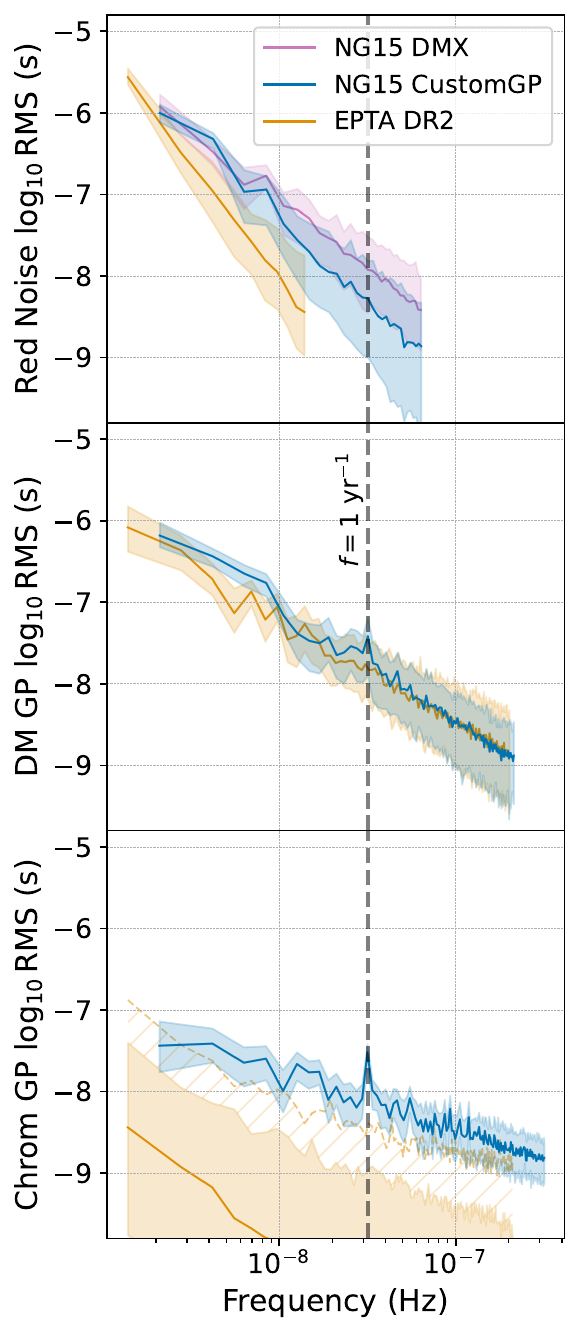}} &
        \hspace{-4mm}\multirow{3}{*}[35mm]{\includegraphics[width=0.445\linewidth]{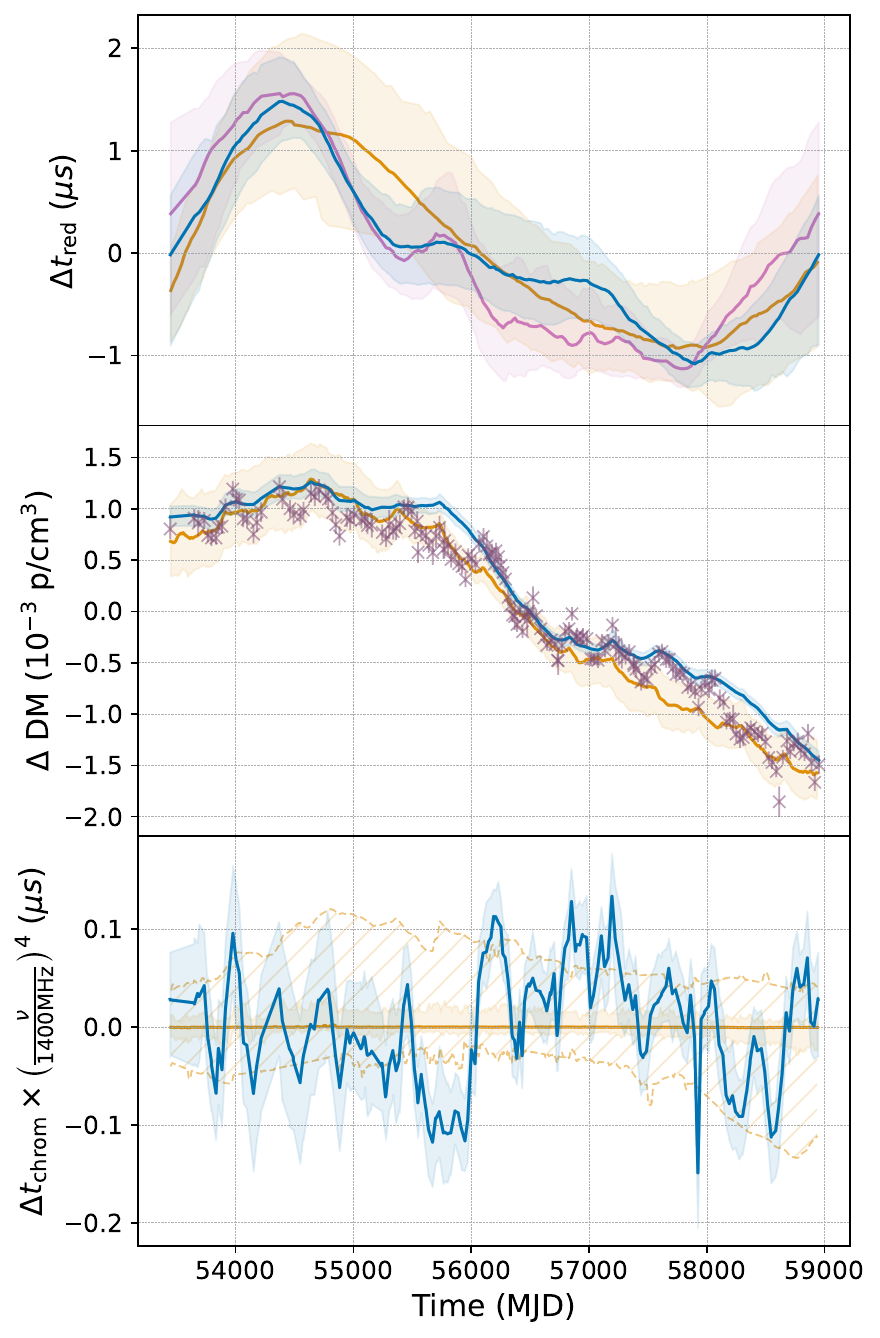}} \\
        \includegraphics[width=0.24\linewidth]{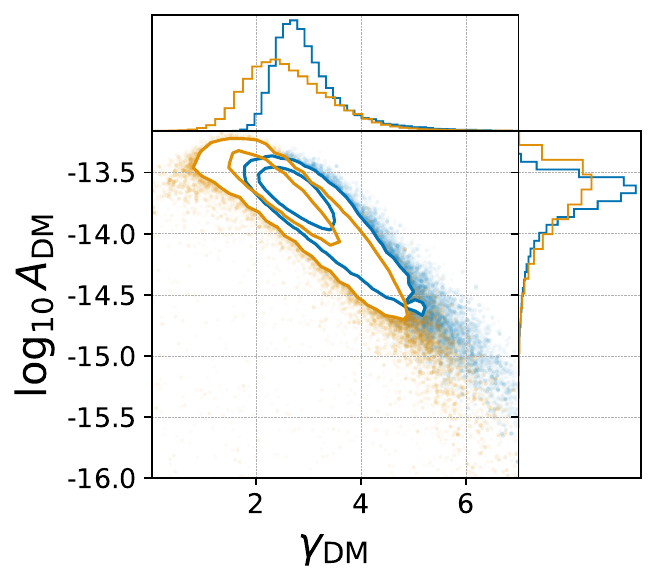} & & \\
        \includegraphics[width=0.24\linewidth]{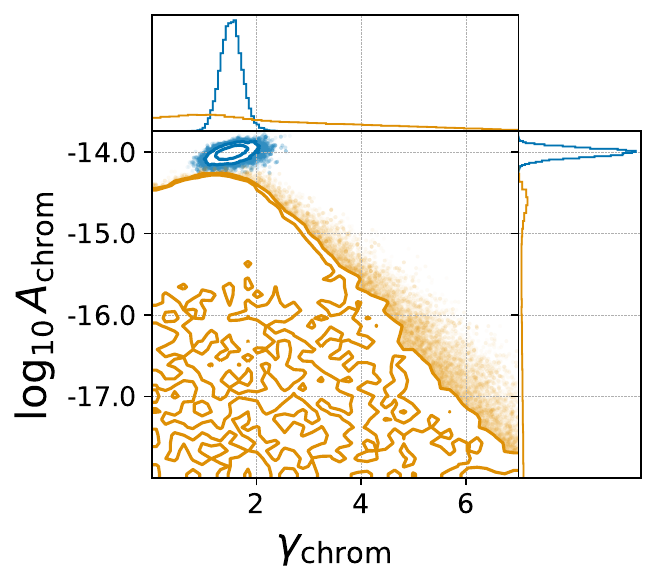} & &
    \end{tabularx}
    %\vspace{-2\baselineskip} 
    \caption{\textbf{PSR J0613$-$0200 -- A variation in the achromatic red noise, uniquely observed using \citetalias{NG15}, is mitigated using \texttt{CustomGP}.}
    Noise properties are displayed by column: posterior parameter distributions (left), spectra (middle), and time-domain GP realizations (right).
    Noise types are displayed by row: ARN (top), DM variations (middle), and \Comment{scattering-like chromatic} noise with $\chi=4$ (bottom).
    Data/model in use are displayed by color: \citetalias{NG15} using \texttt{DMX} (purple), \citetalias{NG15} using \texttt{CustomGP} (blue), and \citetalias{EPTA_data} (yellow).
    All chromatic noise spectra are referenced to a delay at 1400 MHz.
    A variation present in the time-domain ARN realizations near MJD 56000 using \texttt{DMX} is now absorbed by the chromatic noise model using \texttt{CustomGP}.
    This ARN variation is not present in \citetalias{EPTA_data}.
    The ARN spectrum is steeper and more consistent with \citetalias{EPTA_data} when applying \texttt{CustomGP}.
    An annual trend in the time series of DMX parameters is now absorbed by the \Comment{scattering-like chromatic} model using \texttt{CustomGP}.
    However, the \Comment{scattering-like chromatic} noise amplitude from \citetalias{NG15} is above the upper limit set by \citetalias{EPTA_data}.}
    \label{fig:J0613}
\end{figure*}

We first consider how the recovered ARN parameters are affected by switching from \texttt{DMX} to \texttt{DMGP}.
Each model results in notably different ARN posteriors for PSRs J1012+5307 and J1600$-$3053, but no major differences for the remaining pulsars (Figure~\ref{fig:red_noise}). 
The change to $\log_{10}A_{\text{RN}}$ for PSR J1012+5307 is significant at a $>$1-$\sigma$ level (Table~\ref{tab:noise_params}).
These findings coincide with much larger differences in DM recovery for PSRs J1012+5307 and J1600$-$3053 than the remaining pulsars (Appendix~\ref{appendix:DM}).

Transitioning from \texttt{DMX} to \texttt{CustomGP} yields further changes to the single-pulsar ARN parameters.
In general, switching to \texttt{CustomGP} results in lower $\log_{10}A_{\text{RN}}$ and higher $\gamma_{\text{RN}}$ (Figure~\ref{fig:red_noise}), \Comment{consistent with the effects of using custom chromatic models }
As a counterexample, PSR J1600$-$3053 favors a slightly higher $\log_{10}A_{\rm{RN}}$ going from \texttt{DMGP} to \texttt{CustomGP}.
Overall, the ARN properties of PSR J1909$-$3744 remain the most similar under all three models.
Meanwhile, PSR J1713+0747's change is the most dramatic, as it experiences a significant ($>$1-$\sigma$) decrease to $\log_{10}A_{\text{RN}}$ and supports a much broader range of $\gamma_{\text{RN}}$ values.
\Comment{Furthermore, PSR J1713+0747's ARN parameters using \texttt{DMX} and \texttt{DMGP} are highly constrained near the GWB parameters measured in \citetalias{NG15_gwb}, but become \emph{less consistent} with the measured GWB parameters when using \texttt{CustomGP}, favoring instead a lower amplitude and higher spectral index.
Notably, the alternative noise model used in \citetalias{NG15_gwb}, which included the \texttt{CustomGP} model for PSR J1713+0747 alongside DMGP for the remaining pulsars, also resulted in a shift towards lower amplitude and higher spectral index of the common noise.
As such, our results signpost PSR J1713+0747's noise model as a strong contributor to this change.
However, the GWB parameter inference is dependent on information from 61 additional pulsars not studied here.
As such, a direct quantification of these impacts on GWB characterization will require a more careful analysis using the full PTA, which is the subject of an upcoming work.}

To assess the performance of the models, we next compare the following cases on a pulsar-by-pulsar basis: 1) \texttt{DMX} applied to \citetalias{NG15}, 2) \texttt{CustomGP} applied to \citetalias{NG15}, and 3) \texttt{CustomGP} applied to \citetalias{EPTA_data}.
We do not include the model \texttt{DMGP} in this comparison as it is intermediary to the more disparate models \texttt{DMX} and \texttt{CustomGP} (although we do compare DM estimates using \texttt{DMX} and \texttt{DMGP} in Appendix~\ref{appendix:DM}). 
We present a separate figure for each pulsar (starting from Figure~\ref{fig:J0613}), displaying parameter posterior parameters, spectra, and time-domain GP realizations for ARN, DM variations, and $\chi=4$ \Comment{scattering-like chromatic} noise. 
For model \texttt{DMX} we display only ARN and the time series of DMX parameters.
Time-domain realizations of DM variations using \texttt{CustomGP} include all stochastic and deterministic contributions to DM to allow fair comparison with the DMX time series.
Spectra and time-domain realizations are visualized using medians and 68\% (1-$\sigma$) Bayesian credible intervals, \Comment{each computed from 100 GP realizations.}
\Comment{In cases where the 68\% regions were difficult to make out by eye, we additionally added the 95\% regions, distinguished from the 68\% regions with a different plot style.}
\Comment{Furthermore, since \citetalias{EPTA_data} has a longer timespan than \citetalias{NG15} for several pulsars (Figure~\ref{fig:datasets}), this can result in differing ARN properties between \citetalias{EPTA_data} and \citetalias{NG15} for the same pulsar.
To provide a more useful comparison to \citetalias{NG15}, we generated time-domain realizations from \citetalias{EPTA_data} using the full data timespans for each pulsar, but excised the portion of the realizations before the start of the \citetalias{NG15} timespans.
We then fit out a quadratic in the shortened ARN realizations from \citetalias{EPTA_data} to account for the covariance between ARN and pulsar spindown which could not have been resolved using \citetalias{NG15}.}
All chromatic noise spectra are referenced to a delay at 1400 MHz.

\begin{figure*}[ht!]
    \centering
    \vspace{-0.5\baselineskip}
    \begin{tabularx}{\textwidth}{lll}
        \includegraphics[width=0.24\linewidth]{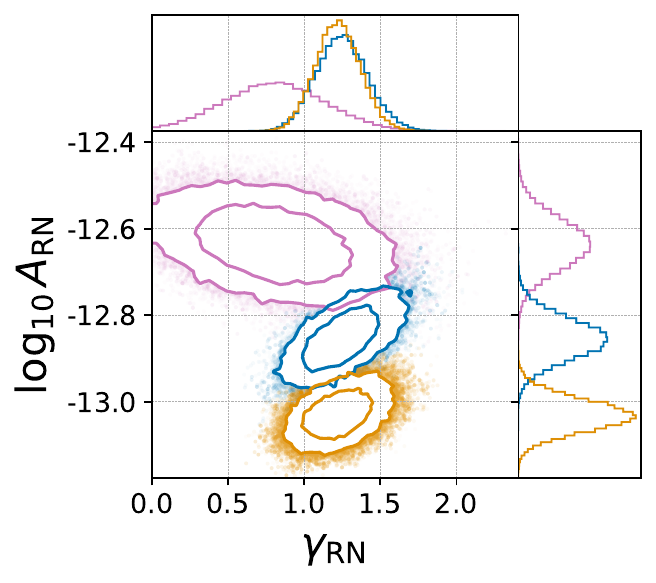} &
        \multirow{3}{*}[35mm]{\includegraphics[width=0.285\linewidth]{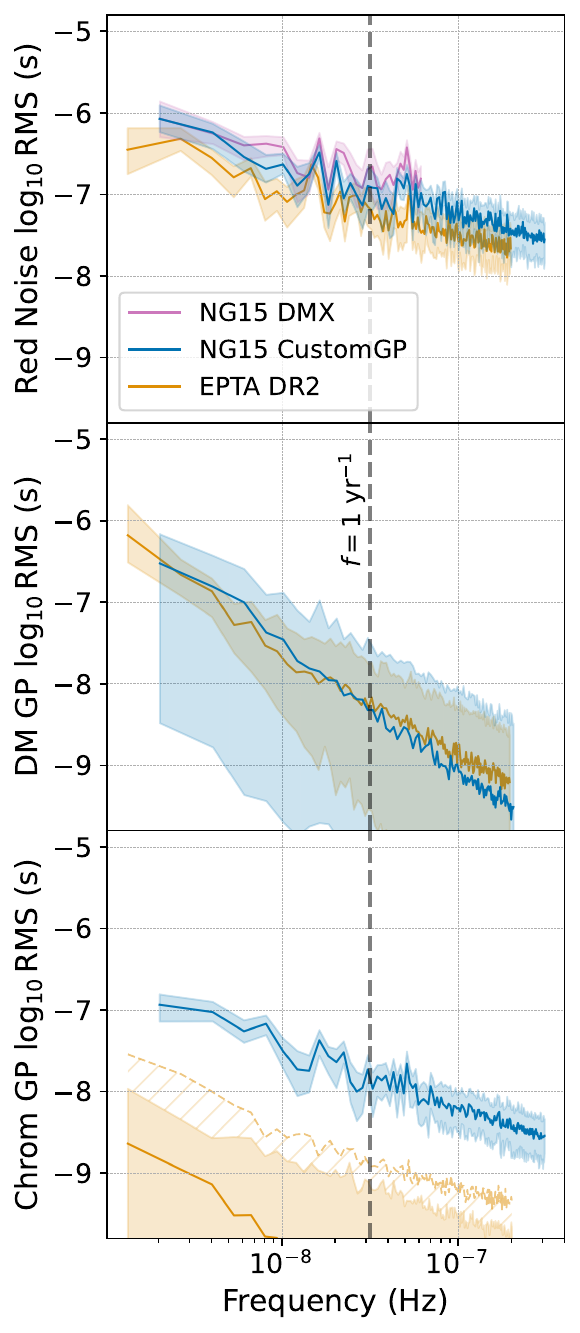}} &
        \hspace{-4mm}\multirow{3}{*}[36mm]{\includegraphics[width=0.44\linewidth]{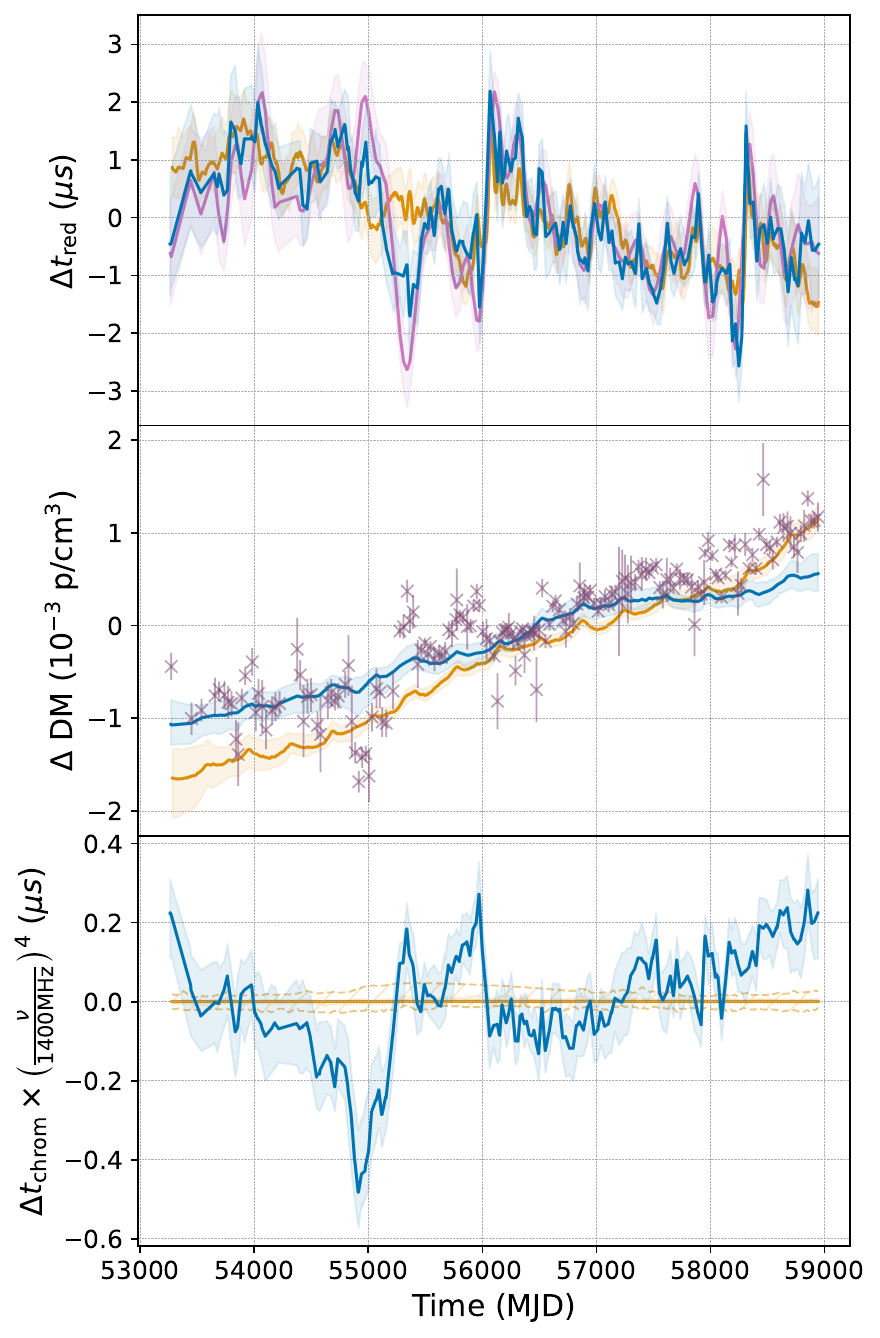}} \\
        \includegraphics[width=0.24\linewidth]{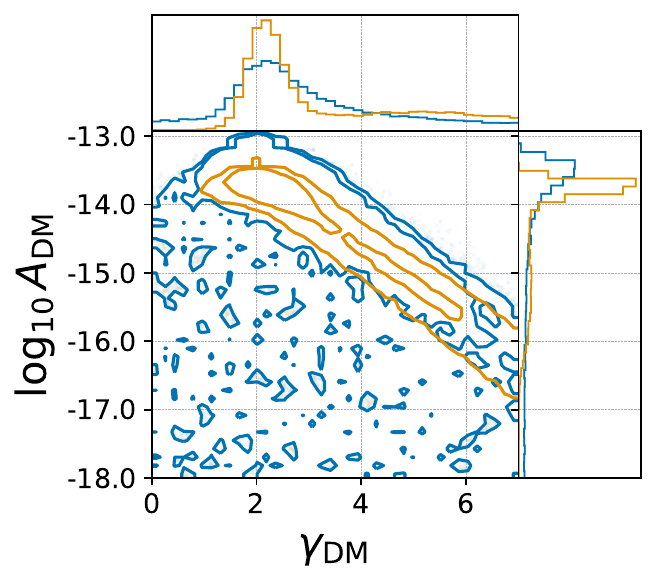} & & \\
        \includegraphics[width=0.24\linewidth]{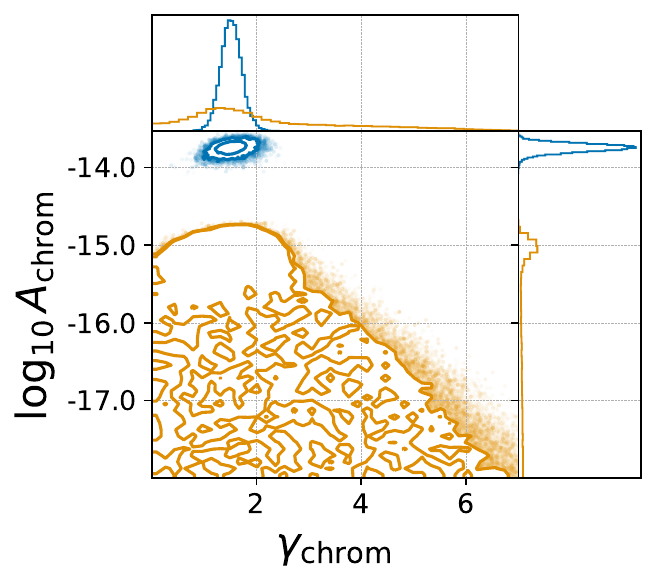} & &
    \end{tabularx}
    %\vspace{-1\baselineskip}
    \caption{\textbf{PSR J1012+5307 -- Each noise model fails to mitigate anti-correlations between chromatic noise and a shallow-spectrum achromatic red noise process in \citetalias{NG15}.}
    J1012+5307 features a peculiar shallow-spectrum ARN process, modeled in \texttt{CustomGP} using $N_{\text{freqs}} = 150$, following \citet{Chalumeau+2022}. 
    Using \texttt{CustomGP}, \citetalias{NG15} and \citetalias{EPTA_data} have the same ARN spectral index and share many common features in the time domain, including sharp spikes nears MJDs 56100 and 58350.
    However, the \citetalias{NG15} ARN spectrum features excess power near 50 nHz, which is not seen in \citetalias{EPTA_data}.
    ARN variations near MJD 55000 and MJD 56000 are anti-correlated with chromatic variations using \citetalias{NG15}, and the \Comment{scattering-like chromatic} noise detection made by \citetalias{NG15} is also well above the upper limit set by \citetalias{EPTA_data}.
    See the Figure~\ref{fig:J0613} caption for panel descriptions.}
    \label{fig:J1012}
    %\vspace{-0.5\baselineskip}
\end{figure*}

\subsection{PSR J0613$-$0200}
\label{subsec:J0613}

Applying \texttt{CustomGP} to PSR J0613$-$0200 results in a steeper ARN spectrum than using \texttt{DMX} (Figure~\ref{fig:J0613}).
In the time-domain ARN realizations, this change corresponds to a variation just before MJD 56000 becoming flat when using \texttt{CustomGP}.
When applying \texttt{CustomGP}, this variation is classified as chromatic, as variations of similar width appear in the DM and \Comment{scattering-like chromatic} variations, with opposite sign. 
It is plausible that this variation was falsely characterized as achromatic using \texttt{DMX}, as no such variation is present in \citetalias{EPTA_data} ARN (\citealt{Chalumeau+2022}; \citetalias{3P+paper}), nor does it appear to be present in ARN from PPTA DR2 \citep{Goncharov2021} or PPTA DR3 (\citealt{PPTA_noise}; \citetalias{3P+paper}). 
Switching to \texttt{CustomGP} for also reduces the Bayes Factor for excess power in PSR J0613$-$0200's ARN free spectrum just above a frequency of $f = 1$/yr (Figure~\ref{fig:FS}, Appendix~\ref{appendix:FS}).
\citetalias{EPTA_noise} notably does not favor the inclusion of any Fourier modes at or above $f = 1$/yr.

\begin{figure*}[ht!]
    \centering
    \vspace{-0.5\baselineskip}
    \begin{tabularx}{\textwidth}{lll}
        \includegraphics[width=0.24\linewidth]{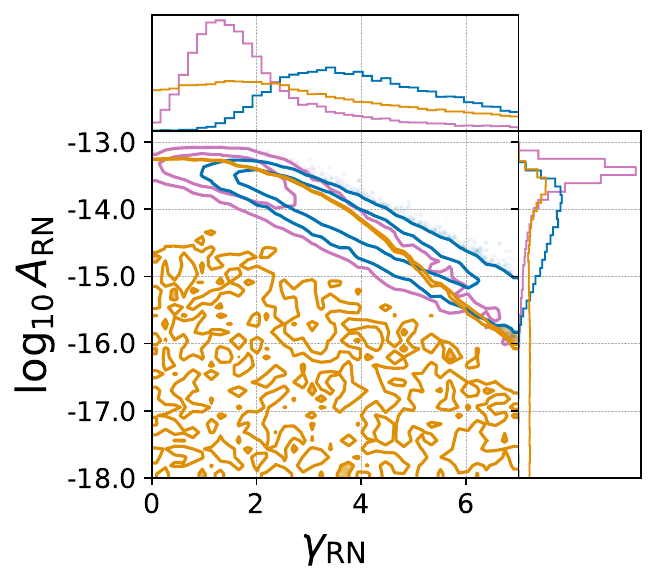} &
        \multirow{3}{*}[35mm]{\includegraphics[width=0.285\linewidth]{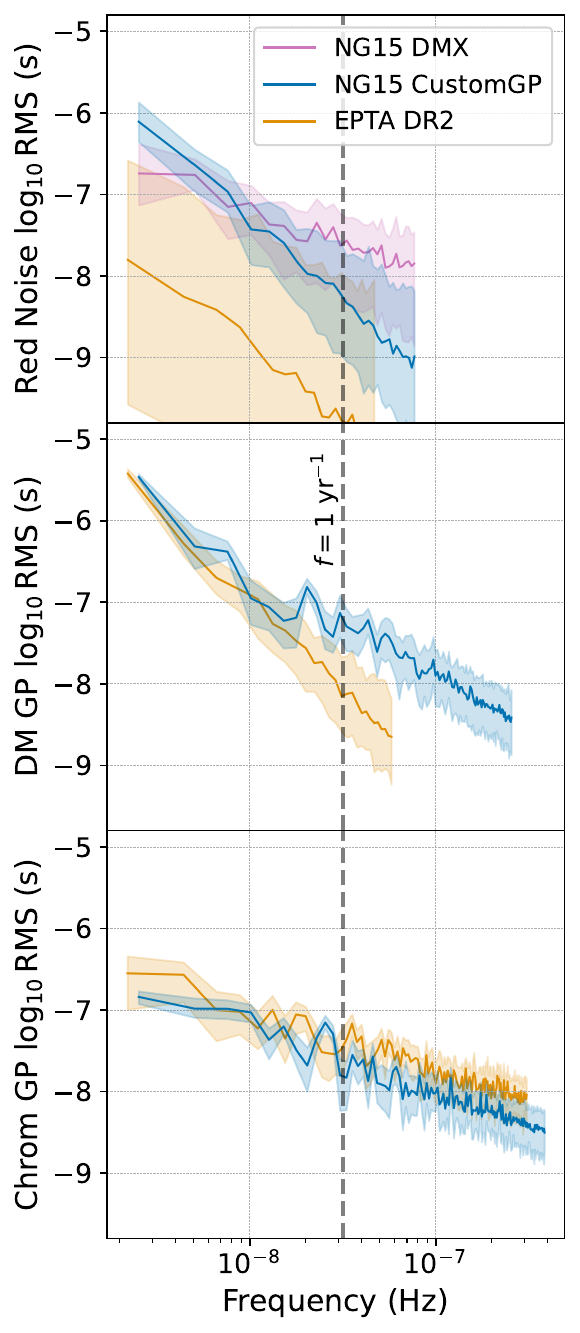}} &
        \hspace{-4mm}\multirow{3}{*}[35mm]{\includegraphics[width=0.445\linewidth]{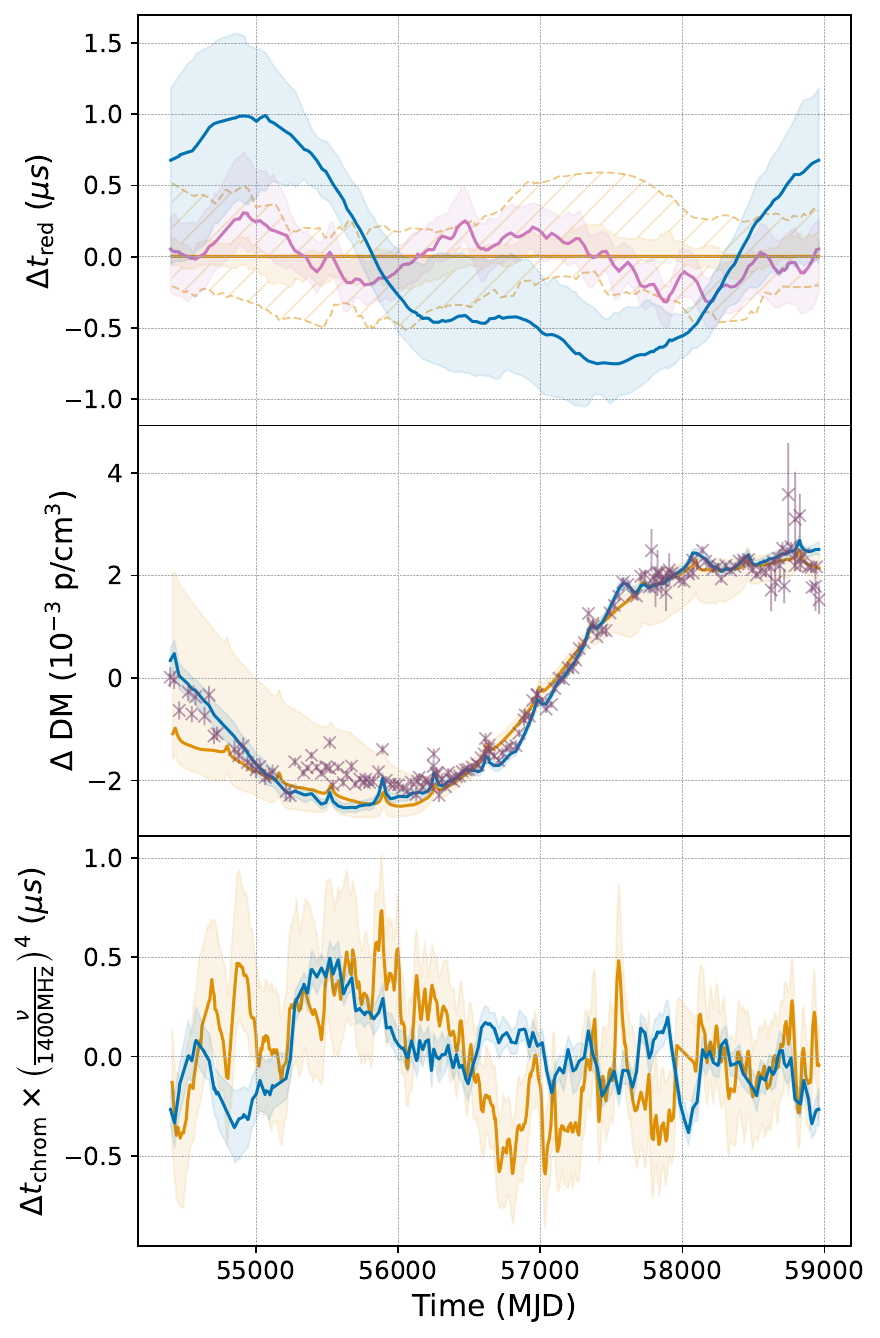}} \\
        \includegraphics[width=0.24\linewidth]{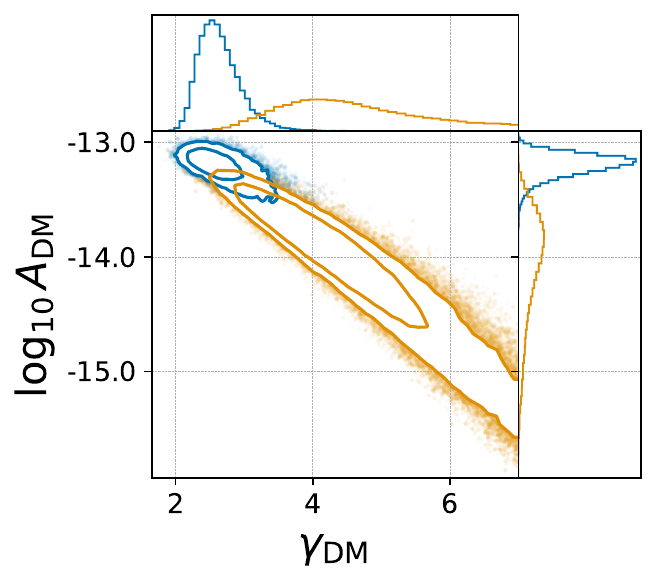} & & \\
        \includegraphics[width=0.24\linewidth]{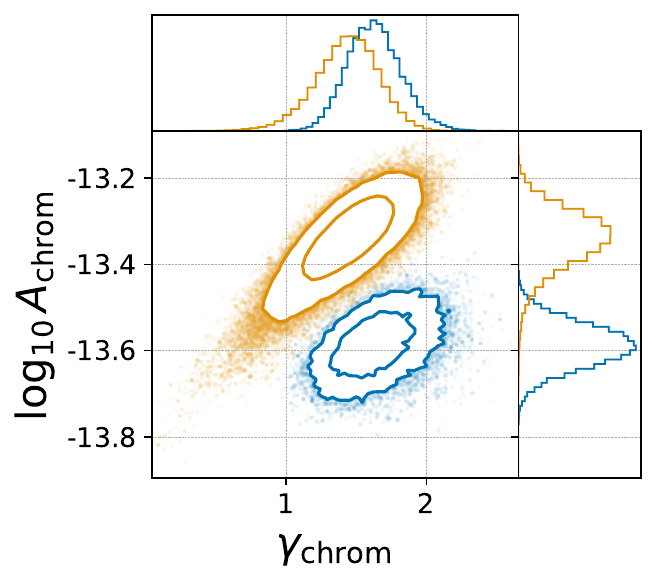} & &
    \end{tabularx}
    %\vspace{-1\baselineskip}
    \caption{\textbf{PSR J1600$-$3053 -- Achromatic red noise is highly sensitive to the chromatic noise parameters, which are inconsistent between \citetalias{NG15} and \citetalias{EPTA_data}.} 
    \texttt{CustomGP} mitigates short-timescale chromatic noise, resulting in a steeper ARN spectrum than \texttt{DMX}. 
    \citetalias{NG15} and \citetalias{EPTA_data} both agree on the presence of strong DM and scattering-like variations, but the \Comment{scattering-like chromatic} noise amplitude and DM noise spectral index are both lower using \citetalias{NG15}.
    In the time-domain, several chromatic features are present in both data sets, but neither data set agrees on whether the DM or \Comment{scattering-like chromatic} term should mitigate them.
    See the Figure~\ref{fig:J0613} caption for panel descriptions.}
    \label{fig:J1600}
\end{figure*}

The DMX time series for PSR J0613$-$0200 shows evidence of annual sinusoidal variations, which could result from a steep DM gradient along the line of sight as the Earth orbits around the Sun \citep{Keith2013, Jones2017}. 
This annual DM trend manifests using both \texttt{DMX} and \texttt{DMGP} (Appendix~\ref{appendix:DM}).
However, when applying model \texttt{CustomGP}, the annual DM trend disappears and instead manifests in the \Comment{scattering-like chromatic} noise.
This is evidenced by a peak at $f = 1$ yr$^{-1}$ in the posterior \Comment{scattering-like chromatic} noise spectrum (Figure~\ref{fig:J0613}). 
An annual scattering trend is supported by measurements of annual scintillation arc variability in this pulsar \citep{Main+2020, Main2023, Liu2023}.
However, no annual chromatic noise in J0613$-$0200 appears to be present using \citetalias{EPTA_data} \citep{Chalumeau+2022}. 
Furthermore, the \Comment{scattering-like chromatic} noise parameters estimated using \citetalias{NG15} lie above the upper limit set by \citetalias{EPTA_data}, as evidenced by the lack of overlap in their \Comment{scattering-like chromatic} noise parameters and spectra.
This discrepancy warrants further investigation, as the inclusion of $\chi=4$ \Comment{chromatic} noise is entirely responsible for the changes to PSR J0613$-$0200's ARN properties.

\subsection{PSR J1012+5307}
\label{subsec:J1012}

PSR J1012+5307 uniquely features a shallow ARN spectrum, present at a much higher amplitude than the GWB across the spectrum \citepalias{NG15detchar}. 
Increasing the number of ARN Fourier modes $N_{\text{freqs}} \to 150$ in model \texttt{CustomGP} \citep{Chalumeau+2022} results in a slightly steeper, more constrained ARN spectrum, since power is now spread across more frequencies (Figure~\ref{fig:J1012}). 
The posteriors for $\gamma_{\text{RN}}$ are almost an exact match for \citetalias{NG15} and \citetalias{EPTA_data}.
In the time-domain ARN realizations there are many features common to both data sets, including sharp and sudden delays near MJDs 56100 and 58350.
However, $\log_{10}A_{\text{RN}}$ is higher for \citetalias{NG15} than \citetalias{EPTA_data}.
Furthermore, \citetalias{NG15detchar} showed that PSR J1012+5307 features excess noise at $f \sim 50$ nHz.
Here this is produced as a bump in the \citetalias{NG15} ARN spectra using both \texttt{DMX} and \texttt{CustomGP}, but this bump is not present in the \citetalias{EPTA_data} ARN spectrum.

\begin{figure*}[ht!]
    \centering
    \vspace{-0.5\baselineskip}
    \begin{tabularx}{\textwidth}{lll}
        \includegraphics[width=0.24\linewidth]{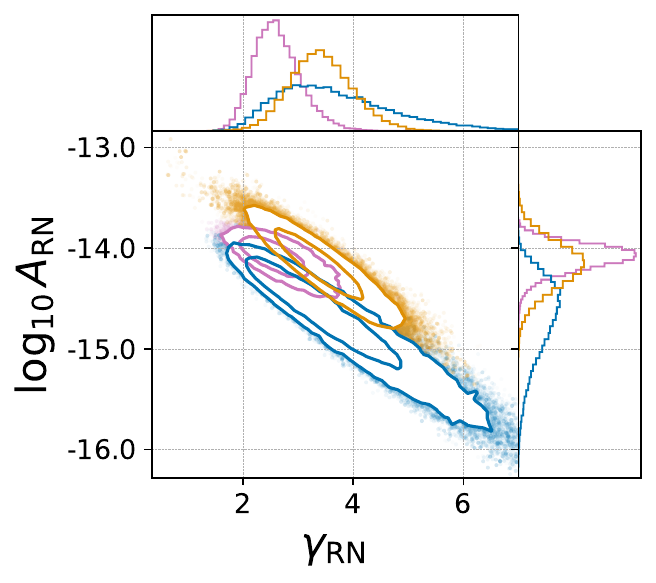} &
        \multirow{3}{*}[35mm]{\includegraphics[width=0.285\linewidth]{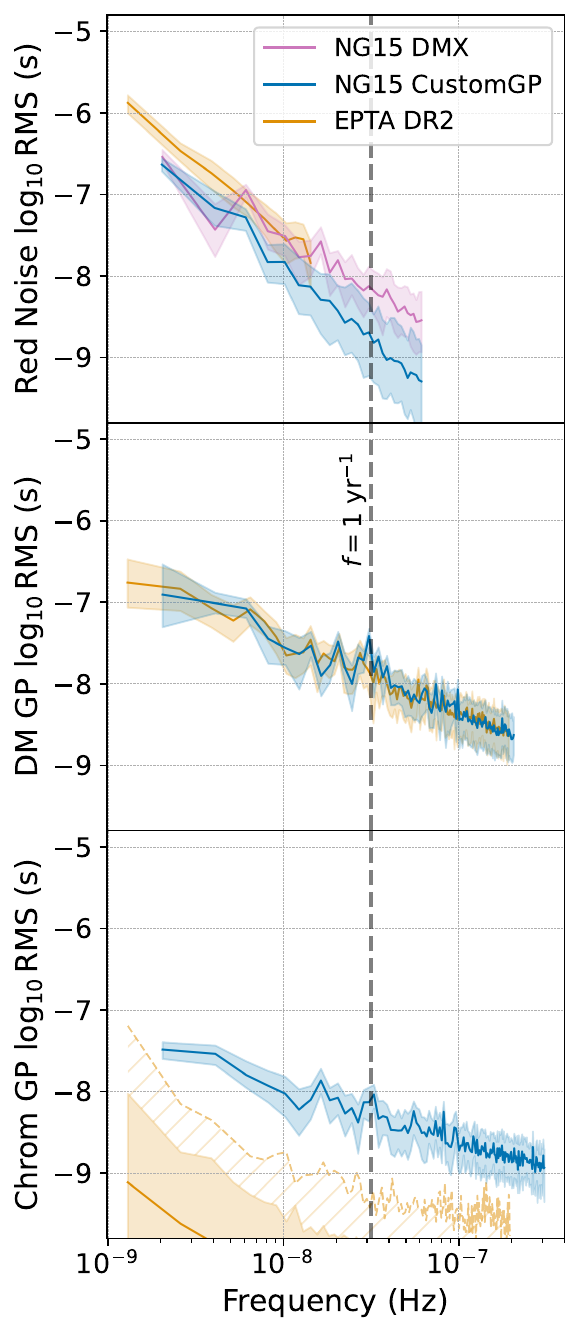}} &
        \hspace{-4mm}\multirow{3}{*}[35mm]{\includegraphics[width=0.445\linewidth]{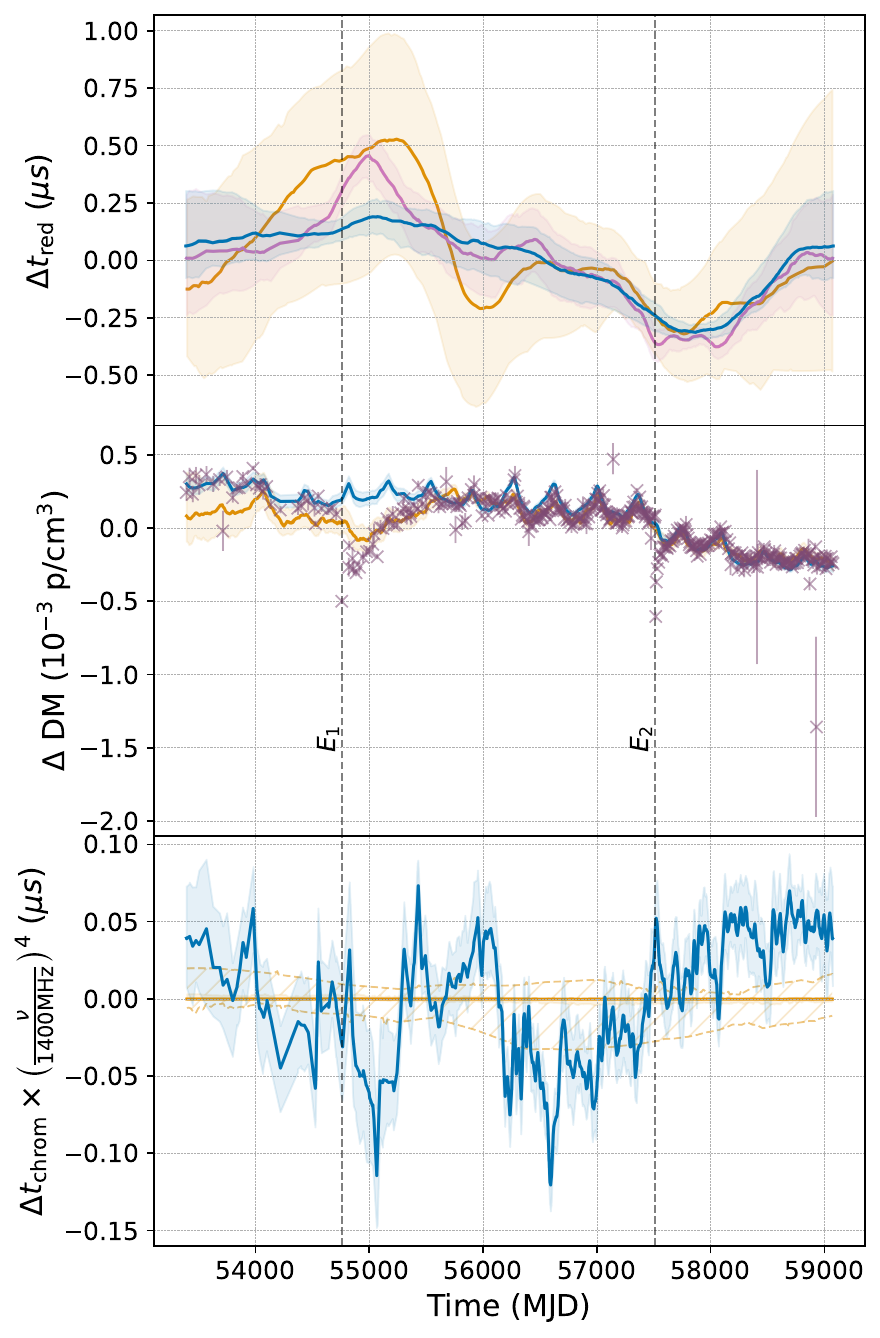}} \\
        \includegraphics[width=0.24\linewidth]{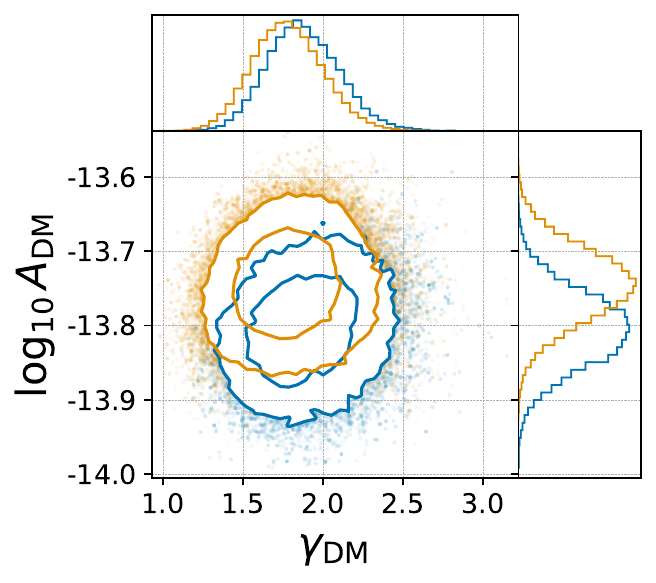} & & \\
        \includegraphics[width=0.24\linewidth]{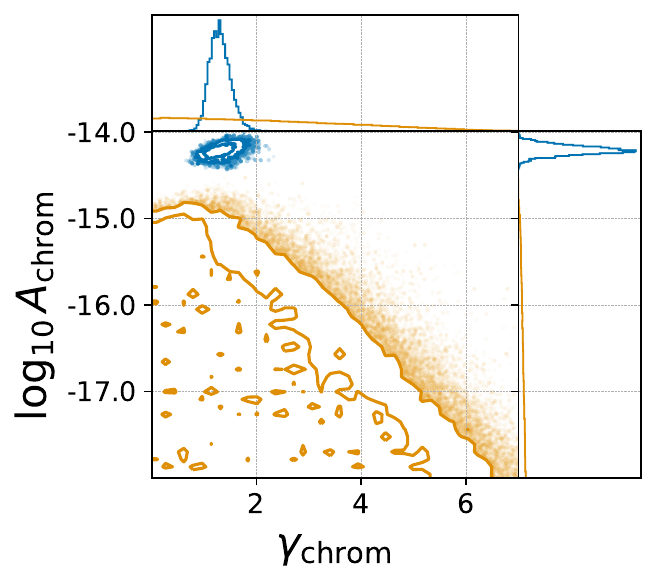} & &
    \end{tabularx}
    %\vspace{-1\baselineskip}
    \caption{\textbf{PSR J1713+0737 -- Simultaneous achromatic red noise and DM variations near the second transient timing event are decoupled using \texttt{CustomGP}.} 
    The times of events $E_1$ and $E_2$ are indicated by the dashed lines.
    Using \texttt{DMX}, $E_2$ manifests as a dip in both the \citetalias{NG15} DMX time series and time-domain ARN realizations.
    Modeling $E_2$ in \texttt{CustomGP} using a free chromatic index successfully decouples this event from both the ARN and DM noise. 
    This lowers the value of $\log_{10}A_{\text{RN}}$ and widens the posterior on $\gamma_{\text{RN}}$.
    However, the ARN and \Comment{scattering-like chromatic} noise parameters in \citetalias{NG15} are in tension with \citetalias{EPTA_data}.
    See the Figure~\ref{fig:J0613} caption for panel descriptions.}
    \label{fig:J1713}
\end{figure*}

Using both \texttt{DMX} and \texttt{CustomGP}, we notice simultaneous, anti-correlated variations in both the time-domain ARN and chromatic noise realizations, particularly near MJD 55000.
Where \texttt{DMX} features simultaneous ARN and DM variations, \texttt{CustomGP} shows simultaneous ARN and $\chi = 4$ \Comment{scattering-like chromatic} variations, with little support for power law DM noise (Table~\ref{tab:noise_params}). 
The anti-correlated nature of these chromatic and achromatic variations is highly unusual, and signifies a high chance of chromatic mismodeling (see Appendix~\ref{appendix:chromatic_modeling}).
To diagnose the issue using \texttt{CustomGP}, we highlight a sudden achromatic delay near MJD 56000, which corresponds to a sudden advance in the \Comment{scattering-like chromatic} noise at the same time.
At $\nu = 800$ MHz, the median chromatic advance near MJD 56000 corresponds to $\Delta t_{\text{chrom}} \sim -3.36$ $\mu$s, which is nearly the same amplitude as the achromatic delay near MJD 56000 ($\Delta t_{\text{ARN}} \sim 3.71$ $\mu$s).
As such, at the time of the event, both noise processes together ($\Delta t_{\text{chrom}} + \Delta t_{\text{ARN}}$) effectively cancel out in the lowest frequency band.
The achromatic delay remains in the higher radio frequency bands, while the inferred chromatic delay decays down to $|\Delta t| < 360$ ns at $\nu > 1400$ MHz.
Interestingly, no evidence for these anti-correlated variations is observed using \citetalias{EPTA_data}. 
Additionally, the \Comment{scattering-like chromatic} noise parameters are in major tension, as $\log_{10}A_{\text{chrom}}$ from \citetalias{NG15} is over an order of magnitude larger than the upper limits set by \citetalias{EPTA_data}.

\subsection{PSR J1600$-$3053}
\label{subsec:J1600}

The choice of chromatic noise model has a noticeable effect on the inferred ARN parameters (Figure~\ref{fig:red_noise}). 
\texttt{DMX} detects a shallow-spectrum ARN process, while the spectrum is much steeper using \texttt{CustomGP}. 
In the time-domain, short-timescale ARN fluctuations with \texttt{DMX} are replaced by short-timescale \Comment{scattering-like chromatic} variations (with opposite sign) using \texttt{CustomGP}.
This is especially clear near MJDs 55000, 56500, 57800, and 58000.
The sign change may result from the DM model overcompensating for an unmodeled scattering delay at low radio frequencies (see Appendix~\ref{appendix:chromatic_modeling} for an example of this). 
Interestingly, the ARN spectrum is not only steeper using \texttt{CustomGP}, but \emph{amplified} at the lowest Fourier mode (Figure~\ref{fig:J1600}, top panel). 
This begins to raise the power law ARN posteriors inferred using \texttt{CustomGP} above the upper limits from \citetalias{EPTA_data}.

\begin{figure*}[ht!]
    \centering
    \vspace{-0.5\baselineskip}
    \begin{tabularx}{\textwidth}{ll}
        \includegraphics[width=0.49\linewidth]{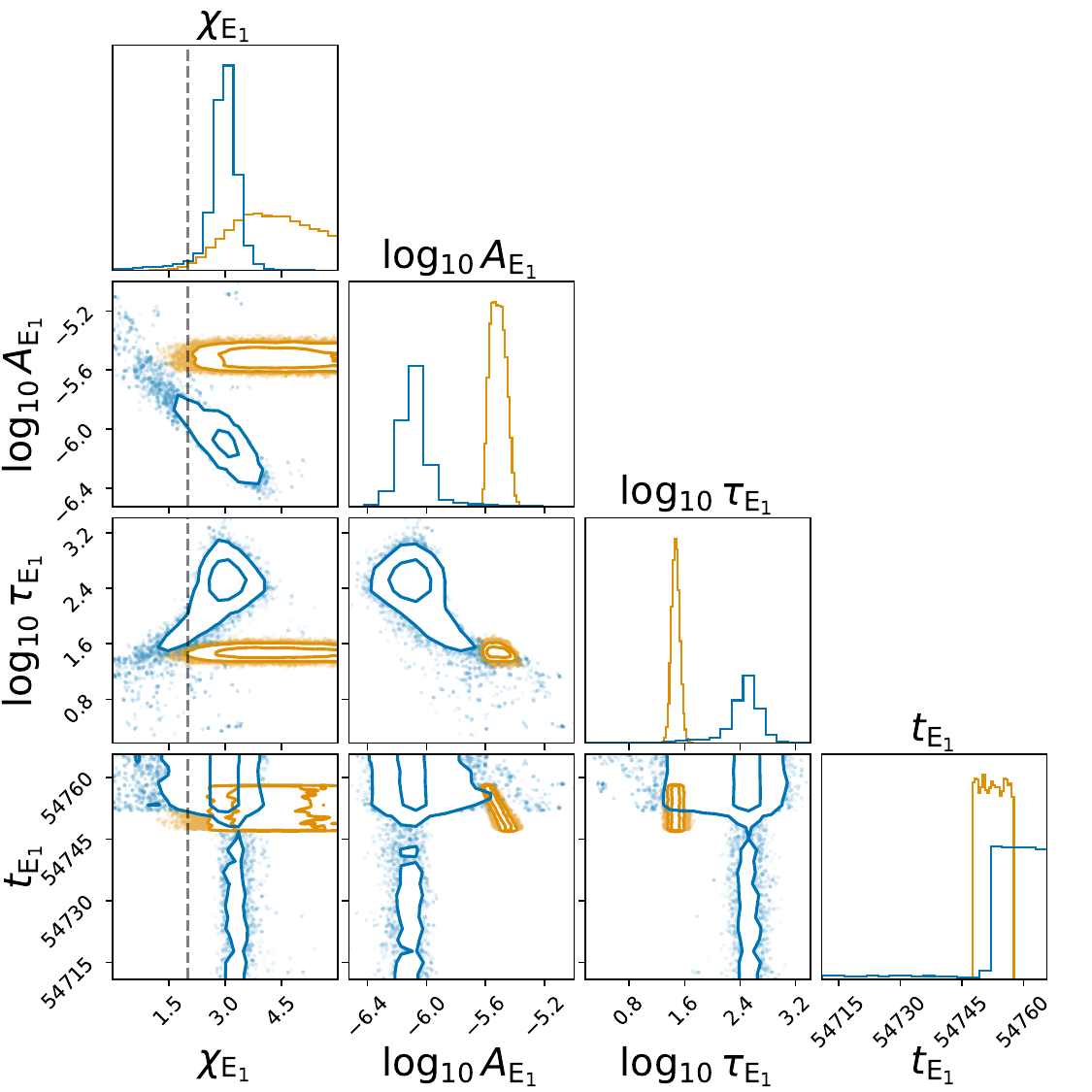} &
        \includegraphics[width=0.49\linewidth]{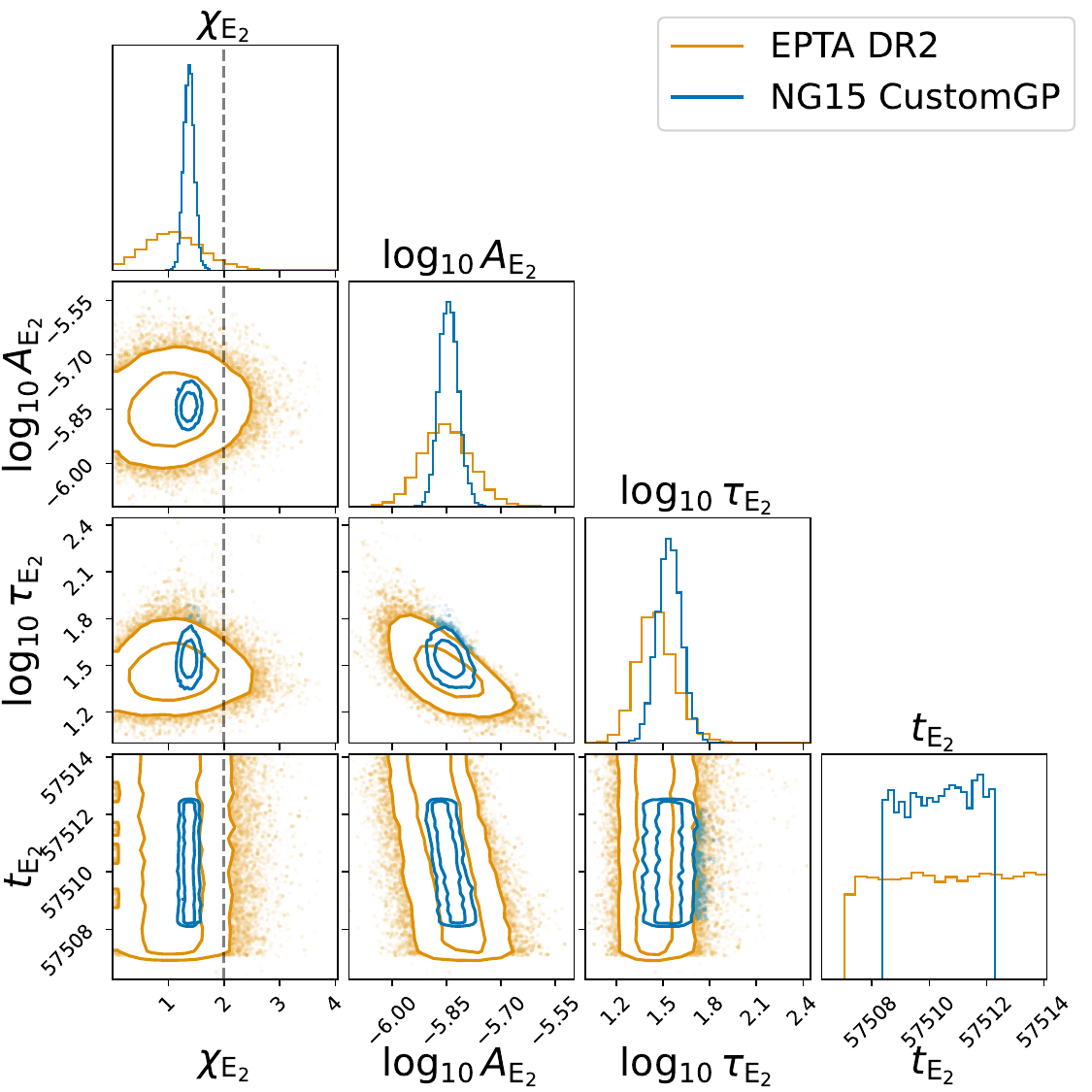}
    \end{tabularx}
    %\vspace{1\baselineskip}
    \caption{Posterior distributions for PSR J1713+0747's exponential dip parameters inferred using model \texttt{CustomGP} with both \citetalias{NG15} and \citetalias{EPTA_data}, \Comment{where the dashed line indicates $\chi = 2$ as expected for DM variations}. Both data sets provide consistent estimates on the second dip ($E_2$) model parameters but differ in characterization of the first dip ($E_1$).}
    \label{fig:exp_dips}
\end{figure*}

Both \citetalias{NG15} and \citetalias{EPTA_data} agree on the presence of $\chi = 4$ \Comment{scattering-like chromatic} noise in PSR J1600$-$3053 with $\log_{10}\mathcal{B}^{\text{chrom}} > 2.2$. 
However, Figure~\ref{fig:J1600} shows the chromatic noise parameters estimated by \citetalias{NG15} and \citetalias{EPTA_data} using \texttt{CustomGP} are not consistent. 
\citetalias{EPTA_data} favors a \Comment{scattering-like chromatic} noise spectrum with a similar spectral index but a higher amplitude than \citetalias{NG15}.
Meanwhile, \citetalias{NG15}'s DM noise spectrum deviates from a pure power law, as it is similar to \citetalias{EPTA_data} below 10 nHz but becomes more shallow past 10 nHz.
In the time-domain, several events (a spike near MJD 57500, a large bump between MJDs 55200 and 56500, and a dip near MJD 56700) are characterized as \Comment{scattering-like chromatic} events by \citetalias{EPTA_data} but as DM events by \citetalias{NG15}, or vice versa.
\Comment{These discrepancies suggest a PSR J1600$-$3053 could benefit from a modified chromatic noise model, e.g., a model with a varied radio-frequency dependence.}

\subsection{PSR J1713+0747}
\label{subsec:J1713}

Using \texttt{CustomGP}, PSR J1713+0747's ARN parameters change significantly ($\log_{10}A_{\text{RN}}$ by $>$1-$\sigma$; Table~\ref{tab:noise_params}) over the use of \texttt{DMX} and \texttt{DMGP}.
We find that allowing $\chi_{E_2}$ to vary as a free parameter in \texttt{CustomGP} is directly responsible for this change.
Figure~\ref{fig:J1713} shows at the time of $E_2$ there is a noticeable dip in both the DMX time series and the time-domain ARN realizations, i.e., the ARN and DM variations are coupled at the time of $E_2$. 
Allowing $\chi_{E_2}$ to vary successfully decouples the event from the ARN and DM variations, as evidenced by the lack of excess noise in the time-domain ARN and DM realizations at the time of $E_2$ using \texttt{CustomGP}.
Furthermore, the power in the 8th frequency bin of PSR J1713+0747's ARN free spectrum experiences a major drop in detection significance when using \texttt{CustomGP} (Appendix~\ref{appendix:FS}, Figure~\ref{fig:FS}).
This may be relevant for GWB characterization, as the 8th frequency bin of the GWB free spectrum has been identified as a driver in pushing the the fit for $\gamma_{\text{GWB}}$ to lower values (\citetalias{NG15_gwb}; \Comment{\citealt{Agazie2024_discreteness}}).

Despite using the same models for \citetalias{NG15} and \citetalias{EPTA_data} with \texttt{CustomGP}, there are several inconsistencies between the two data sets. 
For one, the recovered \Comment{scattering-like chromatic} noise parameters using \citetalias{NG15} are above the upper limit set by \citetalias{EPTA_data}.
Furthermore, the recovered ARN amplitude is lower in \citetalias{NG15} than \citetalias{EPTA_data}.
\citetalias{3P+paper} report a similar discrepancy between \citetalias{NG15} and a version of EPTA DR2 without legacy data at the 1.4-$\sigma$ level. 
\Comment{The comparison of ARN realizations shows that the ARN is less consistent between the two datasets prior to MJD 57000, while it is more consistent after MJD 57000.
Understanding the nature of these differences may be useful to improve PSR J1713+0747's noise modeling in the future.}

%\captionsetup[figure]{skip=-8pt}
\begin{figure*}[ht!]
    \centering
    \vspace{-0.5\baselineskip}
    \begin{tabularx}{\textwidth}{lll}
        \includegraphics[width=0.24\linewidth]{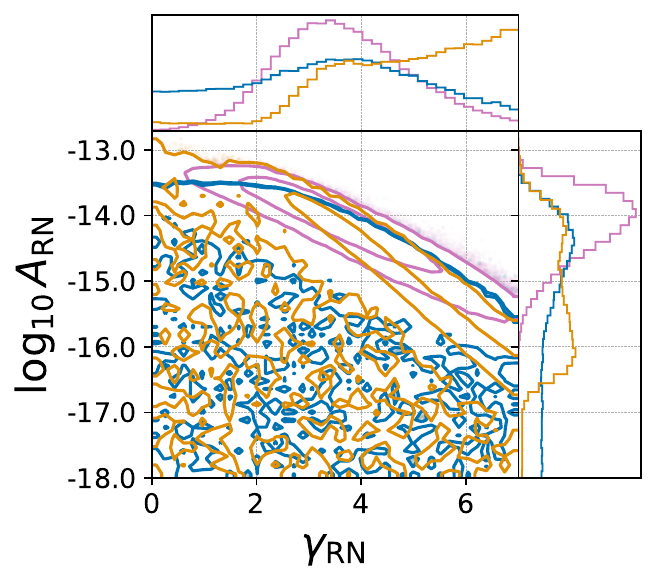} &
        \multirow{3}{*}[35mm]{\includegraphics[width=0.285\linewidth]{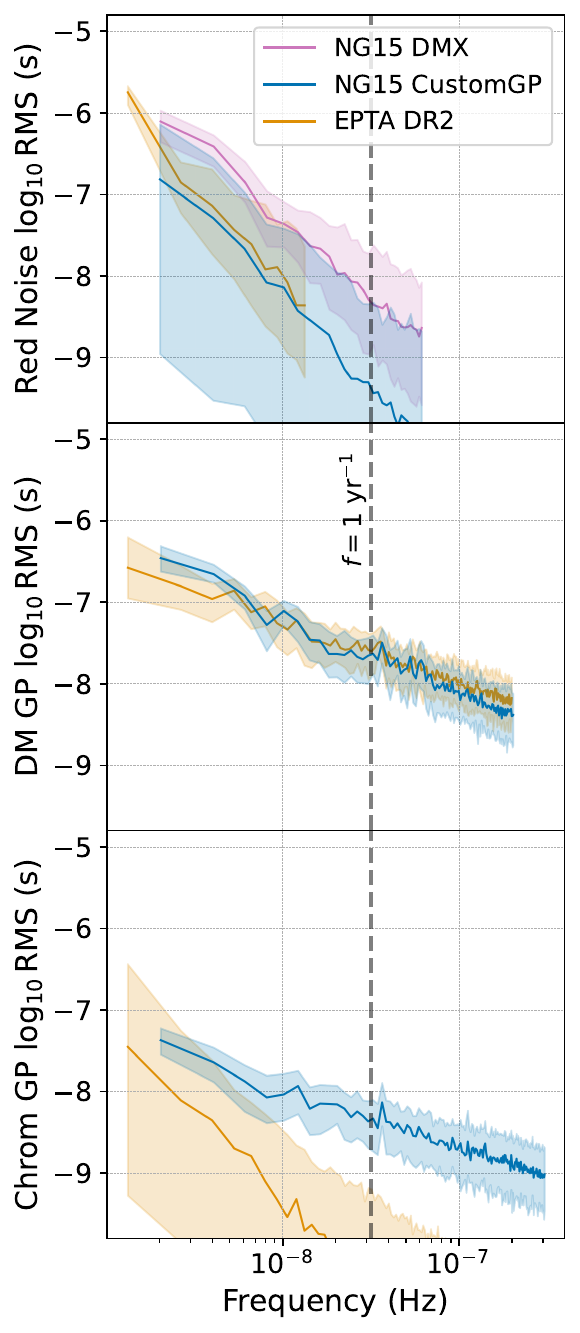}} &
        \hspace{-4mm}\multirow{3}{*}[35mm]{\includegraphics[width=0.44\linewidth]{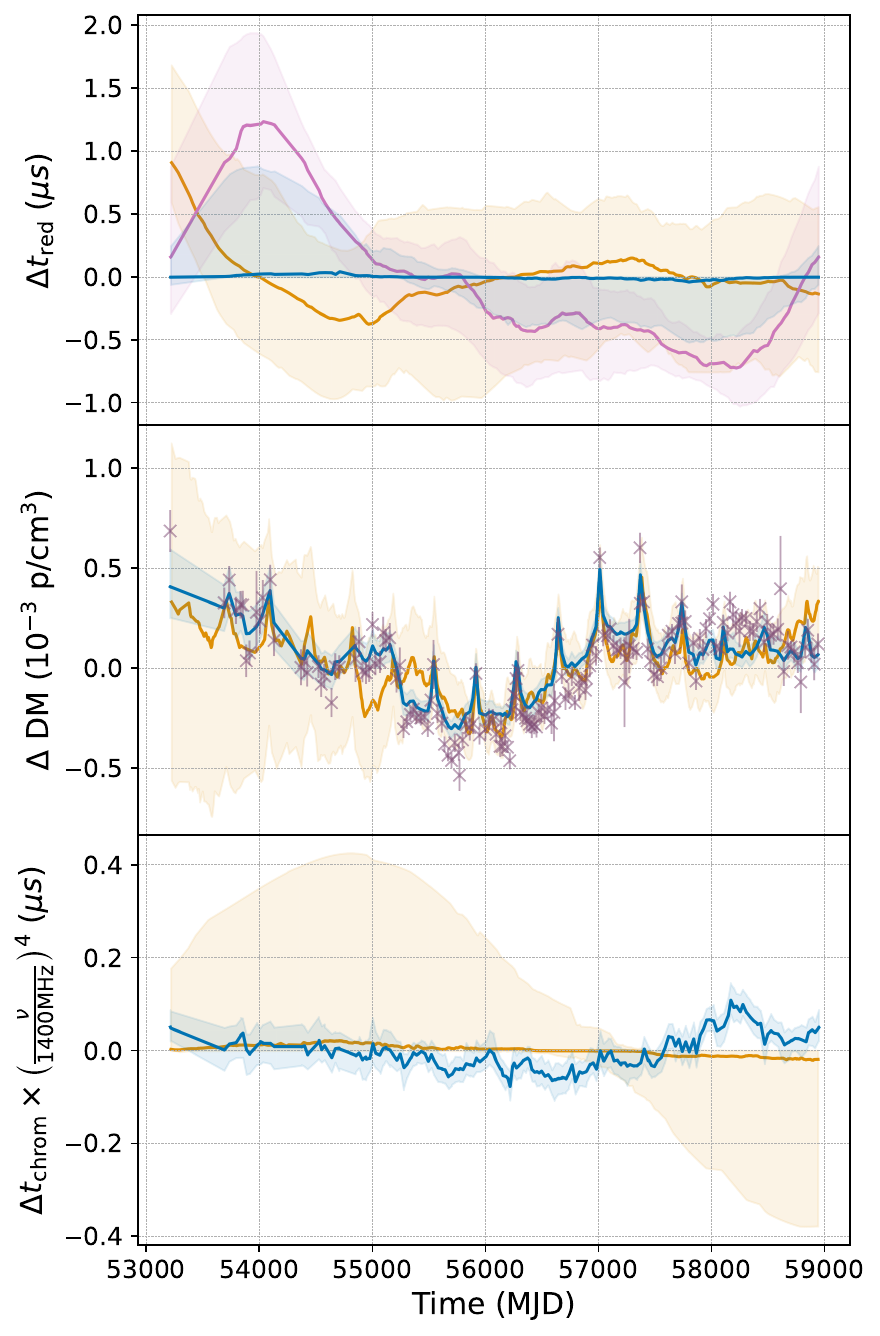}} \\
        \includegraphics[width=0.24\linewidth]{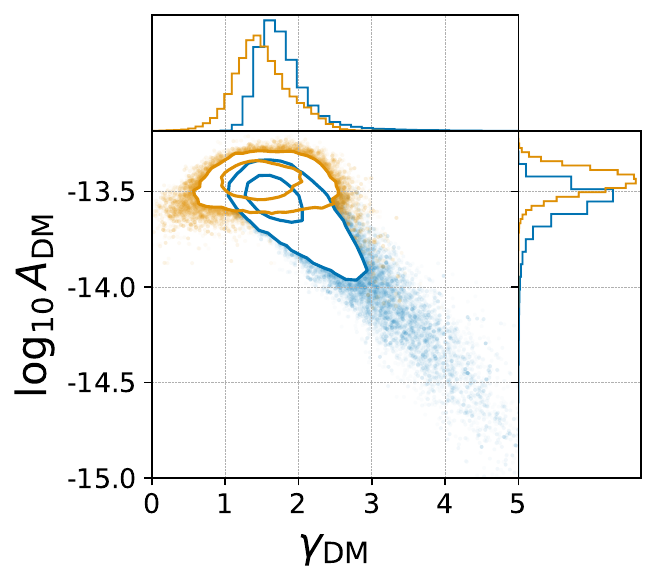} & & \\
        \includegraphics[width=0.24\linewidth]{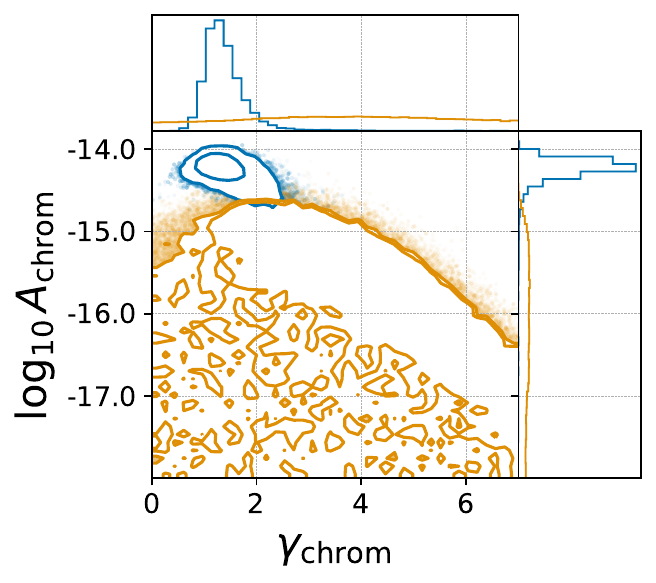} & &
    \end{tabularx}
    \vspace{-0.5\baselineskip}
    \caption{\textbf{PSR J1744$-$1134 -- Use of \texttt{CustomGP} reduces the discrepancy in achromatic red noise characterization between \citetalias{NG15} and \citetalias{EPTA_data}.}
    When applying \texttt{CustomGP} to \citetalias{NG15}, the detection of ARN is reduced to an upper limit.
    This improves the consistency in ARN characterization between data sets, \Comment{as indicated by their ARN spectra.}
    However, the detection of \Comment{scattering-like chromatic} noise in \citetalias{NG15} is above the upper limit set by \citetalias{EPTA_data}.
    See the Figure~\ref{fig:J0613} caption for panel descriptions.}
    \label{fig:J1744}
\end{figure*}

Figure~\ref{fig:exp_dips} shows posterior distributions for the exponential dip model parameters, inferred using both \citetalias{NG15} and \citetalias{EPTA_data}. 
For $E_1$, \citetalias{EPTA_data} favors a high amplitude, short decay timescale, and a chromatic index $\chi_{E_1} = 4.2^{+1.1}_{-1.0}$, whereas \citetalias{NG15} favors a lower amplitude, longer recovery timescale, and a chromatic index $\chi_{E_1} = 3.0^{+0.3}_{-0.3}$. 
The \citetalias{NG15} $E_1$ posteriors also have long tails and are covariant, requiring a larger amplitude and a smaller chromatic index as the decay timescale becomes smaller.
The inconsistent characterization of $E_1$ between data sets is explained by the uneven properties of each data set near $t_{E_1}$.
\citetalias{EPTA_data} features multiple TOAs from the NRT, WSRT, and EFF with an average cadence of $\sim$3.5 days, but only at 1400 MHz and above.
\citetalias{NG15} features lower frequency (800 MHz) TOAs at MJD 54765 from the GBT, but these are not followed up with higher frequency (1400, 2300 MHz) TOAs from the GBT and AO until MJD 54819.

The \citetalias{NG15} and \citetalias{EPTA_data} posteriors for $E_2$ are more consistent with each other, only featuring differences in their variances (Figure~\ref{fig:exp_dips}, right side).
\citetalias{NG15} includes many subbanded TOAs and thus a high radio frequency resolution near $t_{E_2}$, which explains why its posterior on $\chi_{E_2}$ is more constrained than \citetalias{EPTA_data}.
The chromatic index of $E_2$ is $\chi_{E_2} = 1.1^{+0.5}_{-0.5}$ using \citetalias{EPTA_data} and $\chi_{E_2} = 1.37^{+0.09}_{-0.09}$ using \citetalias{NG15}.
These values are also consistent with $\chi = 1.15^{+0.18}_{-0.19}$ estimated by \citet{Goncharov2021} using PPTA DR2, \Comment{supporting the proposal therein that these events originate from the pulsar's magnetosphere.}
Furthermore, the observation that $0 < \chi_{E_2} < 2$ critically explains why this event manifests as both excess ARN and DM noise using the models \texttt{DMX} and \texttt{DMGP}. 

\subsection{PSR J1744$-$1134}
\label{subsec:J1744}

\Comment{We begin by noticing some differences in the ARN properties of \citetalias{EPTA_data} and \citetalias{NG15} when using \texttt{DMX}.
Specifically, in the top middle panel of Figure~\ref{fig:J1744}, the ARN spectrum using \citetalias{NG15} with \texttt{DMX} is at a higher amplitude than the spectrum using \citetalias{EPTA_data} across the region of frequency space where the spectra overlap. 
In the time-domain, there are peaks and troughs in \citetalias{NG15}'s ARN realizations near MJDs 54000 and 58000 which are not present using \citetalias{EPTA_data}.
Note that PSR J1744-1134's ARN is dominated in \citetalias{EPTA_data} by the lowest frequency bin, which would manifest in the time-domain as a single sinusoidal trend if using the \citetalias{EPTA_data} data from before MJD 53000 \citep{Chalumeau+2022}.}

\begin{figure*}[ht!]
    \centering
    \vspace{-0.5\baselineskip}
    \begin{tabularx}{\textwidth}{lll}
        \includegraphics[width=0.24\linewidth]{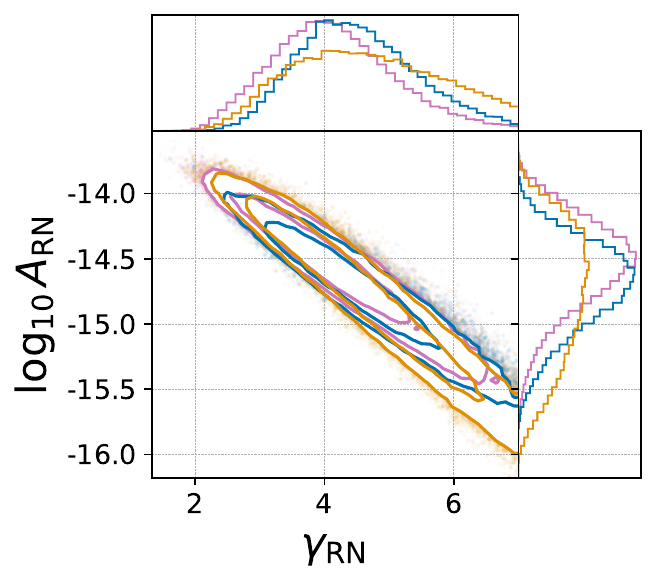} &
        \multirow{3}{*}[35mm]{\includegraphics[width=0.285\linewidth]{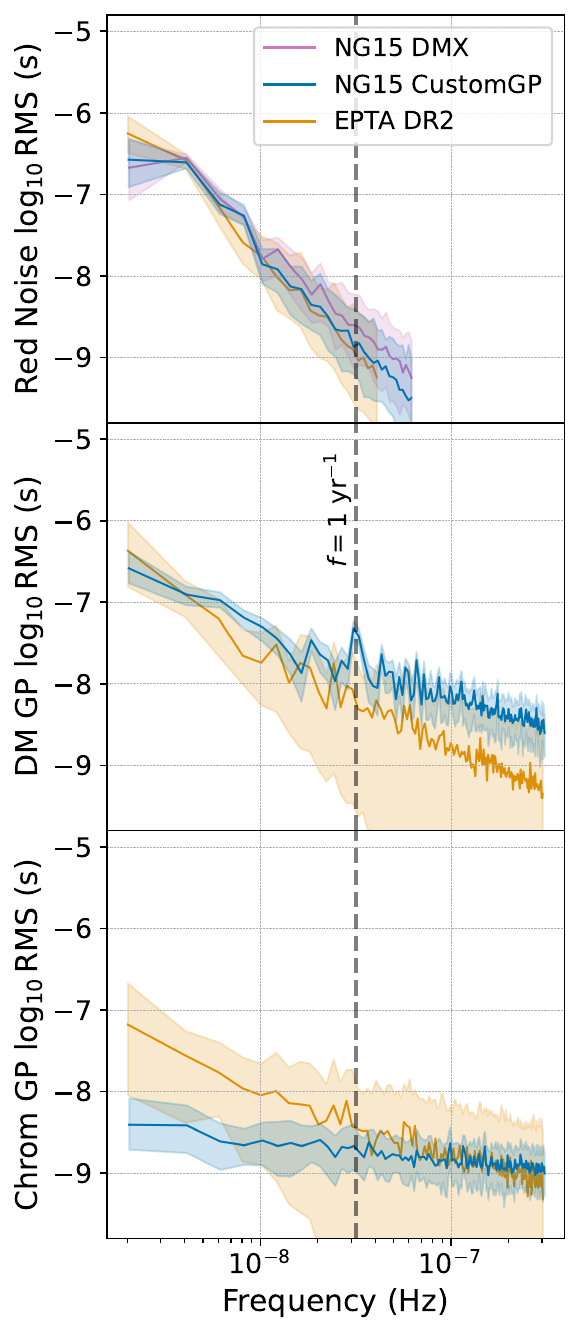}} &
        \hspace{-4mm}\multirow{3}{*}[36mm]{\includegraphics[width=0.44\linewidth]{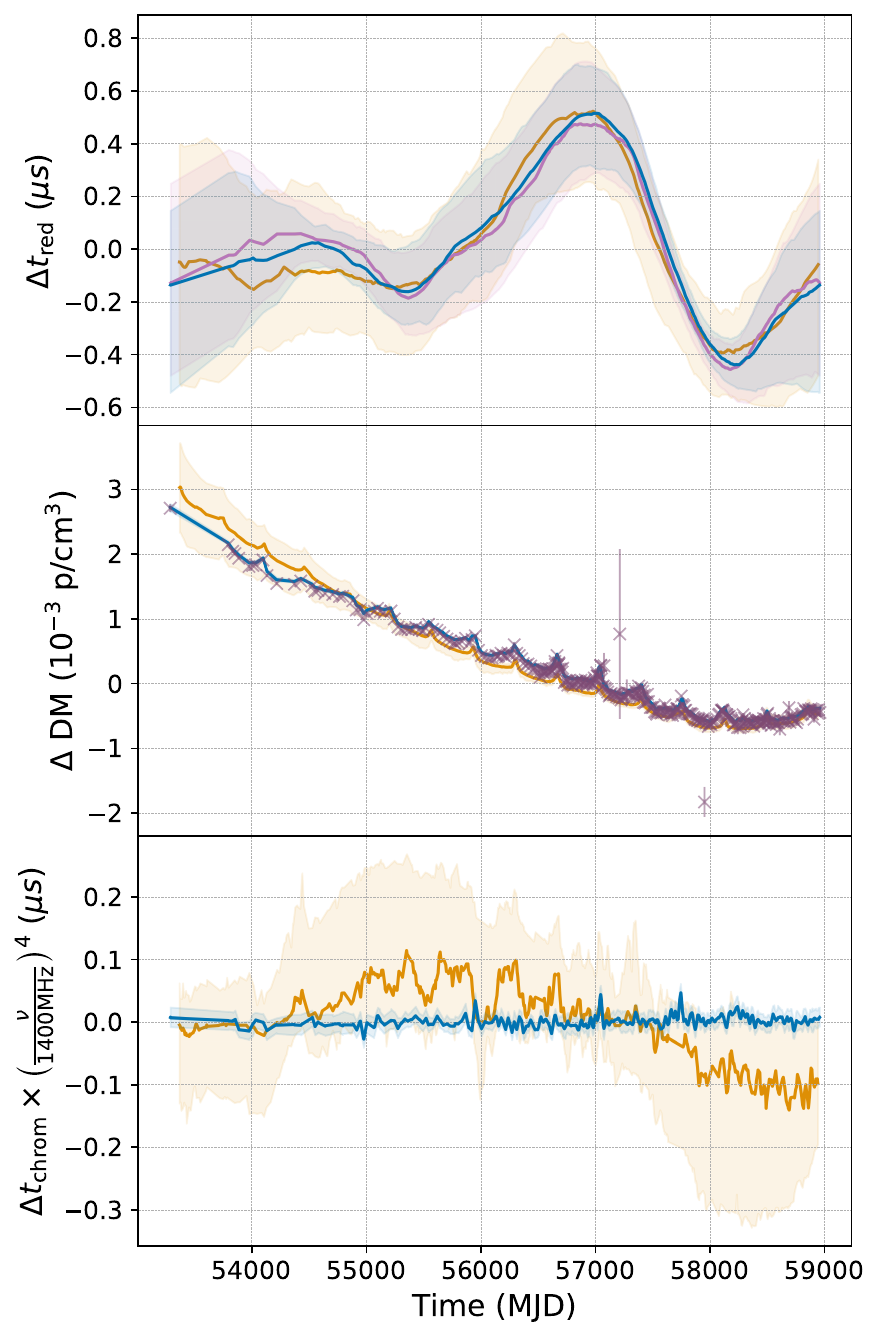}} \\
        \includegraphics[width=0.24\linewidth]{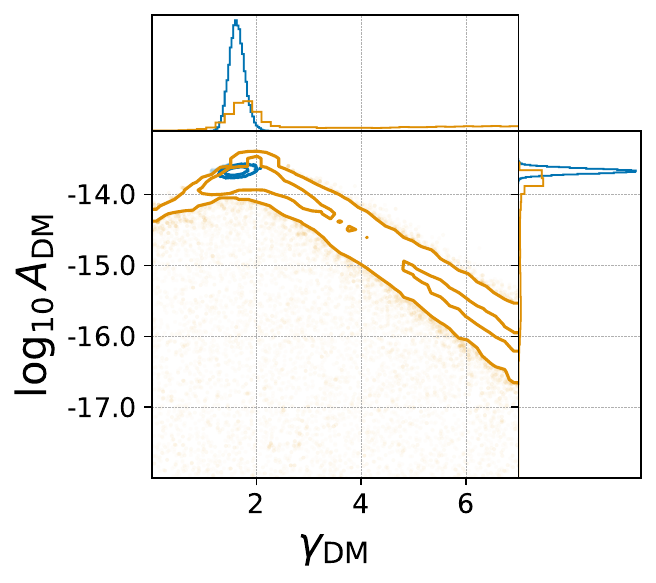} & & \\
        \includegraphics[width=0.24\linewidth]{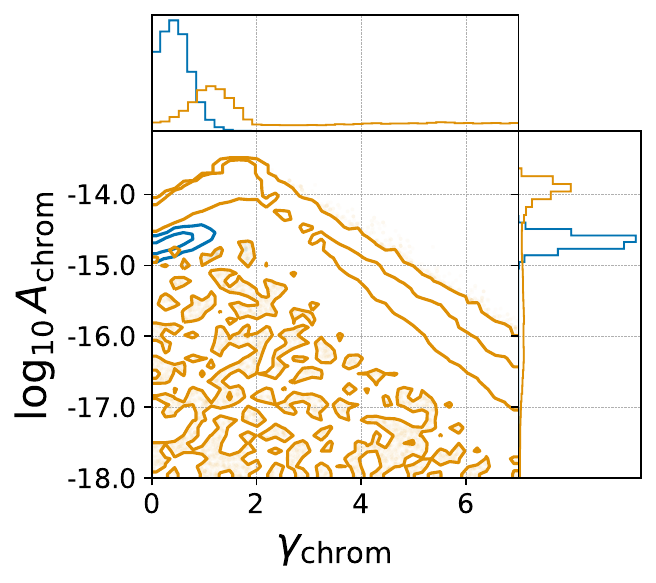} & &
    \end{tabularx}
    \vspace{-0.5\baselineskip}
    \caption{\textbf{PSR J1909$-$3744 -- Achromatic red noise is robust to the choice of chromatic noise model.}
    PSR J1909$-$3744's ARN in \citetalias{NG15} experiences little to no change going from \texttt{DMX} to \texttt{CustomGP}, indicating any errors in chromatic noise mitigation are small and decoupled from ARN.
    ARN characterization is also remarkably consistent between \citetalias{NG15} and \citetalias{EPTA_data}.
    Spikes in the \citetalias{NG15} time series for \Comment{scattering-like chromatic} noise are also observed near annual DM cusps, indicating they likely originate from the heliosphere.
    See the Figure~\ref{fig:J0613} caption for panel descriptions.}
    \label{fig:J1909}
\end{figure*}

Using \texttt{CustomGP} reduces the ARN Bayes Factor and the median amplitude of ARN over both the models \texttt{DMX} and \texttt{DMGP} (Table~\ref{tab:noise_params}). 
\Comment{In particular, the dip in the ARN realizations near MJD 58000 using \texttt{DMX} is now absorbed by the DM variations using \texttt{CustomGP}, with a corresponding bump in the scattering-like chromatic noise to counterbalance.}
As a result, the ARN properties are also now more consistent with \citetalias{EPTA_data} (Figure~\ref{fig:J1744}). 
That said, the new ARN posteriors still overlap entirely with the ARN posteriors inferred under models \texttt{DMX} and \texttt{DMGP} (as well as the GWB posteriors), as such the ARN process detected using \texttt{DMX} is not ruled out. 
\Comment{Information from additional frequency bands in future datasets will help us better decouple and understand these signals.}

The $\chi = 4$ \Comment{scattering-like chromatic} noise detected in \citetalias{NG15} is also above the upper limits set by \citetalias{EPTA_data}.
This discrepancy \emph{appears} to be less significant in PSR J1744$-$1134 than the other pulsars, as the \Comment{scattering-like chromatic} noise spectra of \citetalias{NG15} and \citetalias{EPTA_data} still overlap at the lowest Fourier modes (Figure~\ref{fig:J1744}).
However, the time-domain \Comment{scattering-like chromatic} noise realizations allowed by \citetalias{EPTA_data} do not line up with the \citetalias{NG15} \Comment{scattering-like chromatic} noise realizations.

\subsection{PSR J1909$-$3744}
\label{subsec:J1909}

Unlike the preceding pulsars, there is very little change to PSR J1909$-$3744's ARN parameters when switching from \texttt{DMGP} to \texttt{CustomGP}, aside from a very minute increase in $\gamma_{\text{RN}}$ (Table~\ref{tab:noise_params}).
There is also no change to PSR J1909$-$3744's free spectrum below $f = 1$ yr$^{-1}$ (Appendix~\ref{appendix:FS}, Figure~\ref{fig:FS}).
This indicates the inferred ARN signal in PSR J1909$-$3744 is robust to the choice of chromatic noise model.
The ARN is also extremely consistent with \citetalias{EPTA_data} (Figure~\ref{fig:J1909}).

Using \texttt{CustomGP}, the \citetalias{NG15} DM noise spectrum features an excursion from the power law prior at $f = 1$ yr$^{-1}$, indicating support for an annual DM process. 
We also detect $\chi = 4$ \Comment{scattering-like chromatic} variations in PSR J1909$-$3744, characterized by a nearly flat spectrum ($\gamma_{\text{chrom}} = 0.5^{+0.3}_{-0.3}$).
This was also found by \citet{Srivastava2023} using InPTA DR1 \citep{Tarafdar+2022}.
While \citetalias{EPTA_data} does not detect a significant scattering-like noise process, this introduces no tension with \citetalias{NG15}, since the \citetalias{NG15} \Comment{scattering-like chromatic} noise posteriors are below the \citetalias{EPTA_data} upper limits. 
The \citetalias{NG15} time-domain \Comment{scattering-like chromatic} noise realizations feature three sharp spikes that all align with annual DM cusps from the deterministic solar wind model.
This suggests either that the \Comment{scattering-like chromatic} model is capturing scattering variations from within the heliosphere, or that it is compensating for excess DM from a time-variable solar wind electron density.

\section{Summary and Discussion}
\label{sec:discussion}

% Point: need to use custom noise models for some pulsars
We compare three different chromatic noise models on a subsample of six pulsars from \citetalias{NG15}.
Since five out of these six pulsars ARN processes are spectrally similar to the GWB signal \citepalias{NG15detchar}, we pay special attention to the model-dependence of the ARN.
Out of these five, PSR J1713+0747 is the only pulsar whose ARN parameters change significantly (at a $>$1-$\sigma$ level) using the \texttt{CustomGP} model over the \texttt{DMX} model.
The change is directly linked to the modeling of its non-dispersive chromatic timing event near MJD 57510 \citep{Lam2018, Hazboun2020}.
\Comment{Since the GWB spectrum inferred in \citetalias{NG15_gwb} is similar to the ARN in just PSR J1713+0747 using \texttt{DMX}, this change is likely to impact GWB spectral characterization.}

% Point: Custom pulsar noise models impacts on spectral characterization
\Comment{Alongside PSR J1713+0747,} several of these pulsars favor steeper ARN spectra under the \texttt{CustomGP} models.
This result is unsurprising, since we detect excess non-dispersive chromatic noise in these pulsars, and unmitigated chromatic noise is expected to result in low spectral index ARN (\citealt{CordesShannon2010}; \citetalias{NG15detchar}).
\Comment{However,  both \citetalias{NG15_gwb} and \citet{PPTA_noise} found preference for steeper GWB spectra using alternative pulsar noise models similar to the \texttt{DMGP} and \texttt{CustomGP} models used here.
Since the GWB spectrum is used to inform astrophysical interpretations of the signal (\citealt{Phinney2001}; \citetalias{NG15_astro}; \citealt{Agazie2024_discreteness}), we plan to continue exploring effects of chromatic noise models on the full PTA in an upcoming work.
Furthermore, it will be worth investigating if mitigation of excess chromatic noise in the dataset could improve the measurements of Hellings and Downs cross-correlations between pulsar pairs, as suggested by \citet{diMarco2024}.}

% Point: take advantage of pulsars that are not affected by custom noise models
It is also interesting that the changes to the ARN properties under the \texttt{CustomGP} model are much smaller in some pulsars (e.g., PSR J1909$-$3744) than in others.
As such, it is possible that applying the \texttt{CustomGP} model on some pulsars could have little to no impact on GWB analyses, even if \texttt{CustomGP} provides a better fit to the pulsar's TOAs. 
In a future work, the impact of the choice of noise model for individual pulsars on GWB characterization could be assessed e.g., from factorized likelihood methods \citep{Taylor2022, Lamb2023}.
This could then be used to expand upon existing model selection methods (e.g., \citealt{Lentati2016, HazbounSimon2020, Goncharov2021}) by only using custom noise models for pulsars that measurably impact GWB spectral characterization (or other results of interest).
Since the \texttt{CustomGP} model is more computationally expensive than the \texttt{DMX} model, this could help reduce the computational burden of cross-correlation analyses of the GWB (which require simultaneous analysis of all pulsars) with custom noise models.

% Discuss secondary science from GPs: solar wind models
Custom pulsar noise models are also useful tools to study ISM processes.
For instance, the distribution of $\gamma_{\text{DM}}$ values can be used to assess if the ISM is consistent with the model of Kolmogorov turbulence, which predicts  $\gamma_{\text{DM}} = 8/3$ \citep{Keith2013, Lentati2016, Goncharov2021}. 
Here we find $\gamma_{\text{DM}} < 8/3$ when using \texttt{DMGP} for these six pulsars in \citetalias{NG15}.
However, for PSRs J0613$-$0200 and J1600$-$3053, consistency with $\gamma_{\text{DM}} = 8/3$ improves using \texttt{CustomGP} (Table~\ref{tab:noise_params}).
Separating DM variations into a stochastic GP and a SW component is also very useful for estimating the SW electron density, $n_{\text{Earth}}$ (Appendix~\ref{appendix:SW}, \citealt{Madison2019, Hazboun2022, Kumar2022, PPTA_noise}).
Since the model we use here assumes $n_{\text{Earth}}$ is constant, a time-variable SW will still induce excess noise in the DM noise spectrum at harmonics of $f = 1$ yr$^{-1}$ \citep{Hazboun2022}.
We observe this here for PSR J1909$-$3744, and expect this effect to be much more pronounced for pulsars close to the ecliptic.
Introducing an additional GP to vary $n_{\text{Earth}}$ over time is a promising method to mitigate this effect \citep{Hazboun2022, Nitu2024}. 

% Point: Custom noise models appear to improve consistency with EPTA DR2 in most but not all cases
To validate our results, we compare the inferred ARN properties from both \citetalias{NG15} and \citetalias{EPTA_data} using the \texttt{CustomGP} model, which is based on the favored models from \citetalias{EPTA_noise}. 
This analysis complements \citetalias{3P+paper}, who have also compared pulsar noise properties from recent PTA data sets using the same noise models.
For most pulsars, we find the inferred ARN properties are consistent using \texttt{CustomGP}.
Furthermore, applying \texttt{CustomGP} instead of \texttt{DMX} to three pulsars in \citetalias{NG15} (PSRs J0613$-$0200, J1012+5307, and J1744$-$1134) alleviated some discrepencies in ARN characterization between the two data sets.
This strengthens our confidence the model \texttt{CustomGP} is improving ARN estimation for these pulsars.
However, PSRs J1600-3053 and J1713+0747 is an exception: the ARN detected using \citetalias{NG15} is lower in amplitude than the ARN detected using \citetalias{EPTA_data}.
This discrepancy was already found by \citetalias{3P+paper}, but here we confirm its existence using the \texttt{CustomGP} model. 
\citetalias{EPTA_noise} also recover similarly high ARN amplitudes for PSR J1713+0747 using both alternative versions of \citetalias{EPTA_data}. 
Analysis of a future combined data set (IPTA DR3) may help resolve this inconsistency and assess any potential impacts on GWB inferences.
It would also be useful to extend the comparison to other datasets, e.g., by applying the PPTA DR3 noise models \citep{PPTA_noise} to the \citetalias{NG15} data and comparing the results.

% Point: Scattering noise discrepancy has a plausible explanation
This comparison of data sets reveals another major discrepancy in the \Comment{scattering-like chromatic} noise: four out of six pulsar’s scattering-like noise amplitudes estimated using \citetalias{NG15} are above the upper limits set by \citetalias{EPTA_data}. 
Throughout this work we have used $\Delta t \propto \nu^{-\chi}$ with $\chi = 4$ to describe \Comment{scattering-like chromatic} noise.
However, an index $\chi < 4$ may resolve the above discrepancy.
We suggest this because the \Comment{scattering-like chromatic} noise detected using \citetalias{NG15} is only above the upper limit set by \citetalias{EPTA_data} for pulsars where \citetalias{EPTA_data} contains low radio frequency ($\sim 300$ MHz) TOAs from the WSRT (Figure~\ref{fig:datasets}). 
If these low-frequency TOAs are responsible for ruling out a $\chi = 4$ scattering-like noise process, a lower chromatic index would likely reduce the delay at the lowest radio frequencies and therefore raise the upper limits in \citetalias{EPTA_data}. 
Supporting evidence for scattering scaling indices $\chi < 4$ has been found in several past studies \citep{Lewandowski2015, Levin2016, Turner2021}, and the scaling indices may also vary over time \citep{Bansal2019, Liu2022}.
\Comment{Other chromatic processes, such as low-level profile variations, also need not have $\chi = 4$ dependence or even a power-law dependence on radio frequency.
If the \Comment{scattering-like chromatic} noise GP is absorbing some additional chromatic process, this is also a viable explanation for the discrepancy.
Ultimately, the $\chi = 4$ \Comment{scattering-like chromatic} noise GP is still a valuable phenomenological component of the noise model, as it is still favored with a large Bayes factor for all six pulsars.}

% Point: We learn a lot by looking at GP coupling
In particular, we suggest that cases of chromatic noise processes displaying time-correlations with one another, or ``competing'' with each other, may highlight cases of chromatic mismodeling.
Assuming a two-radio-frequency measurement of the TOA, modeled as an ARN and DM process, an excess achromatic delay will be introduced with the opposite sign of any unmodeled scattering delay (Appendix~\ref{appendix:chromatic_modeling}).
This occurs since the unmodeled scattering delay is primarily absorbed by the DM model, rather than the ARN.
We suggest this is potentially taking place for PSRs J0613$-$0200 and J1600$-$3053 in \citetalias{NG15}, as based on our comparisons of the ARN and chromatic noise realizations using the \texttt{DMX} and \texttt{CustomGP} models.
We similarly find that incomplete modeling of the 2nd chromatic event ($E_2$) in PSR J1713+0747 (which is found to have a chromatic index of $\chi = 1.37^{+0.09}_{-0.09}$) results in excess ARN and DM noise at the exact time of the event.
Furthermore, we find using both \texttt{DMX} and \texttt{CustomGP} that the ARN and \Comment{scattering-like chromatic} noise in PSR J1012+5307 display anti-correlations over time.
The behavior is consistent with the presence of a $\chi < 0$ chromatic process as explored at the end of Appendix~\ref{appendix:chromatic_modeling}.
Its unknown to us what type of physical process this may correspond to, however this idea is supported by \citetalias{EPTA_noise}, who found $\chi = -0.65^{+0.46}_{-0.41}$ after a free-chromatic analysis of PSR J1012+5307 from the joint EPTA DR2 and InPTA DR1 dataset \citep{Tarafdar+2022, EPTA_data}.

% Point: These three models are not exhaustive and future works should improve the models
The GP models used here could benefit from further advances, several of which we will test on more pulsars from \citetalias{NG15} in an upcoming work. 
As discussed, it will be important to assess the evidence of time-variable scattering at chromatic indices other than $\chi = 4$.
This could be assessed using a GP with $\chi$ as a free variable (e.g., \citealt{Goncharov2021, Srivastava2023}; \citetalias{EPTA_noise}).
Furthermore, for pulsars close to the ecliptic, it will be important to implement more sophisticated GP models for time-variability of the SW density (e.g., \citealt{Hazboun2022, Nitu2024}).
GP models could also help mitigate other effects such as frequency-dependent jitter \citep{Lam2019, Kulkarni2024}, frequency-dependent DM variations due to multipath propagation \citep{Cordes2016}, or nonstationary noise \citep{EllisCornish2016}. 
Performing white noise model selection (e.g., \citealt{Srivastava2023, Miles2023}) or including additional ARN processes at high fluctuation frequencies \citep{PPTA_noise} may also reduce the noise floor at higher GW frequencies.
Additionally, PSR J1713+0747 displayed a dramatic pulse profile change in early 2021 \citep{Singha2021}, which introduced timing delays that scale non-monotonically with radio frequency \citep{Jennings2022}.
Mitigating this event at the level of timing residuals would require a more sophisticated chromatic model than those previously used for PSR J1713+0747.

%Point: these pulsars are not exhaustive and we should look at more pulsars
We highlight that the six pulsars we investigate here are only a small subsample of the full \citetalias{NG15} dataset, and are not necessarily representative of the whole.
Many pulsars (e.g., PSRs B1937+21 and J1903+0327) have very different noise properties due either to factors intrinsic to the pulsar or to its location in the ISM \citepalias{NG15detchar}.
Pulsars closer to the ecliptic will be impacted more strongly by the choice of solar wind model \citep{Tiburzi+2021}.
Many pulsars in \citetalias{NG15} also have shorter timespans than the six we study here.
Additionally, all six pulsars here have been observed by the GBT, while many pulsars in \citetalias{NG15} have been primarily observed by the AO.
As such, investigating more pulsars should reveal new discoveries about the implications of chromatic noise modeling choices.

% Point: Future data sets are very promising for noise mitigation
Finally, upcoming data set improvements are very promising for chromatic noise mitigation prospects.
The GBT has been upgraded with the VErsatile GBT Astronomical Spectrometer \citep{Vegas2012} and an ultra wide bandwidth receiver capable of observations up to 3.8 GHz.
Work is underway to install a cyclic spectroscopy backend at the GBT, with the goal to instantly remove scattering effects before any further timing analysis has taken place \citep{Dolch2021, Turner2023}. 
Even if scattering cannot be removed in some pulsars, the high frequency resolution enabled by cyclic spectroscopy can allow more accurate regular measurement of a pulsar’s scintillation bandwidth, and thus the scattering delay \citep{Dolch2021}.
CHIME/Pulsar will also provide observations in a 400-800 MHz bandwidth with daily cadence \citep{Chime2021}. 
Taken together, these developments should allow for highly precise modeling of DM and other chromatic processes in future NANOGrav data sets.
Finally, a future IPTA data set (IPTA DR3) will combine data from all PTAs together to maximize the data cadence, timing baselines, sky coverage, and effective radio frequency coverage achievable using current data sets.
This will allow excellent mitigation of chromatic noise processes and further improve PTA sensitivity to GWs.

%\section*{Acknowledgements}
\emph{Author Contributions:}
This paper is the result of the work of many people and uses data from over a decade of pulsar timing observations.
B.L. performed the analysis, created all figures and tables, and wrote a majority of the text.
C.M.F.M. conceived of the project, supervised the analysis, helped develop the manuscript, and provided advice on figures and interpretation.
C.M.F.M., J.S.H., and D.C.G. provided continuous advice and feedback on results throughout the project.
A.C. assisted with the analysis of EPTA DR2 data and interpretation of the noise analysis results.
J.S.H. and J.S. provided additional expertise on pulsar noise modeling as well as extensive resources and codes which were adapted to perform the analysis.
G.A., A.A, A.M.A., Z.A., P.T.B., P.R.B., H.T.C., K.C., M.E.D, P.B.D., T.D., E.C.F, W.F., E.F., G.E.F., N.G.D., D.C.G., P.A.G., J.G., R.J.J., M.L.J., D.L.K., M.K., M.T.L., D.R.L., J.L., R.S.L., A.M., M.A.M., N.M., B.W.M., C.N., D.J.N., T.T.N., B.B.P.P., N.S.P., H.A.R., S.M.R., P.S.R., A.S., C.S., B.J.S.A., I.H.S., K.S., A.S., J.K.S., and H.M.W. all ran observations and developed timing models for the NANOGrav 15 yr data set.
A.C., D.J.C., I.C., L.G., H.H., M.J.K., K.L., J.W.M., A.Pa., D.P., A.Po., G.M.S., and G.T. contributed observations, timing, and noise models for EPTA DR2.

\emph{Acknowledgements:}
We thank the anonymous reviewer for several useful comments which helped us improve the quality of this paper.
We additionally thank Jacob Turner, Sarah Vigeland, Jeremy Baier, Priyamvada Natarajan, Joris Verbiest, James Andrew Casey-Clyde, London Willson, Jonathan Nay, Ma\l{}gorzata Cury\l{}o, Deven Bhakta, and NANOGrav's Detection and Noise Budget Working Groups for useful input.
B.L. was supported in part by NASA CT Space Grant PTE Federal Award Number 80NSSC20M0129.
C.M.F.M. was supported in part by the National Science Foundation under Grants No. NSF PHY-1748958 and AST-2106552.
A.C. acknowledges support from the Paris Île-de-France Region and financial support provided under the European Union’s H2020 ERC Consolidator Grant “Binary Massive Black Hole Astrophysics” (B Massive, Grant Agreement: 818691).
J.S. is supported by an NSF Astronomy and Astrophysics Postdoctoral Fellowship under award AST-2202388, and acknowledges previous support by the NSF under award 1847938.
P.R.B. is supported by the Science and Technology Facilities Council, grant number ST/W000946/1.
Support for H.T.C. is provided by NASA through the NASA Hubble Fellowship Program grant \#HST-HF2-51453.001 awarded by the Space Telescope Science Institute, which is operated by the Association of Universities for Research in Astronomy, Inc., for NASA, under contract NAS5-26555.
Pulsar research at UBC is supported by an NSERC Discovery Grant and by CIFAR.
K.C. is supported by a UBC Four Year Fellowship (6456).
M.E.D. acknowledges support from the Naval Research Laboratory by NASA under contract S-15633Y.
T.D. and M.T.L. are supported by an NSF Astronomy and Astrophysics Grant (AAG) award number 2009468.
E.C.F. is supported by NASA under award number 80GSFC21M0002.
G.E.F. is supported by NSF award PHY-2011772.
D.R.L. and M.A.M. are supported by NSF \#1458952.
M.A.M. is supported by NSF \#2009425.
The Dunlap Institute is funded by an endowment established by the David Dunlap family and the University of Toronto.
T.T.P. acknowledges support from the Extragalactic Astrophysics Research Group at E\"{o}tv\"{o}s Lor\'{a}nd University, funded by the E\"{o}tv\"{o}s Lor\'{a}nd Research Network (ELKH), which was used during the development of this research.
N.S.P. was supported by the Vanderbilt Initiative in Data Intensive Astrophysics (VIDA) Fellowship.
S.M.R. and I.H.S. are CIFAR Fellows.
Portions of this work performed at NRL were supported by ONR 6.1 basic research funding.
The Nan\c{c}ay radio Observatory is operated by the Paris Observatory, associated with the French Centre National de la Recherche Scientifique (CNRS), and partially supported by the Region Centre in France. I.C., L.G., and G.T. acknowledges financial support from ``Programme National de Cosmologie and Galaxies'' (PNCG), and ``Programme National Hautes Energies'' (PNHE) funded by CNRS/INSU-IN2P3-INP, CEA and CNES, France. I.C., L.G., and G.T. acknowledges financial support from Agence Nationale de la Recherche (ANR-18-CE31-0015), France.
Pulsar research at Jodrell Bank Centre for Astrophysics is supported by an STFC Consolidated Grant (ST/T000414/1).

\emph{Software:} \texttt{ENTERPRISE} \citep{enterprise}, \texttt{enter- prise\_extensions} \citep{enterprise_extensions}, \texttt{PTMCMC} \citep{ptmcmcsampler}, \texttt{la\_forge} \citep{la_forge}, \texttt{PINT} \citep{pint}, \texttt{numpy} \citep{numpy}, \texttt{matplotlib} \citep{matplotlib}, \texttt{corner} \citep{corner}, \texttt{seaborn} \citep{seaborn}.

\appendix
\section{Single-pulsar Solar Wind Electron Density Estimates}
\label{appendix:SW}

\begin{table*}[ht]
    \vspace{-0.5\baselineskip}
    \centering
    \begin{tabular}{ c | c | c c | c c | c c }
        \hline\hline & & \multicolumn{2}{c|}{\citetalias{NG15} \texttt{DMGP}} & \multicolumn{2}{c|}{\citetalias{NG15} \texttt{CustomGP}} & \multicolumn{2}{c}{\citetalias{EPTA_data}} \\% & \multicolumn{2}{c}{Solar Wind} \\
        PSR & ELAT ($^\circ$) & $n_{\text{Earth}}$ (cm$^{-3}$) & $\log_{10}\mathcal{B}^{\text{SW}}$ & $n_{\text{Earth}}$ (cm$^{-3}$) & $\log_{10}\mathcal{B}^{\text{SW}}$ & $n_{\text{Earth}}$ (cm$^{-3}$) & $\log_{10}\mathcal{B}^{\text{SW}}$ \\
        \hline J0613$-$0200 & -25.4 & $2.7_{-1.3}^{+1.3}$ & $-0.2$ & $1.8_{-1.1}^{+1.3}$ & $-0.8$ & $3.3_{-2.1}^{+2.6}$ & $-0.5$ \\
        J1012+5307 & 38.8 & $8.6_{-4.5}^{+4.9}$ & $0.2$ & $4.7_{-3.2}^{+4.3}$ & $-0.4$ & $6.4_{-1.9}^{+1.9}$ & $1.3$ \\
        J1600$-$3053 & -10.1 & $6.2_{-0.8}^{+0.8}$ & $>3$ & $5.2_{-0.9}^{+0.9}$ & $>3$ & $2.9_{-0.7}^{+0.7}$ & $2.6$ \\
        J1713+0747 & 30.7 & $4.9_{-0.6}^{+0.6}$ & $>3$ & $5.9_{-0.8}^{+0.8}$ & $>3$ & $4.0_{-1.1}^{+1.0}$ & $1.7$ \\
        J1744$-$1134 & 11.8 & $4.1_{-0.5}^{+0.5}$ & $>3$ & $4.2_{-0.5}^{+0.5}$ & $>3$ & $3.2_{-0.7}^{+0.7}$ & $2.5$ \\
        J1909$-$3744 & -15.2 & $3.7_{-0.4}^{+0.4}$ & $>3$ & $3.8_{-0.4}^{+0.4}$ & $>3$ & $3.3_{-0.4}^{+0.4}$ & $>3$ \\
        \hline \hline
    \end{tabular}
    \caption{Comparisons of solar wind electron density parameters and Bayes factors from each pulsar and model/data set combination: \citetalias{NG15} with \texttt{DMGP}, \citetalias{NG15} with \texttt{CustomGP}, and \citetalias{EPTA_data}.
    Parameters are presented using the median and 68.3\% Bayesian credible intervals (referenced here as 1-$\sigma$ regions), and Bayes factors are calculated from our posterior distributions using the Savage-Dickey approximation.}
    %\vspace{-1\baselineskip}
    \label{tab:sw_params}
    \vspace{-1.5\baselineskip}
\end{table*}

We estimate the SW electron density at 1 AU ($n_{\text{Earth}}$) independently for each pulsar using our time-independent, $1/r^2$ density profile SW model.
PTA estimates of $n_{\text{Earth}}$ have been performed more comprehensively elsewhere (e.g., \citealt{Madison2019, Hazboun2022, PPTA_noise}), but we report our own estimates of $n_{\text{Earth}}$ to detail the similarities and differences among the models and data sets considered here.
Table~\ref{tab:sw_params} compares $n_{\text{Earth}}$ medians and 68.3\% credible intervals (1-$\sigma$ regions) as well as log-scaled SW Bayes factors ($\log_{10}\mathcal{B}^{\text{SW}}$) estimated using \citetalias{NG15} (under models $\texttt{DMGP}$ and $\texttt{CustomGP}$) and \citetalias{EPTA_data} for each pulsar.
For reference, we also show the ecliptic latitude (ELAT) value for each pulsar, since the SW introduces a larger DM correction for pulsars closer to the ecliptic.
PSRs J0613$-$0200 and J1012+5307 are relatively far from the ecliptic, and low values of $\log_{10}\mathcal{B}^{\text{SW}}$ indicate the SW is not well-detected using these pulsars. 
Meanwhile, PSR J1713+0747 is more precisely timed and has more TOAs, so it can still constrain the SW despite having ELAT$=30.7^\circ$.

Switching from \texttt{DMGP} to \texttt{CustomGP} results in slight changes to the estimated $n_{\text{Earth}}$ parameters.
These changes are most significant for PSRs J1600$-$3053 and J1713+0747, as the median value of $n_{\text{Earth}}$ estimated under \texttt{CustomGP} lies just outside the 68.3\% credible interval under \texttt{DMGP}.
The $n_{\text{Earth}}$ estimates are fairly consistent using \citetalias{NG15} and \citetalias{EPTA_data}, although for PSRs J1600$-$3053, J1713+0747, J1744$-$1134, and J1909$-$3744, $n_{\text{Earth}}$ estimates are slightly higher using \citetalias{NG15}.
Estimates of $n_{\text{Earth}}$ from \citetalias{NG15} and \citetalias{EPTA_data} are the most different for PSR J1600$-$3053.
However, this difference in PSR J1600$-$3053 could be related to the different chromatic noise properties estimated using \citetalias{NG15} and \citetalias{EPTA_data} (section~\ref{subsec:J1600}).

\section{Comparison of DM time series}
\label{appendix:DM}

\begin{figure*}
    \centering
    \vspace{-0.5\baselineskip}
    \begin{tabularx}{\textwidth}{ll}
        \hspace{-1mm}\includegraphics[width=0.49\linewidth]{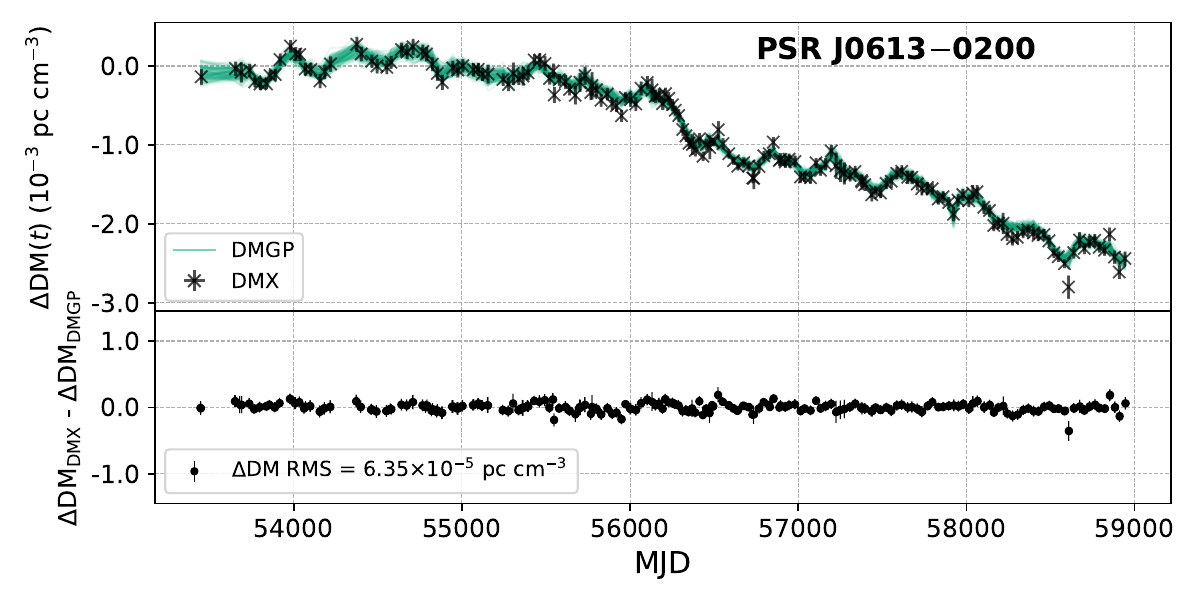} &
        \hspace{-1mm}\includegraphics[width=0.49\linewidth]{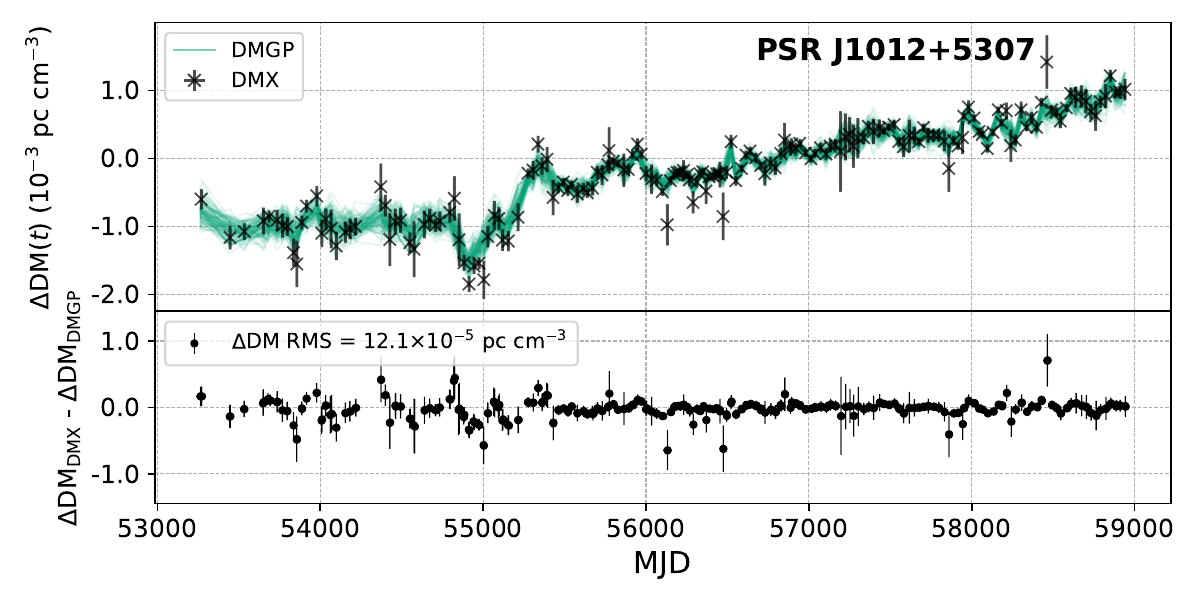} \\
        \hspace{-1mm}\includegraphics[width=0.49\linewidth]{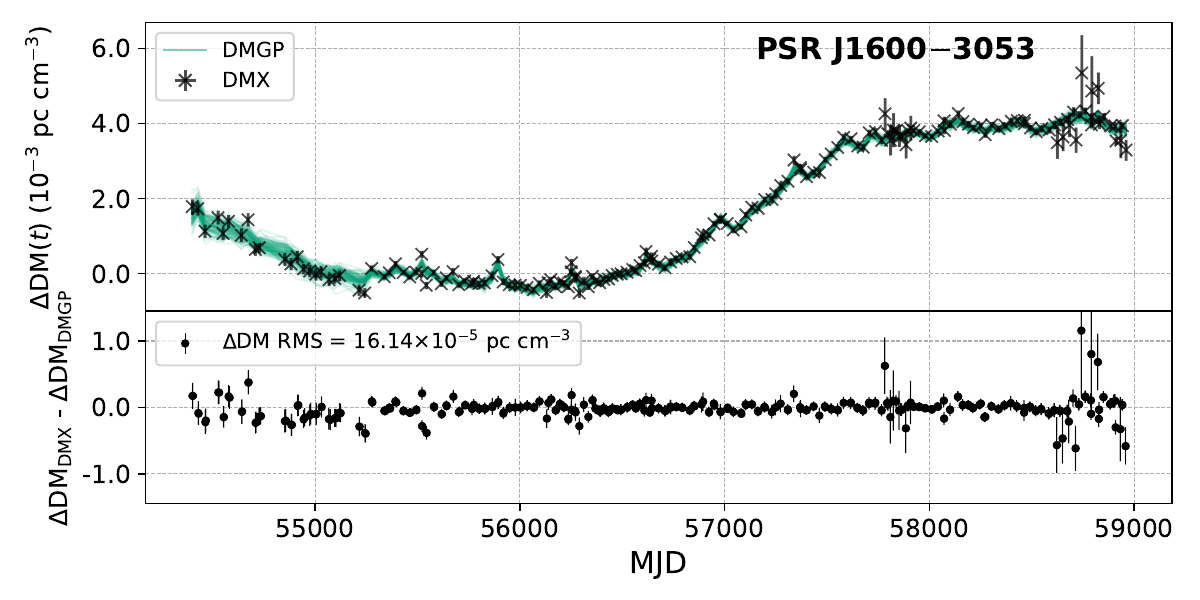} &
        \hspace{-1mm}\includegraphics[width=0.49\linewidth]{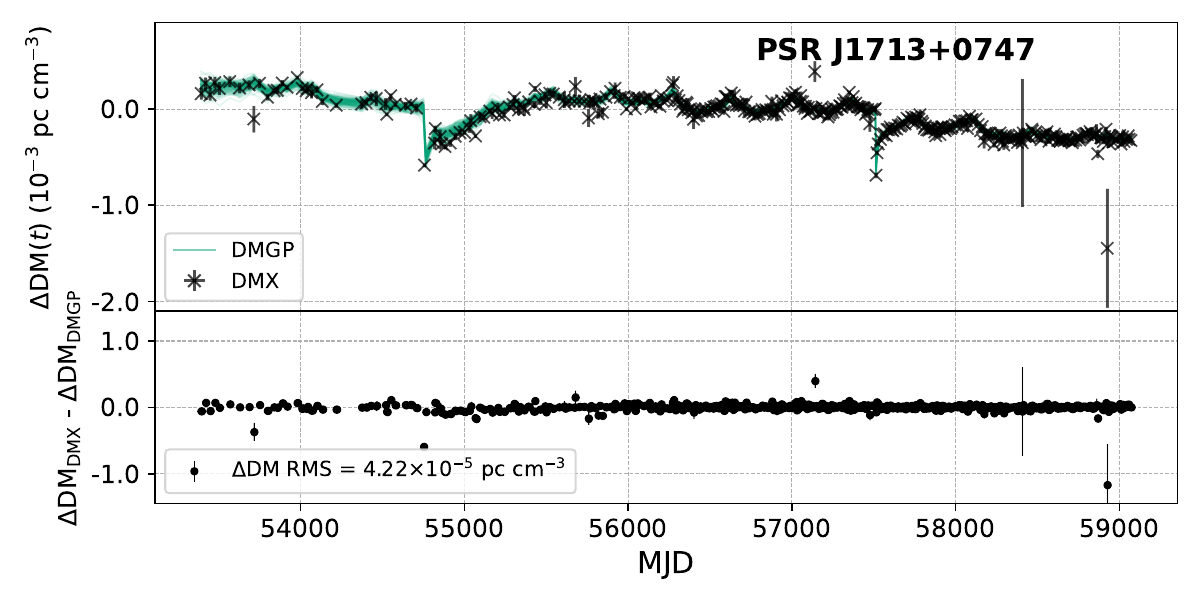} \\
        \hspace{-1mm}\includegraphics[width=0.49\linewidth]{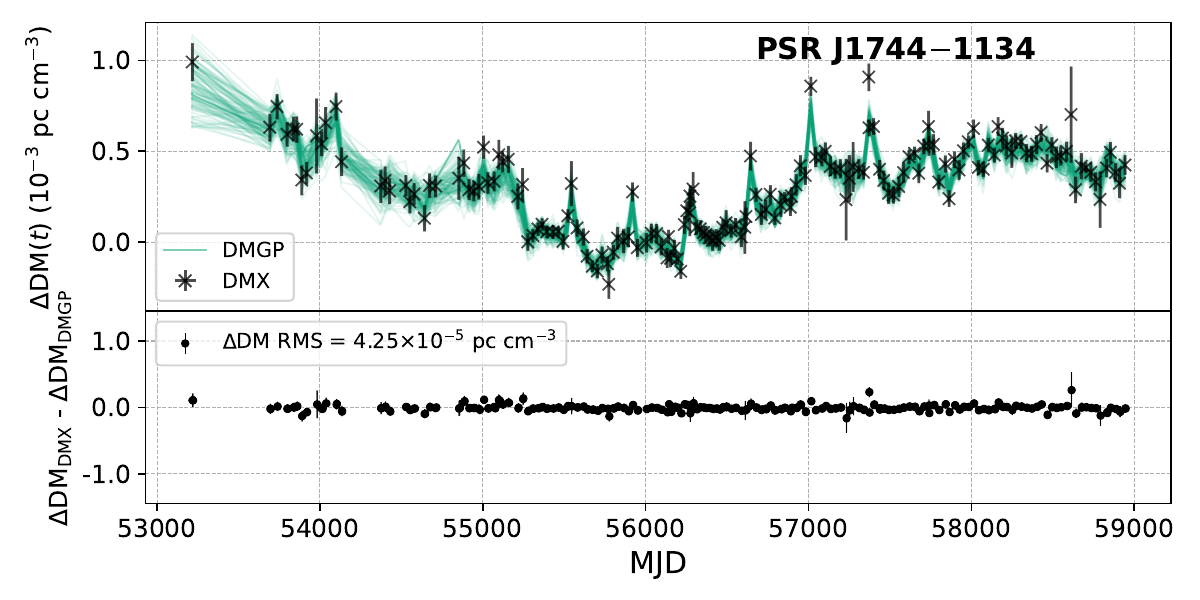} &
        \hspace{-1mm}\includegraphics[width=0.49\linewidth]{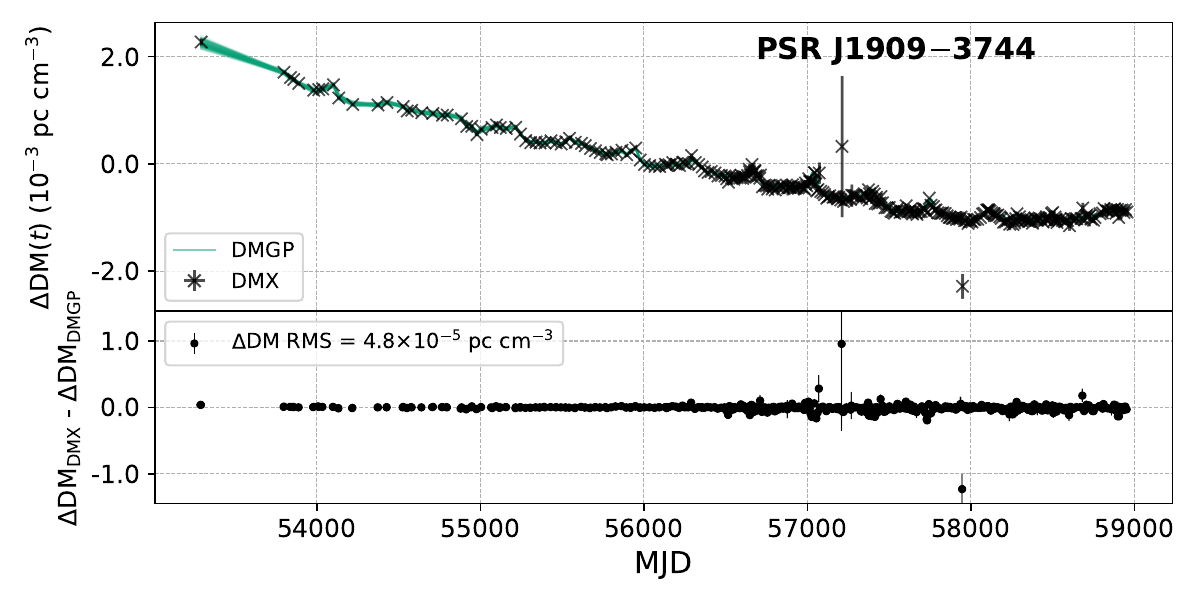}
    \end{tabularx}
    \vspace{-2\baselineskip}
    \caption{Comparison of DM variations ($\Delta$DM) recovery using the two models \texttt{DMX} and \texttt{DMGP}, for each pulsar.
    Both models recover qualitatively similar trends in DM for all pulsars, however the difference in estimated DM values is largest for PSRs J1012+5307 and J1600$-$3053.
    \emph{Top panels:} Time series of DMX parameters (black) superimposed with 100 DM GP realizations (turquoise). 
    \emph{Bottom panels:} Difference in estimated DM over time and root-mean-squared difference between $\Delta$DM values.}
    \label{fig:DM_compare}
    \vspace{-0.5\baselineskip}
\end{figure*}

In Figure~\ref{fig:DM_compare} we compare DM($t$) values as estimated under \texttt{DMX} and \texttt{DMGP} for the six pulsars in \citetalias{NG15}.
The top panels show the estimated deviations to each pulsar's fiducial DM value over time, in units of $10^{-3}$ pc cm$^{-3}$.
These are represented by the time series of DMX parameters using the \texttt{DMX} model, and 100 GP realizations of DM variations using the \texttt{DMGP} model.
Qualitatively, both DM models produce similar trends for all pulsars, especially once transient events in PSR J1713+0747 are accounted for using \texttt{DMGP} \citep{Lam2018, Wang2019, Hazboun2020}.

The bottom panels of Figure~\ref{fig:DM_compare} show the difference between the medians of each estimated DM value over time, $\Delta$DM.
These are calculated by subtracting the medians of the time-domain DM realizations from the DMX time series.
We report the root-mean-square $\Delta$DM for each pulsar.
We find $\Delta\text{DM RMS} > 10^{-4}$ pc cm$^{-3}$ for PSRs J1012+5307 and J1600$-$3053, while $\Delta\text{DM RMS} < 10^{-4}$ pc cm$^{-3}$ for the remaining pulsars.
DM estimation errors are known to induce deviations to TOAs at infinite radio-frequency \citep{CordesShannon2010, Lam2015}, therefore the larger $\Delta$DM RMS values in PSRs J1012+5307 and J1600$-$3053 help explain why only only these two pulsars have significantly different ARN parameters using \texttt{DMX} and \texttt{DMGP} (Table~\ref{tab:noise_params}, Figure~\ref{fig:red_noise}).
There are also some subtle time-correlations in each pulsar's $\Delta$DM, which may be useful to study in future analyses.

\section{Achromatic Red Noise Free Spectra}
\label{appendix:FS}

\begin{figure*}
    \centering
    \vspace{-0.5\baselineskip}
    \begin{tabularx}{\textwidth}{ll}
        \hspace{-1mm}\includegraphics[width=0.49\linewidth]{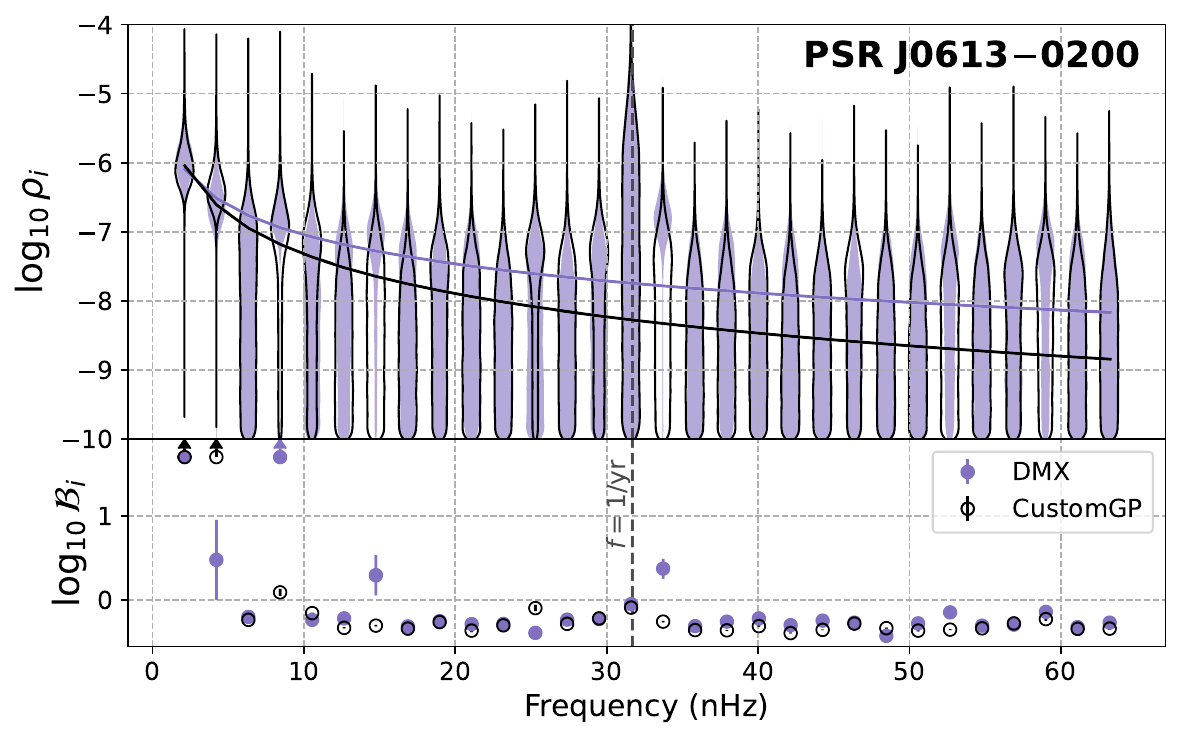} &
        \hspace{-1mm}\includegraphics[width=0.49\linewidth]{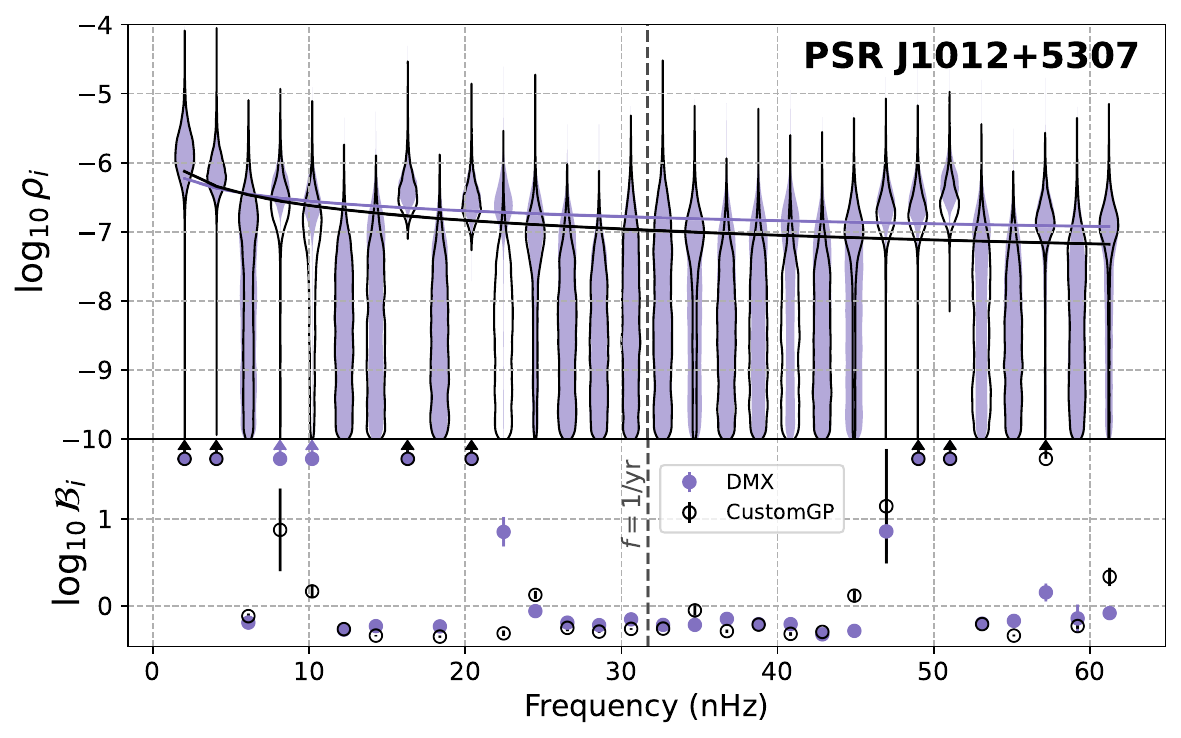} \\
        \hspace{-1mm}\includegraphics[width=0.49\linewidth]{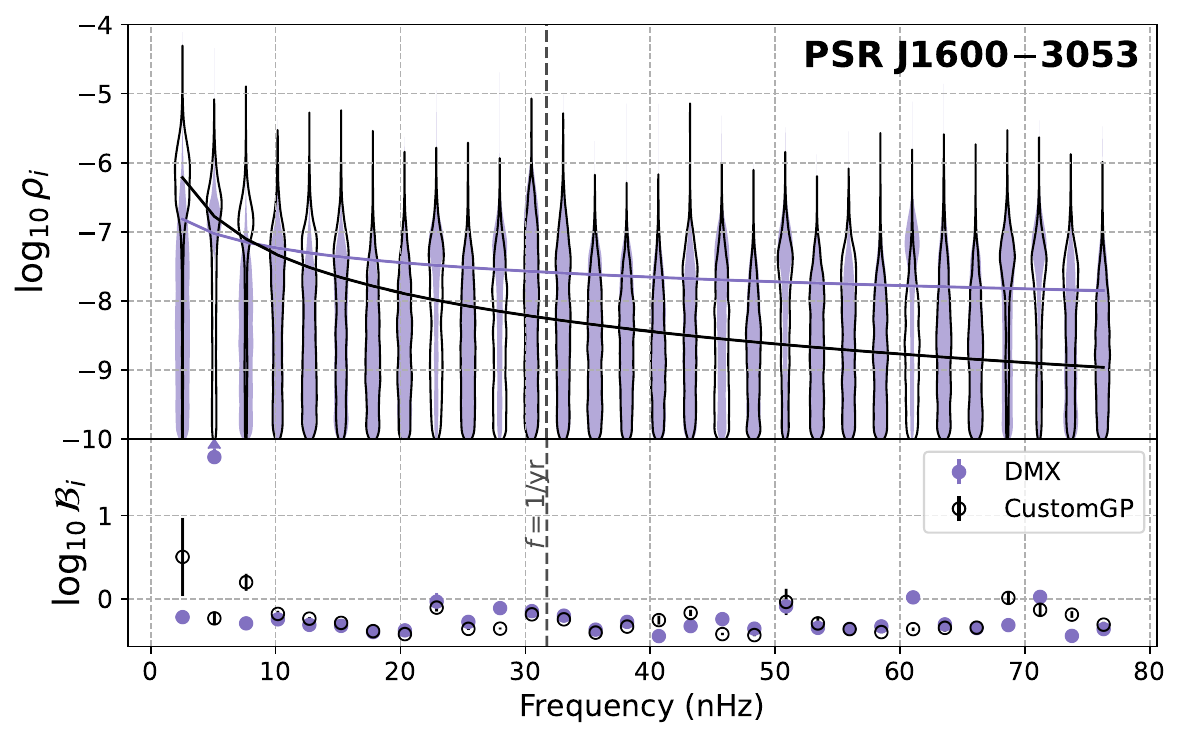} &
        \hspace{-1mm}\includegraphics[width=0.49\linewidth]{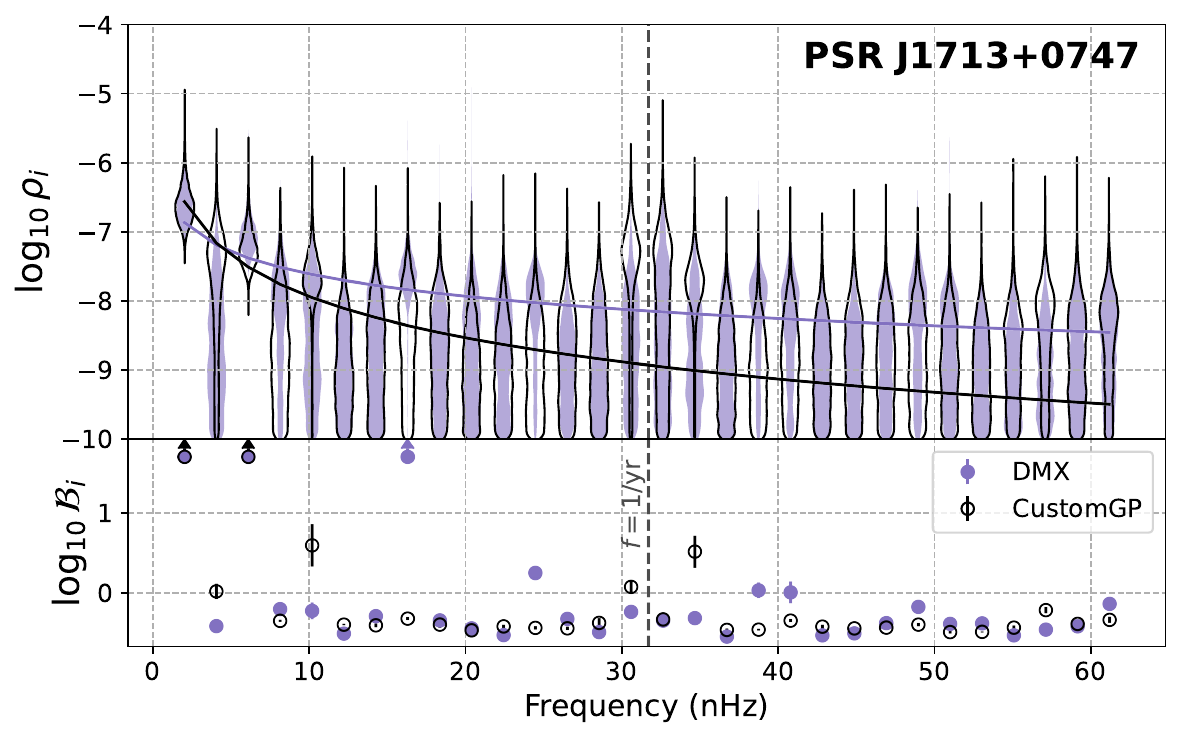} \\
        \hspace{-1mm}\includegraphics[width=0.49\linewidth]{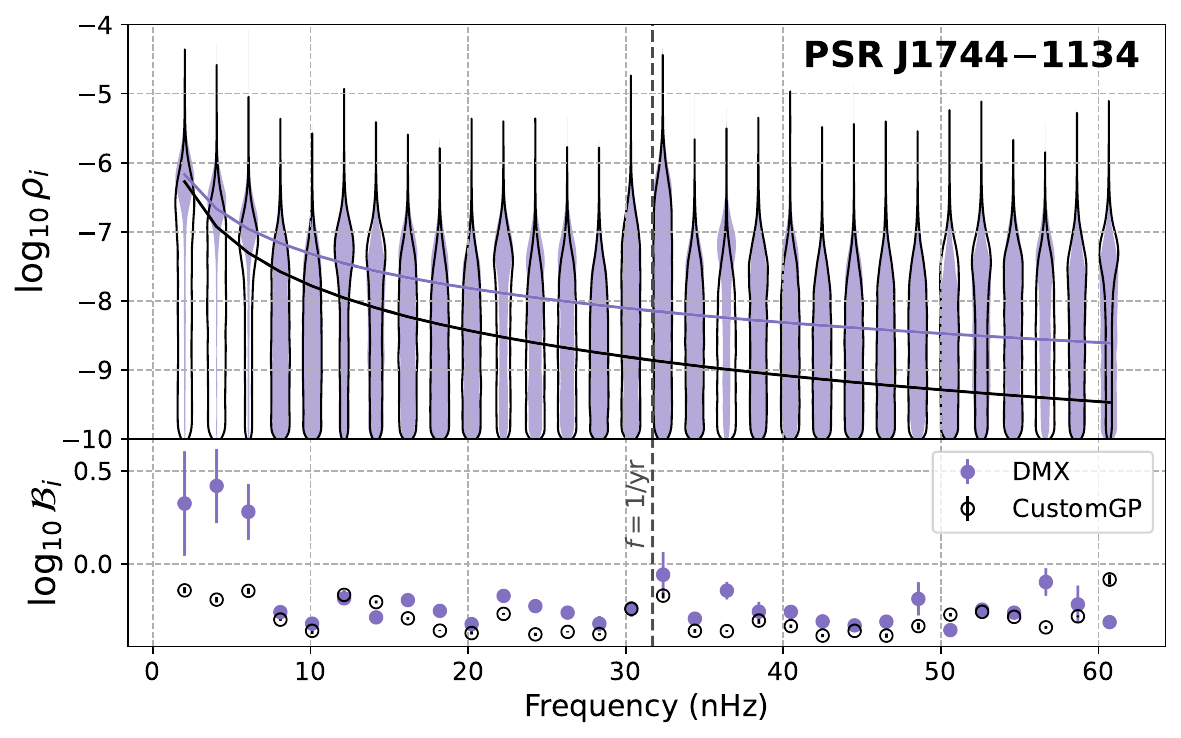} &
        \hspace{-1mm}\includegraphics[width=0.49\linewidth]{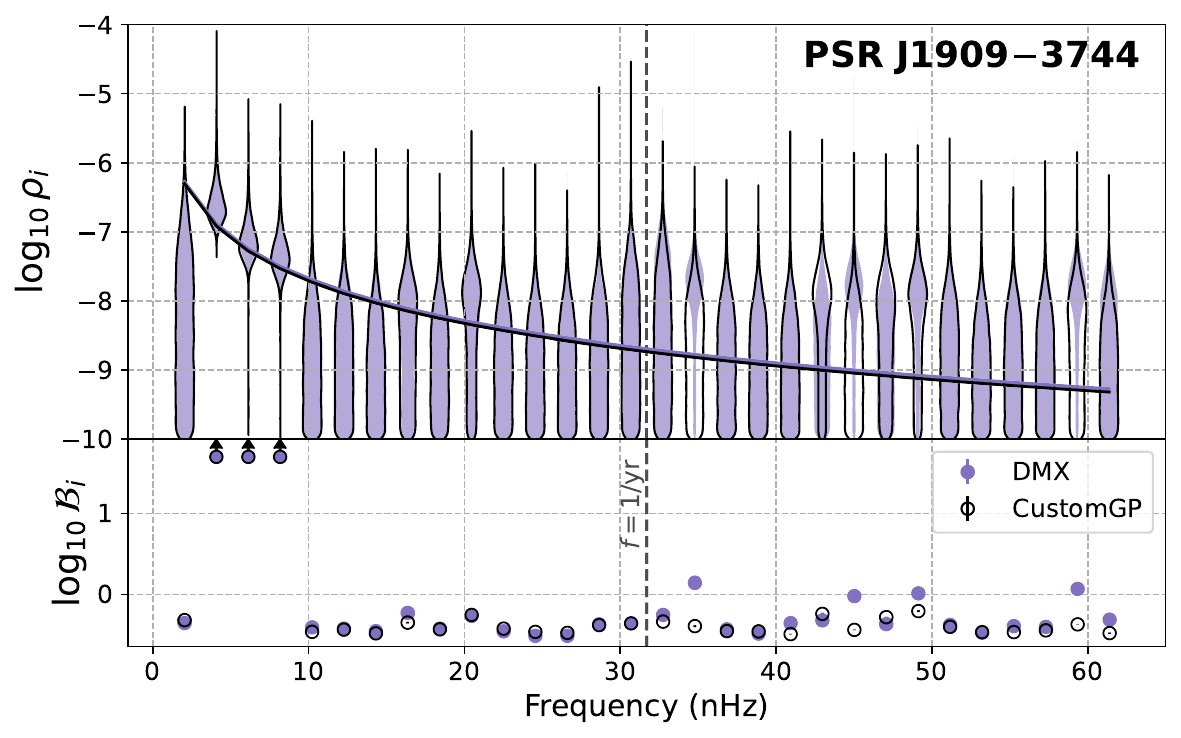}
    \end{tabularx}
    \vspace{-1\baselineskip}
    \caption{Power spectra and Bayes Factors using a free-spectral PSD for each pulsar, under the two models: \texttt{DMX} and \texttt{CustomGP}.
    Top panels show the posteriors for the log-scaled residual power $\log_{10}\rho_i$ at each frequency, in units of excess timing delay.
    Bottom panels show log-scaled Bayes factors $\log_{10}\mathcal{B}_i$ indicating the statistical significance of excess power in each frequency bin.
    For each pulsar, changes to $\rho_i$ are isolated to a few key frequencies, but still influence the inferred power law noise parameters (Figure~\ref{fig:red_noise}).}
    \label{fig:FS}
    \vspace{-0.5\baselineskip}
\end{figure*}

We generate Bayesian power spectra for each pulsar using a free-spectral PSD, which allows the power in each frequency bin to vary as a free parameter \citep{Lentati2013}.
In Figure~\ref{fig:FS} we compare the power spectra for all six pulsars using \texttt{DMX} and \texttt{CustomGP}.
The top panels show the posteriors for the log-scaled RMS timing residual power $\log_{10}\rho_i$ at each frequency $f_i$, alongside a power law using the maximum likelihood values of $\log_{10}A_{\text{RN}}$ and $\gamma_{\text{RN}}$ from each model.
The bottom panels show the log-scaled Bayes factors $\log_{10}\mathcal{B}_i$ for the presence of excess power in each frequency bin, measured using the Savage-Dickey density ratio \citep{Dickey1971}.
These may be interpreted as a measure of the consistency of each $\rho_i$ with zero excess power.
In cases where there are a lack of samples consistent with $\log_{10}\rho_i = -10$, we set a lower limit of $\log_{10}\mathcal{B} > 1.7$.

The choice of chromatic noise model influences the free spectra of all six pulsars.
These changes are most interesting for PSRs J1713+0747 and J1909$-$3744 since they each have the lowest residual RMS power overall.
For PSR J1713+0747, the changes to $\log_{10}\mathcal{B}_i$ indicate the change to $\rho_8$ (the 8th frequency bin) is most consequential for the changes to this pulsar's ARN properties.
For PSR J1909$-$3744, the free-spectra inferred using both models are nearly identical at low frequencies below $f = 1/\text{yr}$; above $f = 1/\text{yr}$, \texttt{CustomGP} reduces the RMS power at a few frequencies.

% What happens when you try to use DM and red noise to model a pulsar affected by DM and scattering $\chi > 2$ noise?
\section{Systematic errors from unmodeled chromatic effects}
\label{appendix:chromatic_modeling}

We demonstrate how unmodeled chromatic effects may bias estimates of ARN and DM variations in a simplified analytic case (see also \citealt{Lentati2016, Lam2020, Sosa2024}).
For simplicity, we will assume a TOA is measured only at two radio frequencies, $\nu_0$ and $\nu_1$, where $\nu_1 > \nu_0$.
Let us define the true frequency-dependent timing delay $\Delta t(\nu)$ from ARN, DM, and scattering as
\begin{eqnarray*}
    \Delta t(\nu) &=& \Delta t_{\text{ARN}} + \Delta t_{\text{DM}}\Tilde{\nu}^{-2} + \Delta t_{\text{chrom}}\Tilde{\nu}^{-\chi},
\end{eqnarray*}
where $\Tilde{\nu}$ is a dimensionless frequency scaled to some reference frequency $\nu_{\text{ref}}$, $\Delta t_{\text{ARN}}$ is the delay from achromatic processes (i.e., spin noise or GWs), $\Delta t_{\text{DM}}$ is the delay due to DM at the reference frequency $\nu_{\text{ref}}$, $\Delta t_{\text{chrom}}$ is the delay due to scattering at the reference frequency $\nu_{\text{ref}}$, and $\chi$ is the scattering scaling index.
We may assume that $\chi = 4$, however this calculation works for any $\chi > 2$.
We will assume our two frequencies $\nu_0$ and $\nu_1$ are widely separated, such that $\Delta t_{\text{ARN}}$ is small (in comparison to chromatic errors) at $\nu_0$,
\begin{eqnarray*}
    \Delta t(\nu_0) &=& \Delta t_{\text{ARN}} + \Delta t_{\text{DM}}\Tilde{\nu}_0^{-2} + \Delta t_{\text{chrom}}\Tilde{\nu}_0^{-\chi} \\
    &\cong& \Delta t_{\text{DM}}\Tilde{\nu}_0^{-2} + \Delta t_{\text{chrom}}\Tilde{\nu}_0^{-\chi},
\end{eqnarray*}
and $\Delta t_{\text{chrom}}$ is small at $\nu_1$,
\begin{eqnarray*}
    \Delta t(\nu_1) &=& \Delta t_{\text{ARN}} + \Delta t_{\text{DM}}\Tilde{\nu}_1^{-2} + \Delta t_{\text{chrom}}\Tilde{\nu}_1^{-\chi} \\
    &\cong& \Delta t_{\text{ARN}} + \Delta t_{\text{DM}}\Tilde{\nu}_1^{-2}.
\end{eqnarray*}
Now we estimate the time delay by modeling it as the sum of only a DM and an achromatic process.
We will define the total estimated delay as
\begin{eqnarray*}
    \delta t(\nu) &=& \delta t_{\text{ARN}} + \delta t_{\text{DM}}\Tilde{\nu}^{-2},
\end{eqnarray*}
where $\delta t_{\text{ARN}}$ and $\delta t_{\text{DM}}$ are the estimated ARN and DM delays.
Again we will write this down at our two frequencies $\nu_0$ and $\nu_1$, assuming $\delta t_{\text{ARN}}$ is small at $\nu_0$,
\begin{eqnarray*}
    \delta t(\nu_0) &=& \delta t_{\text{ARN}} + \delta t_{\text{DM}}\Tilde{\nu}_0^{-2} \\
    &\cong& \delta t_{\text{DM}}\Tilde{\nu}_0^{-2},
\end{eqnarray*}
while at $\nu_1$ we have exactly
\begin{eqnarray*}
    \delta t(\nu_1) &=& \delta t_{\text{ARN}} + \delta t_{\text{DM}}\Tilde{\nu}_1^{-2}.
\end{eqnarray*}
If we have only measured the TOA at $\nu_0$ and $\nu_1$, then our two model parameters $\delta t_{\text{DM}}$ and $\delta t_{\text{RN}}$ can perfectly fit our data such that $\delta t(\nu_0) = \Delta t(\nu_0)$ and $\delta t(\nu_1) = \Delta t(\nu_1)$.
We can then use the measurement at $\nu_0$ to determine how the estimated DM delay $\delta t_{\text{DM}}$ relates to the true delays,
\begin{eqnarray*}
    \delta t(\nu_0) &=& \Delta t(\nu_0), \\
    \delta t_{\text{DM}}\Tilde{\nu}_0^{-2} &=& \Delta t_{\text{DM}}\Tilde{\nu}_0^{-2} + \Delta t_{\text{chrom}}\Tilde{\nu}_0^{-\chi}, \\
    \delta t_{\text{DM}} &=& \Delta t_{\text{DM}} + \Delta t_{\text{chrom}}\Tilde{\nu}_0^{-(\chi-2)}.
\end{eqnarray*}
Meanwhile, we can use the measurement at $\nu_1$ to determine how the estimated achromatic delay relates to the true delays,
\begin{eqnarray*}
    \delta t(\nu_1) &=& \Delta t(\nu_1), \\
    \delta t_{\text{ARN}} + \delta t_{\text{DM}}\Tilde{\nu}_1^{-2} &=& \Delta t_{\text{ARN}} + \Delta t_{\text{DM}}\Tilde{\nu}_1^{-2}, \\
    \delta t_{\text{ARN}} &=& \Delta t_{\text{ARN}} + \Delta t_{\text{DM}}\Tilde{\nu}_1^{-2} - (\Delta t_{\text{DM}} + \Delta t_{\text{chrom}}\Tilde{\nu}_0^{-(\chi-2)})\Tilde{\nu}_1^{-2} \\
    &=& \Delta t_{\text{ARN}} - \Delta t_{\text{chrom}}\Tilde{\nu}_0^{-(\chi-2)}\Tilde{\nu}_1^{-2}.
\end{eqnarray*}
This shows if there is an unmodeled scattering delay with $\chi > 2$, the DM will be shifted by $\Delta t_{\text{chrom}}\Tilde{\nu}_0^{-(\chi-2)}$, while the achromatic delay will be shifted by $-\Delta t_{\text{chrom}}\Tilde{\nu}_0^{-(\chi-2)}\Tilde{\nu}_1^{-2}$.
In other words, unmodeled scattering variations may manifest as an excess ARN process proportional to the true scattering-induced delay with an opposite sign. 
Additionally, the excess achromatic delay will tend to zero as $\Tilde{\nu}_{1}$ becomes very large. 
While this result is based on a simplified model of the TOA with measurements at only two radio frequencies, it is possible this effect may still arise in real data.
A more rigorous quantification will be left for future work.

One can also perform a similar calculation in the case of an unmodeled chromatic process with $0 < \chi < 2$.
The result is the DM and achromatic delays will each be overestimated as:
\begin{eqnarray*}
    \delta t_{\text{DM}} &=& \Delta t_{\text{DM}} + \Delta t_{\text{chrom}}\Tilde{\nu}_0^{2-\chi}, \\
    \delta t_{\text{ARN}} &=& \Delta t_{\text{ARN}} + \Delta t_{\text{chrom}}\Tilde{\nu}_1^{-2}\left(\Tilde{\nu}_1^{2-\chi} - \Tilde{\nu}_0^{2-\chi}\right).
\end{eqnarray*}
The difference $\Tilde{\nu}_1^{2-\chi} - \Tilde{\nu}_0^{2-\chi}$ will always be positive for $\chi < 2$.
For a hypothetical chromatic process with $\chi < 0$, 
\begin{eqnarray*}
    \delta t_{\text{DM}} &=& \Delta t_{\text{DM}} - \Delta t_{\text{chrom}}\Tilde{\nu}_1^{-\chi}\Tilde{\nu}_0^{2}, \\
    \delta t_{\text{ARN}} &=& \Delta t_{\text{ARN}} + \Delta t_{\text{chrom}}\Tilde{\nu}_1^{-\chi}.
\end{eqnarray*}
This shows that an unmodeled chromatic process with $0 < \chi < 2$ may manifest as excess DM and achromatic delays with the same sign, whereas if $\chi < 0$ then the excess DM and achromatic delays will again have opposite sign.

\bibliography{refs.bib}

\end{document}